\documentclass{aa}

\usepackage{graphicx}
\usepackage[dvipsnames]{xcolor, colortbl}
\usepackage{xkcdcolors}
\usepackage{tikz}
\usepackage{siunitx}
\usepackage{txfonts}
\usepackage{natbib}
\usepackage{orcidlink}

\bibpunct{(}{)}{;}{a}{}{,} 

\usepackage{hyperref}
\hypersetup{colorlinks, allcolors=blue}
\usepackage{subcaption} 

\begin{document}

   \title{The velocity field of the Scorpius-Centaurus OB association}

   \subtitle{I. Method and general properties}

   \author{S.~Hutschenreuter\inst{1}\orcidlink{0000-0002-6952-9688}
          \and
          J.~Alves \inst{1}
          \and
          L.~Posch \inst{1}
          \and
          J.~Gro\ss schedl\inst{2}
          \and 
          M.~Piecka \inst{1}
          \and
          N.~Miret-Roig \inst{3,4}
          \and 
          S.~Ratzenb\"ock \inst{5}
          \and
          C.~Swiggum \inst{1}
          }

   \institute{University of Vienna, Department of Astrophysics, T\"urkenschanzstra{\ss}e 17, 1180 Vienna, Austria
              \email{sebastian.hutschenreuter@univie.ac.at}
              \and
               Astronomical Institute of the Czech Academy of Sciences, Boční II 1401, 141 31 Prague 4, Czech Republic  
               \and Departament de Física Quàntica i Astrofísica (FQA), Universitat de Barcelona (UB), Martí i Franquès, 1, 08028 Barcelona, Spain
                \and
                Institut de Ciències del Cosmos (ICCUB), Universitat de Barcelona (UB), Martí i Franquès, 1, 08028 Barcelona, Spain
               \and Center for Astrophysics | Harvard \& Smithsonian, 60 Garden St., Cambridge, MA 02138, USA
             }

   \date{}

  \abstract{
We present a non-parametric reconstruction of the three-dimensional velocity field of the Scorpius-Centaurus OB association (Sco-Cen). 
Using Gaia DR3 astrometry and radial velocities, we infer the velocity field using information field theory on a $70 \times 70 \times 50$ grid at 3 pc resolution. 
Our model suggests the existence of a primary stellar velocity field with a secondary field that accounts for an additional young kinematic component in Upper Scorpius and Lupus. 
We find clear tracers of a feedback-driven expansion of the association, while Galactic rotation appears to play a subordinate role. 
The results confirm the existence of cluster chains and reveal coherent large-scale expansion with characteristic speeds of 1--2 km\,s$^{-1}$ and local maxima of about 10 km\,s$^{-1}$. 
Power spectra indicate an excess of small-scale structure and slopes shallower than Kolmogorov, consistent with energy injection from stellar feedback. 
Maps of the divergence reveal net positive values, implying an approximate dispersal timescale of 10--15 Myr. 
A comparison with molecular gas in Lupus and Ophiuchus shows broadly consistent patterns but systematic velocity offsets of several km\,s$^{-1}$, suggesting partial decoupling for optically visible young stars and gas. 
The framework presented provides a physically motivated description of the Sco-Cen velocity field and a basis for quantifying the dynamical state and feedback history of OB associations in the local Galaxy.}


   \maketitle
\defcitealias{2023Ratzenboeck1}{R23a}
\defcitealias{2023Ratzenboeck2}{R23b}
\defcitealias{2025MiretRoig}{MR25}
\defcitealias{2023Posch}{P23}
\defcitealias{2025Posch}{P25}
\defcitealias{2025Grossschedl}{G25}

\section{Introduction}
\label{sec:intro}

The three-dimensional (3D) velocity structure of the interstellar medium (ISM) is a crucial element in understanding the Milky Way's dynamics. 
Its importance stems from its tight coupling to both the gravitational (via the Boltzmann equation) and magnetic forces (via the MHD equations, cf. flux freezing). 
Hence, the velocity field(s) of gas, dust, and plasma encode the history and future development of the ISM.
Furthermore, these interactions, along with cooling and heating mechanisms, determine the typical lifetimes of structures in the ISM in the order of $10^7$ years \citep{2007McKee, 2023Chevance} and hence provide a natural limit to the inference of the history of the ISM from its own dynamics.
Observationally, determining the 3D velocities of any component of the ISM is notoriously difficult, as tangential velocities are mostly inaccessible to direct measurement (with notable exceptions \citep{2025Piecka}), whereas radial velocities (RVs) are only available for cooler ISM components via Doppler shifts of spectral lines. 
Thus, the dynamical state and history of the ISM remain largely inaccessible directly due to theoretical and observational limits, apart from special short-lived scenarios such as supernova remnants \citep[e.g.,][]{1999ATruelove}. 

Young stars offer a promising solution to this problem for two reasons. 
Firstly, their 3D motion can be measured reliably with spectroscopic and astrometric methods.
Furthermore, their initial movement through space depends largely on the conditions of the interstellar medium (ISM) in which they formed, and they only slowly disperse via gravitational interactions \citep[e.g.,][]{2023Zucker}. 
In the case of very young stellar populations, this may allow for the inference of the present-day cloud motions, as demonstrated, for example, in \citet{2021Grossschedl}. 
But even as they age, the relatively slow spatial dispersion allows inference of the past location of molecular clouds up to several hundreds of Myr \citep{2024Swiggum} with a theoretical lower limit at 100~Myr \citep{2025Arunima}.

To connect the dynamics of stars and the ISM, it is necessary to develop a representation of the stellar kinematics that efficiently encodes its spatial distribution and temporal evolution.   
Traditionally, such analysis is often performed on the cluster level \citep[e.g.,][]{2023Hunt}, i.e. stellar positions and motions are used to identify distinct groups in the position-velocity phase diagram and then averaged for these clusters, with the idea that these bulk motions capture the original motion of the clusters at their formation, and hence can be used reliably for trace-back calculations and isochronal age determination.

In this work, we will take a different point of view and work on the (vector) field level\footnote{We note that this refers to the physical field concept, and not to the common term `field stars'}, meaning that we view the stars as `particles' constituting a flow which can be treated in the context of fluid dynamics.  
This has the advantage that we can relate this field directly to ISM-related quantities such as energy and momentum densities, and thereby contextualize our results in the picture of the ISM life-cycle in our ever-evolving Galaxy. 
Furthermore, the field picture allows for an efficient representation of spatial correlations also on small scales, which might get lost in the cluster representation. 
The value of velocity fields has already been exemplified in cosmology, where the field serves as an important tracer of forces and structure formation processes \citep[e.g.,][]{2025Stiskalek}.   
We make use of Information Field Theory (IFT) \citep{2019Ensslin}, a statistical framework that has yielded results on many ISM-related quantities and is especially suited for large non-parametric inference problems and noisy data, as, for example, demonstrated in  
\citet{2019Leike, 2020Leike, 2022Hutschenreuter, 2023Edenhofer, 2023Scheel-Platz, 2024Hutschenreuter, 2024Westerkamp, 2025Soeding}. 

Of particular interest for the dynamics of the ISM are large complexes of star-forming regions, which can dominate the ISM structure destruction/formation in their immediate surroundings via feedback from supernovae, stellar radiation, and stellar winds \citep{1992deGeus,2018Krause,2025Alves}. By mapping the stellar velocity field, we aim to establish a connection between young stars and their surrounding interstellar medium and to place tighter constraints on the history and future of a large star-forming region.

In this study, we investigate the velocity field of young stars within the Scorpius–Centaurus OB association (Sco-Cen). This nearby and massive star-forming complex provides an important laboratory for studying the interaction between stellar populations and the interstellar medium (ISM). 
Sco-Cen is currently in an advanced stage of its evolution. 
Most of its original molecular gas has been consumed or dispersed, with ongoing star formation limited to a few residual regions \citep{2025Alves}. 
Over the past $\sim 20$~Myr, the region has experienced multiple episodes of star formation and supernova activity, events that likely contributed to the formation and shaping of the Local Bubble \citep{2006Fuchs, 2016Breitschwerdt, 2022Zucker}. 
The influence of Sco-Cen extends well beyond its immediate environment, with flows of material reaching as far as the Solar System and beyond \citep{1995Frisch, 2008Redfield, 2024Piecka}.

Given its proximity and richness in young stars, Sco-Cen has been extensively studied \citep{1914Kapteyn, 1946Blaauw, 1964Blaauw, 1989deGeus, 1992deGeus, 1999deBruijne, 2007MakarovA, 2007AMakarovB, 2010Poeppel, 2016Pecaut, 2018Krause, 2018Wright, 2019Damiani,  2021Forbes, luhman_census_2022, 2025Armstrong}. 
Its stellar population exhibits clear evidence of sequential star formation. 
The oldest subgroups, particularly in the Upper Centaurus–Lupus (UCL) region, have ages of approximately 20~Myr \citep[][hereafter \citetalias{2023Ratzenboeck2}]{2023Ratzenboeck2}. 
From UCL, stellar age gradients suggest that star formation propagated outward along several filamentary structures or “cluster chains”. These are extending toward the regions of Corona Australis (CrA) \citep[][hereafter \citetalias{2023Posch}]{2023Posch}, Upper Scorpius (USco), Ophiuchus, Lower Centaurus Crux (LCC) \citep[][hereafter \citetalias{2025Posch}]{2025Posch}, and the TW Hydrae Association (TWA) \citep[][hereafter \citetalias{2025MiretRoig}]{2025MiretRoig}; see also \citet{2022Miret-Roig} and \citet[hereafter \citetalias{2025Grossschedl}]{2025Grossschedl}. 

The ISM structure of Sco-Cen has also been the subject of extensive investigation, covering the cold and dense phases \citep{1989Loren, 1993Harju, 2021Zucker, 2024Edenhofer}, diffuse atomic and ionized components \citep{2008Nehme, 2018Krause}, and magnetic field morphology \citep{2018Robitaille}. 
These studies point to a dynamic and possibly turbulent past, but the sequence of events leading to the present-day configuration remains uncertain.

Together, stellar and ISM studies emphasize the importance of Sco-Cen in understanding star formation and feedback processes in the Galactic environment. 
By mapping the velocity field of its young stellar population, we aim to constrain the dynamical evolution of the association and its interaction with the surrounding ISM. 
This analysis provides new insight into the star formation history of Sco-Cen and its broader role in shaping the local ISM structure.
 
Given the range of topics addressed and the novelty of our approach to stellar population analysis, we present this work in three parts: 
(1) a description of the methodology and general properties of the reconstructed velocity field (Paper I, this paper); 
(2) an analysis of energy densities and momentum maps (Paper II); 
and (3) a study of the internal rotation of stellar clusters (Paper III).

\section{Data}
\label{sec:data}

\begin{figure*}
\centering
\begin{tikzpicture} 
    \node at (0,0) {\includegraphics[trim=45pt 45pt 45pt 45pt, clip, width=0.3\linewidth]{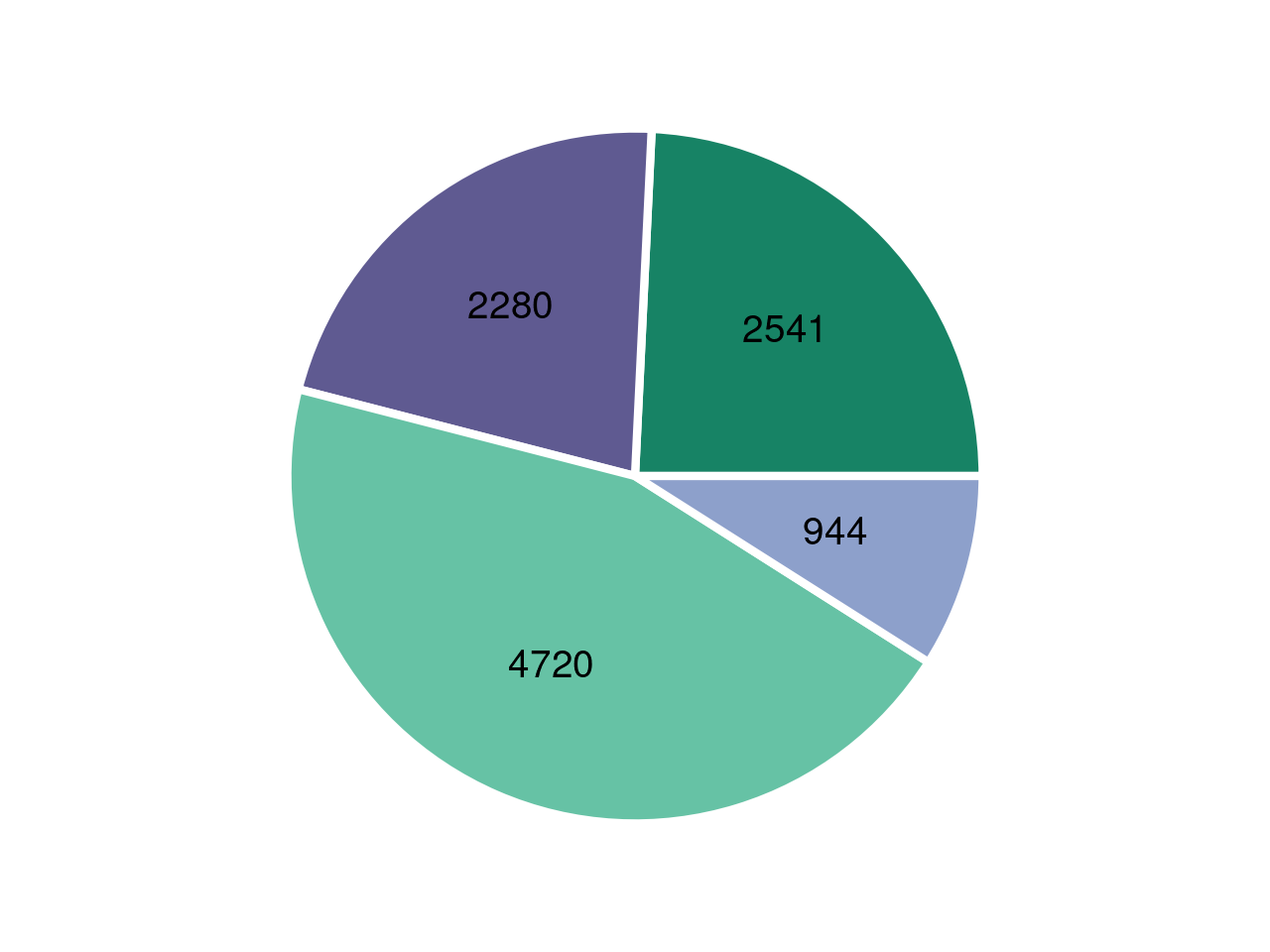}};
    \node at (10,0) {\includegraphics[trim=45pt 45pt 45pt 45pt, clip, width=0.3\linewidth]{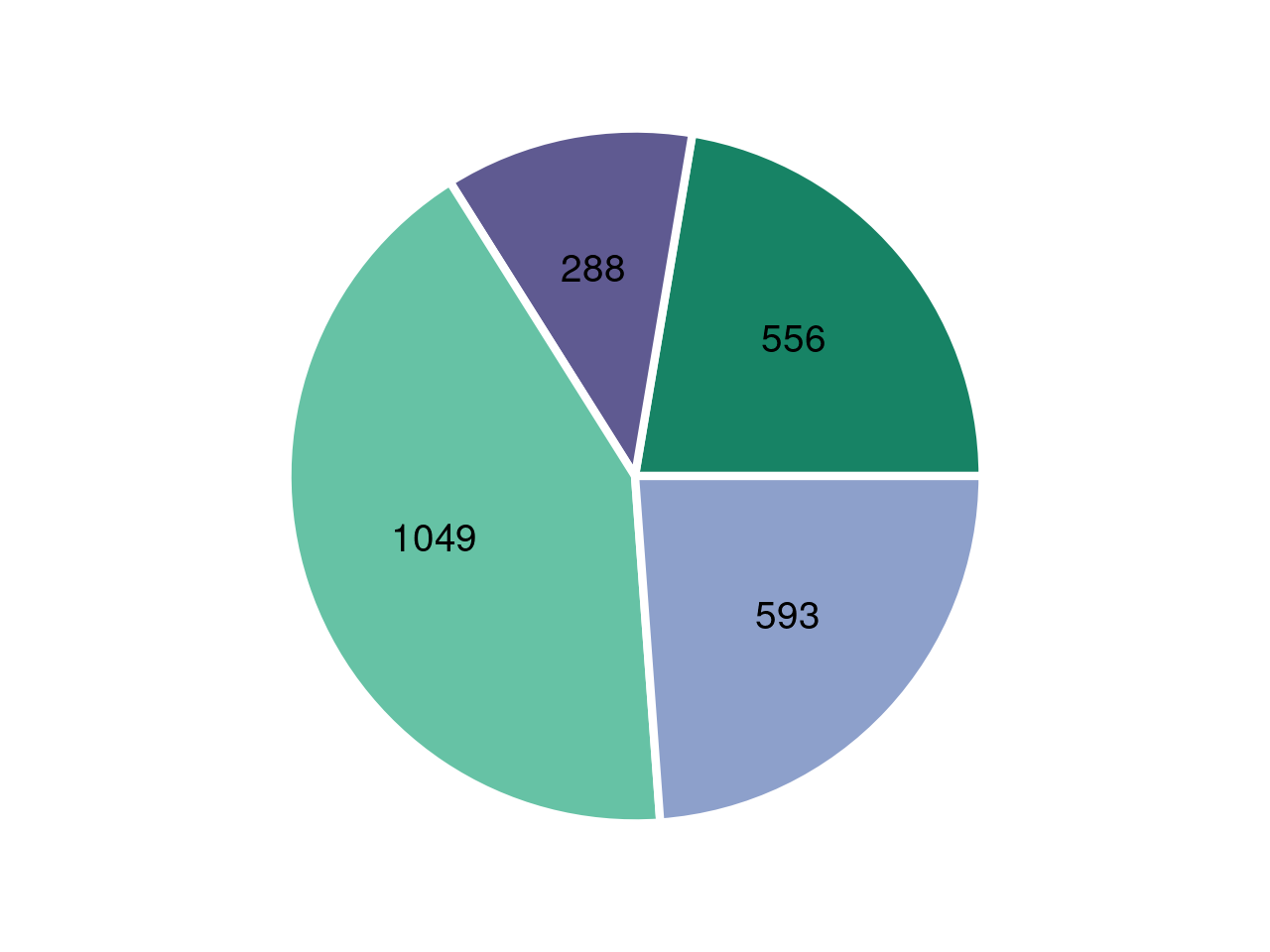}};
    \node at (5, 0) {\includegraphics[width=0.25\linewidth]{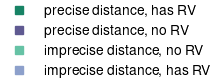}}; 
    \node at (-3.5,2.5) {\underline{$\texttt{main}$ selection}}; 
    \node at (-3.8,2.13) {10485 stars}; 
    \node at (6,2.5) {\underline{$\texttt{secondary}$ selection}}; 
    \node at (5.4,2.13) {2487 stars}; 
\end{tikzpicture}
\caption{Illustration of the different categories within our stellar selection. \label{fig:data-pie}}
\end{figure*}

\subsection{Stellar selection, proper motions  \& distances}

A key requirement for our method is a reliable selection of Sco-Cen members. 
Since field stars are not expected to share the association’s motion, strong contamination would bias results and compromise post-processing. 
Membership classification was provided by the machine-learning clustering algorithm Significant Mode Analysis \citep[\texttt{SigMA},][hereafter \citetalias{2023Ratzenboeck1}]{2023Ratzenboeck1}, which detects stellar populations as significant over-densities in 5D phase space ($x, y, z, v_{\alpha}, v_{\delta}$). 
It applies hierarchical, density-based clustering with a modality test to decide whether density peaks should be merged or retained.

\texttt{SigMA} uses a non-parametric density estimator without explicitly treating heteroscedastic errors. 
Instead, low-S/N astrometry is removed before clustering, including a strict parallax S/N cut ($\varpi/\sigma_{\varpi} > 4.5$). 
Systematically unreliable parallaxes are filtered with the \verb|fidelity_v2| parameter from \cite{2022Rybizki}, reducing the sample from $\sim5.5$ million to $\sim980\,000$ {\it Gaia} DR3 sources in the Sco-Cen volume. 
From this cleaned set, \texttt{SigMA} identifies 13\,103 candidates.

\texttt{SigMA} des not work on the full phase space since it excludes RVs in its analysis and depends on user-defined parameters (e.g., smoothing scale, significance threshold), making some degree of contamination from field stars likely. 
\citetalias{2023Ratzenboeck1} estimate 5--8\%, supported by color–magnitude analysis, but this level is low enough to robustly recover Sco-Cen’s structure and kinematics.

For this sample, we adopt {\it Gaia} DR3 proper motions and parallaxes \citep{2023GaiaColl-Vallenari-DR3}. 
No further cuts were applied, as strict criteria are already part of \texttt{SigMA}. 
Distances are obtained by inverting parallaxes, with distance uncertainties accounted for in our modeling (see Sect.\ref{subsec:model} and Appendix\ref{app:likelihood}).
We have not used alternative distance estimators \citep[e.g][]{2018Queiroz}, as this would lead to inconsistencies with the \texttt{SigMA} results, which were derived from {\it Gaia} parallaxes. 

\subsection{Radial velocities}

We complement the Gaia-based sample with RVs from the compilation of \citetalias{2025Grossschedl}, which merges 22 surveys and literature catalogs (see. Appendix~\ref{app:data}). In contrast to \citetalias{2025Grossschedl}, we use an earlier version of SDSS (\citet{2017Majewski}, DR17 instead of DR19) and we have additionally added the RV’s of \citet{2017Frasca} in the same manner as outlined in \citetalias{2025Grossschedl}. The used line-of-sight velocities (spectroscopic RVs) have generally larger uncertainties than the proper motions and may suffer from systematics between surveys or from binarity, which can affect RVs more strongly than astrometry.
Appendix~\ref{app:data} compares the S/N of tangential and radial velocity tracers, highlighting the consistently higher quality of {\it Gaia} proper motions.

We filtered RVs using the {\it Gaia} re-normalized unit weight error (RUWE; \citealt{2024Castro-Ginard}), an indicator of binarity and other systematics. 
Even after this cut, some RVs may remain affected, which is addressed in our statistical model. 
The final sample contains 4090 stars with acceptable RVs. 
Proper motions were not RUWE-filtered, as binarity affects them to a lesser extent.

Based on RV availability and distance reliability, we classify the data into four subsets, summarized in Fig.~\ref{fig:data-pie}, each requiring specific treatment in the modeling.

\section{Method}
\label{sec:method}

\subsection{Coordinate system and reference frames} 
\label{subsec:coordinates_and_reference}

In the following, all model components and priors are defined in the heliocentric Cartesian velocity frame, the reference frame of the {\it Gaia} velocity data. 
Changing the frame in the model offers no statistical benefit but adds computational cost, while transforming the data into a Sco-Cen frame would alter the error statistics non-trivially.

For our analysis, however, a Sco-Cen-centered frame is required, since quantities like energy and momentum densities are only meaningful there. 
We therefore converted all posterior field samples in post-processing into the Sco-Cen frame of \citetalias{2025Grossschedl}, defined in the barycentric frame by $(v_{x, sc},v_{y, sc},v_{z, sc}) = (-6.2, -20.0, -5.4)$\,km\,s$^{-1}$. 
This frame reflects the mean velocity of the oldest Sco-Cen clusters, assumed to trace the association’s bulk Galactic motion and be least affected by recent feedback.

Unless noted otherwise, all quantities are expressed in heliocentric Galactic Cartesian coordinates $x,y,z$ on a $70 \times 70 \times 50$ grid with voxel size $3^3\, \mathrm{pc}^3$, anchored at $(x_{o},y_{o},z_{o}) = (-5, -190, -65)$\,pc. 
Fields on this grid are denoted in boldface, in contrast to stellar models with subscript $\star$. 
The cube’s extent ensures inclusion of all \texttt{SigMA} stars, and the voxel size reflects typical distance errors in our sample (Sect.~\ref{sec:data}), which set the spatial scale down to which the data can be expected to be informative. 
Inferring smaller scales would be possible, but these would likely be constrained mostly by the prior.

We define $\mathcal{P}_{\mathrm{voxel}}$ as the probability that a star lies inside the voxel implied by its parallax and errors, detailed in Appendix~\ref{app:data}. 
A position is flagged as ill-determined if $\mathcal{P}_{\mathrm{voxel}} < 0.682$ (the 1-$\sigma$ bound). 
This threshold balances computational cost from modeling parallax errors with improved accuracy. 
Under this criterion, $\sim40$\% of the sources have reliable distances. 
In the likelihood description below, we explain how we incorporate uncertainties from the remaining `bad’ sources.

\begin{figure*}
    \centering
    \begin{tikzpicture}
    \node at (0,0) {\includegraphics[trim={50 0 80 600pt}, clip, width=1.\linewidth]{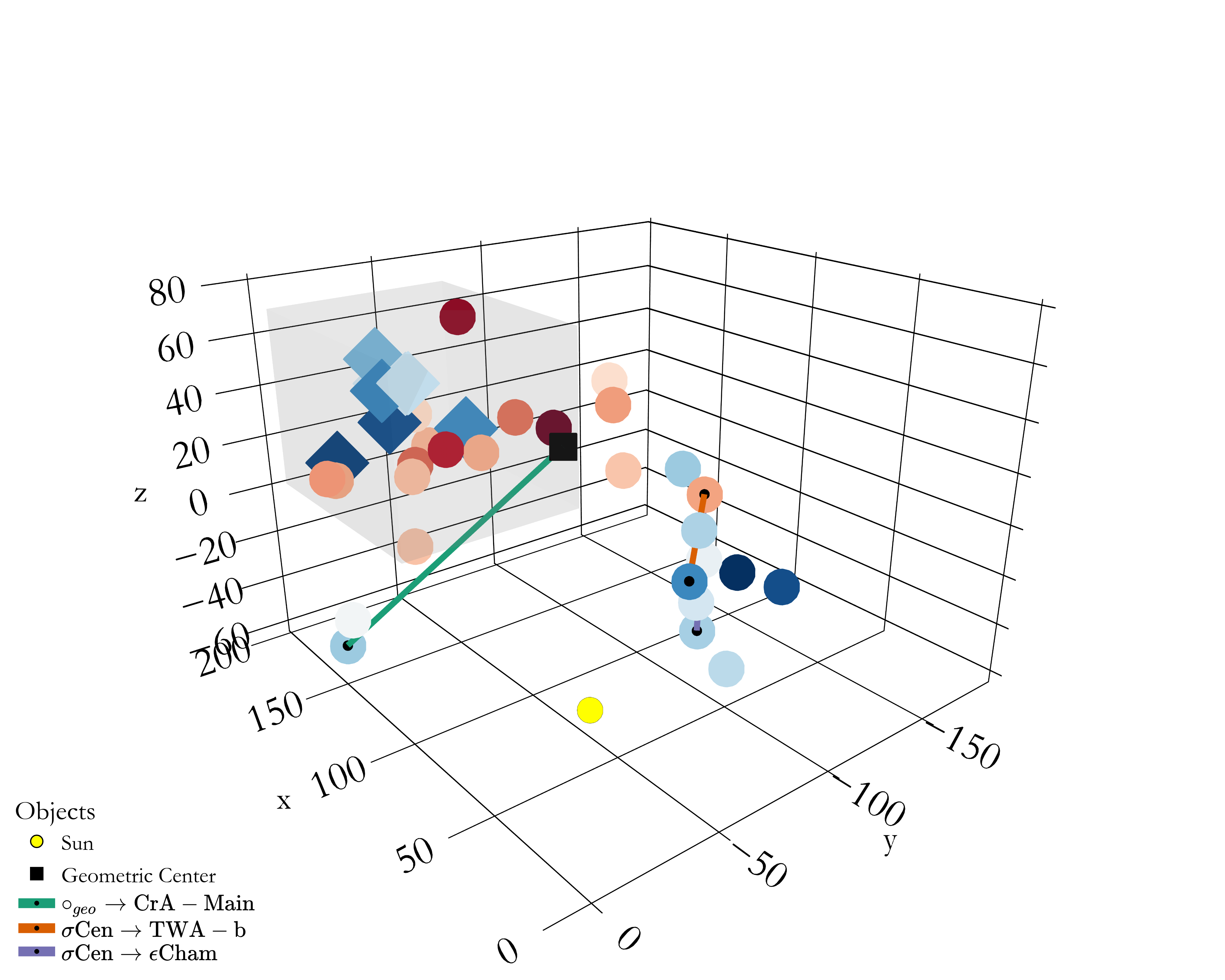}};
    \node at (8,0) {\includegraphics[width=.1\linewidth]{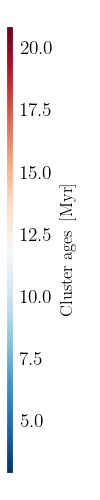}};
    \end{tikzpicture}
    \caption{
    Orientation plot. 
    This image shows the location and ages of the Sco-Cen clusters \citepalias{2023Ratzenboeck1, 2023Ratzenboeck2, 2025MiretRoig}. 
    The clusters that contain the stars that we refer to as the \texttt{main} sub-selection are shown as spheres, while the \texttt{secondary} clusters are shown as diamonds. 
    We shade the volume where the \texttt{secondary} selection was performed in gray.
    The black square indicates the geometric center of Sco-Cen \citepalias{2025Grossschedl}.
    Additionally, we show three cluster chains as lines. 
    A 3D interactive version of this plot is available \href{https://shutsch.github.io/sco_cen_data.html}{online}.
    We note that not all features mentioned are visible in this projection; we refer the reader to the online version for best visibility. 
    Furthermore, individual clusters are identifiable in the online version.
    \label{fig:orientation}}
\end{figure*}

\subsection{The age structure of Sco-Cen}
\label{subsec:ages_scocen}

The ideal picture of OB associations as expanding structures  \citep{1964Blaauw, 2024QuintanaArxiv} may lead to the expectation of a `simple' (i.e., non-overlapping) corresponding stellar flow field. 
Complications can arise due to the presence of field stars or the relative acceleration of sub-selections leading to, e.g., velocity caustics.
We give a more detailed discussion on the physical field assumption in  Appendix~\ref{app:field_assumption}. 

Figure~\ref{fig:orientation} shows an orientation plot over the clusters in Sco-Cen as found by \texttt{SigMA} \citepalias{2023Ratzenboeck1}. 
The plot clearly reveals the inside-out pattern of star formation in Sco-Cen, with the oldest clusters centered in the middle (around the cluster e-Lup) and clusters becoming progressively younger in all directions from there on.  
As detailed in Sect.~\ref{sec:intro}, in the specific case of Sco-Cen, the literature gives solid evidence for an accelerated outward expansion of the stellar population for a large volume fraction of the association.
The only sub-region that a-priori calls for special attention in this regard is the area towards USco, as the mentioned simple expansion pattern fails here \citepalias{2025Posch, 2025Grossschedl}. 
The results of \citetalias{2023Ratzenboeck2} indicate a more complex age pattern with old and young clusters spatially overlapping. 
This is a potential problem for our modeling, as this may be indicative of dynamical mechanisms that have accelerated groups of stars towards each other, leading to overlapping flows.  
A distinction of clusters older and younger than 12~Myr leads to a recovery of spatially distinct age patterns in this region \citepalias{2025Posch}. 
We have hence flagged eight younger clusters in this part of Sco-Cen (as annotated in Fig.~\ref{fig:orientation}) as the `\texttt{secondary}' selection containing 2487 stars, while we refer to the rest of the stars as the `\texttt{main}' selection.
We note that the analysis of isochronal cluster ages is a general (non-kinematic) recipe for the pre-analysis of stellar associations with field-based methods to mitigate the intricacies of collisionless flows (see discussion in Appendix~\ref{app:field_assumption}).  

These clusters were also selected as outliers based on their peculiar motions in \citetalias{2025Grossschedl}, with the sole exception of the cluster L134/L138. 
The fact that selections based on age and peculiar motion in the region of USco give almost the same result further hints at a source of additional momentum that has impacted this younger selection of stars. 
It should further be noted that the low mass cluster L134/L138 is spatially disconnected from USco, and both its age and motion determination rely on a small number of data points, making even this discrepancy of low significance.
We would like to emphasize that we have not performed the 12~Myr age cut in the rest of Sco-Cen, as there is no a-priori indication of overlaying velocity structures there.
We give details on the clusters within the \texttt{secondary} selection in Table~\ref{tab:summary_secondary_selection}.  

\subsection{Model}
\label{subsec:model}

\subsubsection{Velocity field}
\label{subsubsec:field_model}

The aim is to fit velocity fields $\mathbf{V}_{\mathtt{secondary}}$ and $\mathbf{V}_{\mathtt{main}}$ to the proper motions and RVs of stars conditional on their 3D positions $p_\star = \left( x_\star, y_\star, z_\star\right)$.  
As the motion of the \texttt{secondary} selection is likely the result of more recent events (see discussion on ages in Sect.~\ref{subsec:ages_scocen}), we model the velocity field of these stars via a sum of the main velocity field and an additional field $\mathbf{V}_{\delta}$ that models the difference between the \texttt{main} field $\mathbf{V}_{\mathtt{main}}$ and the \texttt{secondary} field $\mathbf{V}_{\mathtt{secondary}}$, i.e.:
\begin{equation}
    \label{eq:secondary}
    \mathbf{V}_{\mathtt{secondary}}  = \mathbf{V}_{\delta}  + \mathbf{V}_{\mathtt{main}}.  
\end{equation}
In case there is no second flow present in the true velocity field, the model above would still fit the data, but the $\mathbf{V}_{\delta}$ component would get as small as the noise level. 
For both fields $f \in (\delta, \mathtt{main})$ we write
\begin{equation}
    \mathbf{V}_{f}\left(x, y, z\right) = \left(\begin{array}{c}
         \mathbf{v}_{x, f}  \\
         \mathbf{v}_{y, f}  \\ 
         \mathbf{v}_{z, f}
         \end{array}\right)
    \end{equation}
on a Cartesian grid in Galactic coordinates and in the barycentric velocity frame (see discussion in Sect.~\ref{subsec:coordinates_and_reference}).
The components $\mathbf{v}_{x, f}, \mathbf{v}_{y, f},$ and $\mathbf{v}_{z, f}$ are each modeled as independent Gaussian random fields with unknown correlation structure.
We model this correlation structure via independent power-spectra for all three components for both the $\mathbf{V}_{\mathtt{main}}$ and $\mathbf{V}_{\delta}$ vector fields.
We make use of the field model presented in \citet{2022Arras} and we have for each scalar field component $\mathbf{v}_{i, f}$ with $i \in \left(x, y, z\right)$:
\begin{equation}
    \mathbf{v}_{f} = m_{i, f} + \mathcal{F}\left(\mathbf{A}_{i, f}(k)\boldsymbol{\xi}_{i, f}(k)\right).
\end{equation}
Here, $m_{i, f}$ denotes the mean over the full volume, while $\mathbf{A}_{i, f}(k)$ is the amplitude of the scalar component-field in Fourier space (indicated by the wavenumber $k$), which is directly related to the square root of the power-spectrum.
For each component field, this amplitude field is multiplied with the excitation field $\boldsymbol{\xi}_ {i, f}(k)$, which encodes the specific realization of the component, and this product is then Fourier transformed via $\mathcal{F}$.
The power-spectrum itself is modeled via the model presented in \citet{2022Arras}, which parametrizes the log-spectrum via the sum of a power-law and a stochastic process.
We chose the priors of this model such that the dynamical range both in the means $m_{i, f}$ and the fluctuations around this mean easily encompasses several tens of km\,s$^{-1}$.
Additionally, we assume a Gaussian prior on the slope of the power-spectrum of the correlation structure of $-7 \pm 3$.
We give further details of the field model prior, as well as choices on hyperparameters, in Appendix~\ref{app:prior}.

\subsubsection{From fields to stars}
\label{subsubsec:response}
The heliocentric Galactic Cartesian velocity vector of a star belonging to a sub-selection $s$ can be extracted from the respective velocity field via
\begin{equation}
    V_{\star, \mathrm{Gal, cart}} = \mathcal{S}_\star\mathbf{V}_{s},
\end{equation}
where $\mathcal{S}_\star$ is a simple selection operator that returns the value of the field at the position of the star.
This is akin to a piece-wise constant interpolation of the voxels. 
In accordance with our discussion on the effective spatial resolution set by the parallax data (see Sect.~\ref{sec:data}), we have refrained from employing more complex interpolation techniques, as these would only have an impact below the 3\,pc scale.  
To be comparable to observations, this vector is converted into the spherical celestial coordinate system (i.e., the International Celestial Reference System, ICRS) via
\begin{equation}
    V_{\star, \mathrm{ICRS, sph}} = \left(\begin{array}{c}
         v_{\star, \mathrm{RA}}  \\
         v_{\star, \mathrm{DEC}} \\
         v_{\star, \mathrm{R}}    \end{array}\right)=R_{\mathrm{cart\rightarrow sph}}\, R_{\mathrm{Gal \rightarrow ICRS}}\,V_{\star, \mathrm{Gal-cart}}.
\end{equation}
The two rotation matrices denote the coordinate transformation from Cartesian to spherical ($R_{\mathrm{cart \rightarrow sph}}$) coordinates and the rotation between the ICRS and Galactic reference frames ($R_{\mathrm{Gal \rightarrow ICRS}}$).
The proper motion model of the star can then be derived from the two tangential components of $V_{\star, \mathrm{ICRS, sph}}$ via
\begin{equation}
    \label{eq:mu_model}
    \mu_\star = \frac{\varpi_\star}{4.74047}\left(\begin{array}{c}
         v_{\star, \mathrm{RA}} \\
         v_{\star, \mathrm{DEC}} 
    \end{array}\right).
\end{equation}

The RV model of a star $\mathrm{RV}_{\star}$ is the third component of $V_{\star, \mathrm{ICRS-sph}}$.
With models for the three stellar velocity observables at hand, the last remaining step is the description of the likelihood, incorporating the fact that the position of some stars is not known with sufficient certainty.

\subsubsection{Likelihood}
\label{subsubsec:likelihood}
We connect the  $\mu_{\star}$ and $\mathrm{RV}_{\star}$ models to the data in the likelihood, where we have to consider the different cases in our data set outlined in Sect.~\ref{sec:data} and Fig.~\ref{fig:data-pie}, namely whether the stars belong to the \texttt{secondary} sub-selection, whether they have RV measurements, and whether they have a precisely determined distance w.r.t.~our computational grid (see Sect.~\ref{subsec:coordinates_and_reference}). 
The simplest cases arise for the stars where we have precise distance data, and can replace the model for the distance of the star $p_\star$ with the observed position $p$ determined from parallaxes, $|p| = 1/\varpi$, and sky positions $\left(\mathrm{RA}, \mathrm{DEC}\right)$. 
In particular, for the cases where we have only proper motion data and under the assumption of Gaussian noise, the likelihood is
\begin{align}
    \label{eq:case_1_likelihood}
      \mathcal{P}_{s, 1}\left(\mu| \vec{V}_{s}, \varpi\right) 
      = \mathcal{G}\left(\mu - \mu_\star(\vec{V}_{s}, p), C_\mu\right). 
\end{align}
The symbol $\mathcal{G}(x-m_x, C_x)$ denotes a multivariate Gaussian distribution in quantity $x$, with mean $m_x$ and covariance $C_x$.  
In the equation above, $C_\mu$ specifically is the noise covariance matrix reported by \emph{Gaia}. 
If an RV is observed for a star, we can exploit the independence of proper motion and RV measurements and write:
\begin{align}
    \label{eq:case_2_likelihood}
    &\mathcal{P}_{s, 2}\left(\mu, \mathrm{RV}| \vec{V}_s,  \eta_\mathrm{RV}, \varpi \right) =
      \mathcal{P}\left(\mathcal{D}_\mu | \vec{V}_s, \varpi \right)   \mathcal{P}\left( \mathcal{D}_\mathrm{RV} | \vec{V}_s,  \eta_\mathrm{RV}, \varpi \right) = \nonumber \\ 
      &= \mathcal{G}\left(\mu - \mu_\star(\vec{V}_s, p), C_\mu\right)\mathcal{G}\left(\mathrm{RV} - \mathrm{RV}_\star(\vec{V}_s, p),  \eta_\mathrm{RV}\sigma^2_\mathrm{RV}\right) 
\end{align}
For the observational errors of the RVs, we have introduced an additional model parameter $\eta_{\mathrm{RV}}$ per RV measurement to modify the noise term of the RV likelihood:
\begin{equation}
    \label{eq:noise_est}
   \widetilde{\sigma}^2_{\mathrm{RV}} =  \eta_\mathrm{RV}\sigma^2_{\mathrm{RV}}.
\end{equation}
This factor has an inverse-gamma prior chosen such that the most likely value is unity. 
This implies that we a-priori trust the observational errors, but the heavy tail of the inverse-gamma distribution allows for an effective down-weighting of outliers caused, for instance, by binarity (see discussion in Sect.~\ref{sec:data}), if necessary. 
We discuss hyperparameters of this part of the model in Appendix~\ref{app:prior} and refer the reader to \citet{2012Oppermann} for the motivation of the inverse-gamma prior.

For the subset of stars with large distance errors compared to the voxel size (see Sect.~\ref{subsec:coordinates_and_reference}), we need to incorporate the uncertainty stemming from the parallax measurements. 
We then include the parallax uncertainty by marginalizing over the positions of the stars conditional on the parallax data.
The dependence of the velocity field estimate on the uncertainty of the stellar position is generally not analytically tractable, as its impact depends on the structure of the velocity field itself. 
To tackle this, we follow the prescription of \citet{2019Leike, 2023Edenhofer} to derive an additional noise term in the likelihood; we give a derivation specific to our setup in the Appendix~\ref{app:likelihood}.
This works under the assumption that we can approximate the additional parallax error term via sampling the model at different distances according to the respective parallax error and estimating a (Gaussian) error term from that, leading to the following expressions for the likelihood, again for the cases without measured RV 
\begin{equation} 
\label{eq:case_3_likelihood}
\mathcal{P}_{s, 3}\left(\mu|  \vec{V}_{s}, \varpi \right) \approx \mathcal{G} \left( \mu - \langle\mu_\star\rangle_\varpi, C_{\mu} + \langle C_{\mu} \rangle_{\varpi}\right) 
\end{equation}
and for stars with measured RV
\begin{equation}  
\label{eq:case_4_likelihood}
\mathcal{P}_{s, 4}\left(\mu, \mathrm{RV}|  \vec{V}_{s}, \varpi\right) \approx \mathcal{G} \left(\begin{pmatrix} \mu \\ \mathrm{RV}  \end{pmatrix} - \begin{pmatrix} \langle\mu_\star\rangle_\varpi\\  \langle \mathrm{RV}_\star\rangle_\varpi \end{pmatrix}, C_{\mu, \mathrm{RV}} + \langle C_{\mu, \mathrm{RV}} \rangle_{\varpi}\right). 
\end{equation}
The notation $\langle x_\star\rangle_\varpi$ and $\langle C_{\mu, \mathrm{RV}} \rangle_{\varpi}$ indicates mean and covariance estimated from the model realization for a quantity $x$ evaluated at different stellar distances; we give their definition in Appendix~\ref{app:likelihood}.
The error estimate from the parallax uncertainties couples the RV and proper motion likelihoods, hence making $\langle C_{\mu, \mathrm{RV}} \rangle_{\varpi}$ a full $3\times3$ matrix per star.
We note that the asymmetry of the distance distribution induced by inverting the parallax sample is fully accounted for in our model.
The only strong assumption is the Gaussianity of the additional error term, which is a first order approximation.  
The statistics of this term depend on the distance sampling relative to the typical scales of the velocity field, which is a-priori unknown.  

{\it Gaia} does not only report cross-correlations between the proper motion components of the stars, but also between proper motion and parallax. 
The latter term can be ignored in our setup, as we are not fitting the parallaxes but marginalizing over them, which implicitly incorporates any model cross-correlation between these quantities, while the data cross-correlation is not informative due to the marginalization. 

Given all the above considerations, the full likelihood is then the product of all eight sub-selection likelihoods: 
\begin{equation}
    \mathcal{P_{\mathrm{full}}}\left(\mu, \mathrm{RV}|  \mathbf{V}, \varpi  \right) = \prod_{\mathrm{s}}  \prod_{i=1}^4 \mathcal{P}_{s, i} \left(\mu, \mathrm{RV} | \mathbf{V}_\mathrm{s}, \varpi\right)
\end{equation}
The subscript $s$ runs over the \texttt{main} and \texttt{secondary} selection of stars.
We give the details of the derivation in Appendix~\ref{app:likelihood}.

\subsection{Inference}
\label{subsec:inference}

The model described above has 5\,027\,299 degrees of freedom, entailing 4\,800\,006 field parameters, 222\,600 power spectrum parameters, and 4\,634 noise estimation parameters.
We note that the number of field and power spectrum parameters mostly depend on the spatial and spectral resolution chosen in our model. 
The number of noise estimation parameters is the number of stars with reliable RV measurement; see discussion in Sect.~\ref{sec:data}.

A full evaluation of such a high-dimensional and non-Gaussian posterior probability distribution is computationally extremely challenging.
We hence employ a variational inference scheme, which means we infer a simpler and more tractable distribution to approximate the high-dimensional and likely very complex structure of the posterior.
Specifically, we employ geometric variational inference \citep[GeoVI,][]{2021Frank}, in which the posterior is approximated by a multivariate Gaussian distribution augmented with a coordinate transformation constructed from Riemannian geometry and the Fisher information metric, which captures non-Gaussian aspects of the true posterior.
For a detailed discussion of the method and comparison to other ways to evaluate posterior distributions, we refer the reader to \citet{2021Frank}.

\section{Results}
\label{sec:results}

In the following, we present the results of the inference setup laid out in the sections above. 
We evaluate the resulting approximated posterior via drawing samples from it and report the corresponding empirical mean and standard deviations for the model and some of its components online\footnote{\label{note_maps} \href{https://zenodo.org/records/17107581}{https://zenodo.org/records/17107581}}.  
Additionally, we publish the latent space samples as well as the computational model online, for full reproducibility of our results.
The full statistical uncertainty is only captured in the latent space samples. For example, cross-voxel correlations are lost if only the statistical standard deviation on the field level is considered. 
We hence recommend the usage of the latent samples if the full statistical information is needed for further analysis.

Results on the velocity fields, the power spectra and the divergence of the \texttt{main} field are shown in the following sections, while results on the noise estimation factors are given in Appendix~\ref{app:noise_estimation}, on the component fields in Appendix~\ref{app:components}, and on some of the derivative fields in Appendix~\ref{app:additional_derivatives}.
All two and 3D plots depict the posterior mean; we have refrained from visualizing the statistical uncertainties in these cases. 
These can be derived from the latent samples and the model\footref{note_maps}. 
Interactive versions of the plots are available online\footnote{\label{note_plots}\href{https://shutsch.github.io/sco_cen_data.html}{https://shutsch.github.io/sco\_cen\_data.html}}.
For one-dimensional plots and numerical values, we show the propagated statistical uncertainties.

\begin{figure*}
    \centering
    \begin{subfigure}{0.96\textwidth}
        \includegraphics[trim=200 0 80 600, clip, width=\textwidth]{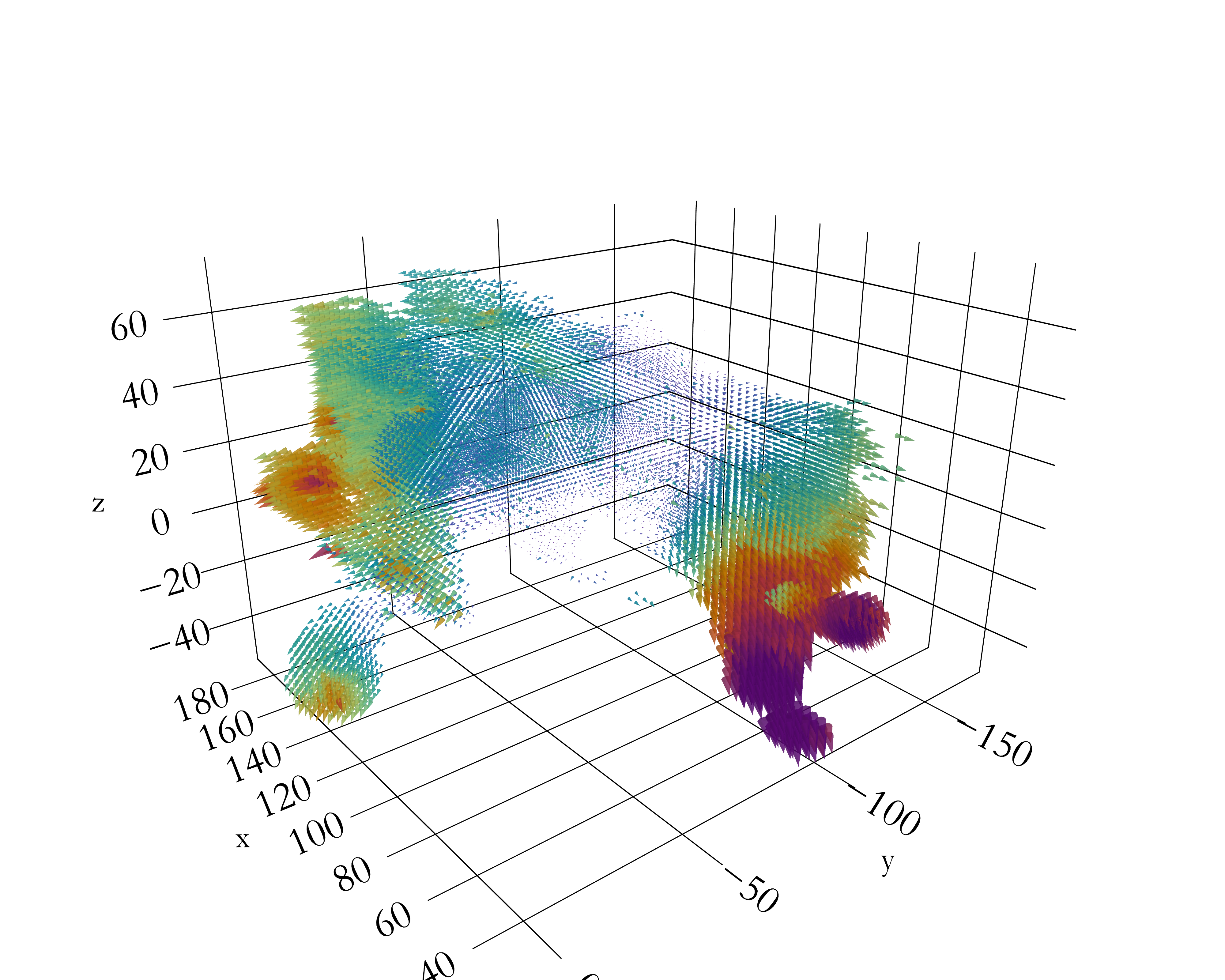}
        \caption{\label{fig:flow_field_main}}
    \end{subfigure}
    \begin{subfigure}{0.96\textwidth}
        \includegraphics[width=\textwidth]{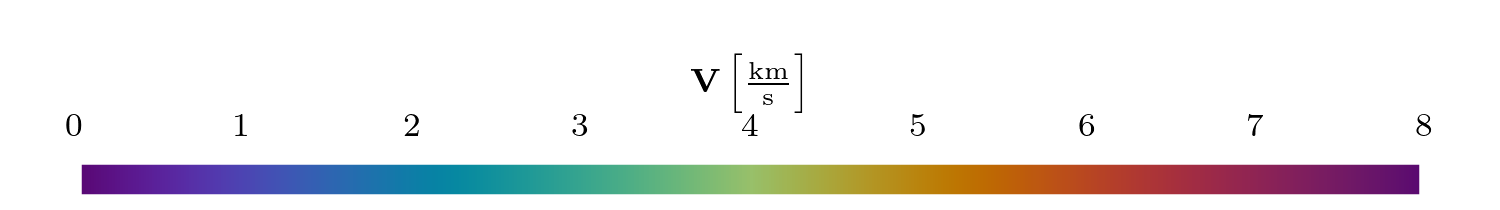}
    \end{subfigure}
    \begin{subfigure}{0.48\textwidth}
        \includegraphics[trim=200 0 80 600, clip, width=\textwidth]{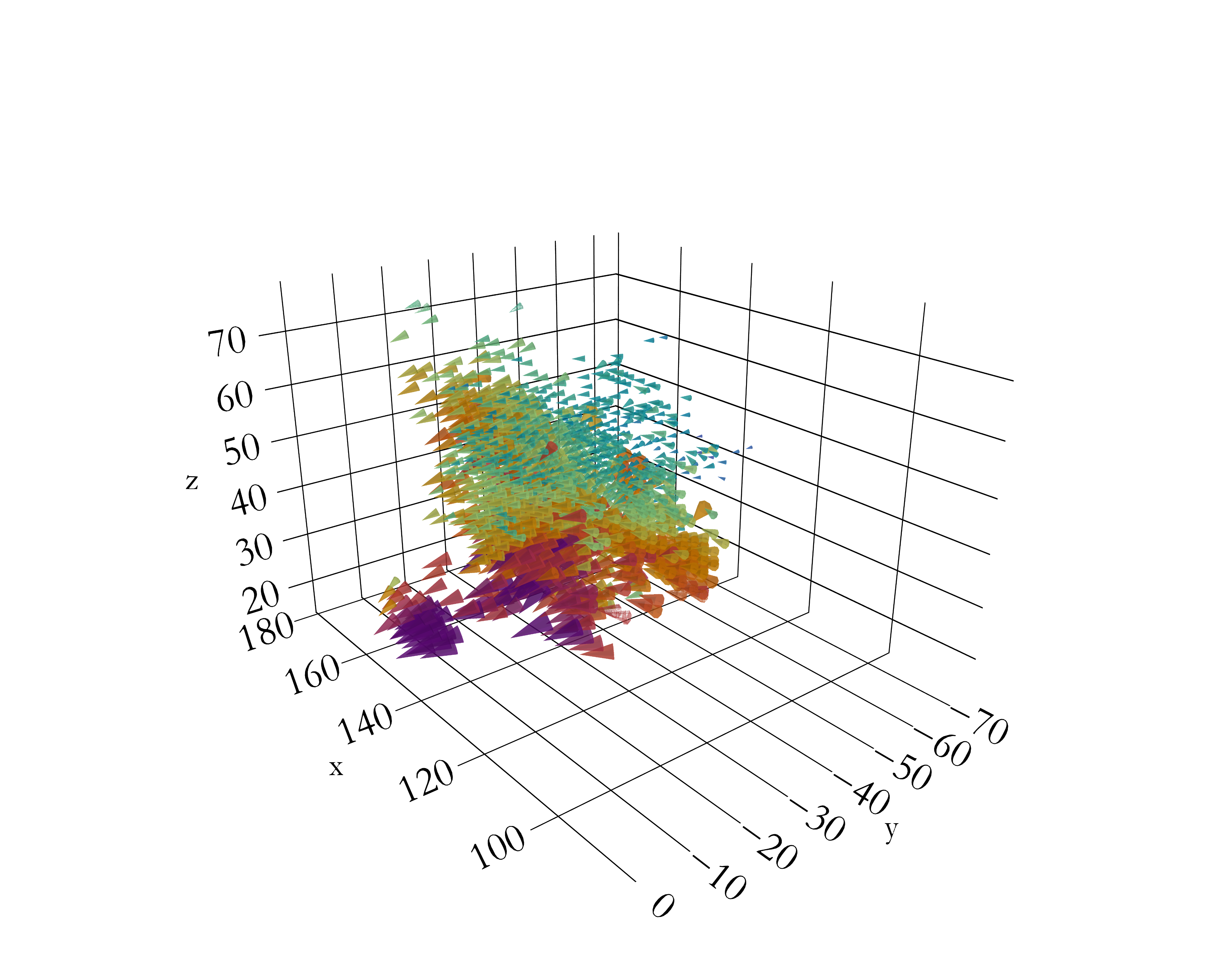}
        \caption{\label{fig:flow_field_us}}      
    \end{subfigure}
    \begin{subfigure}{0.48\textwidth}
        \includegraphics[trim=200 0 80 600, clip, width=\textwidth]{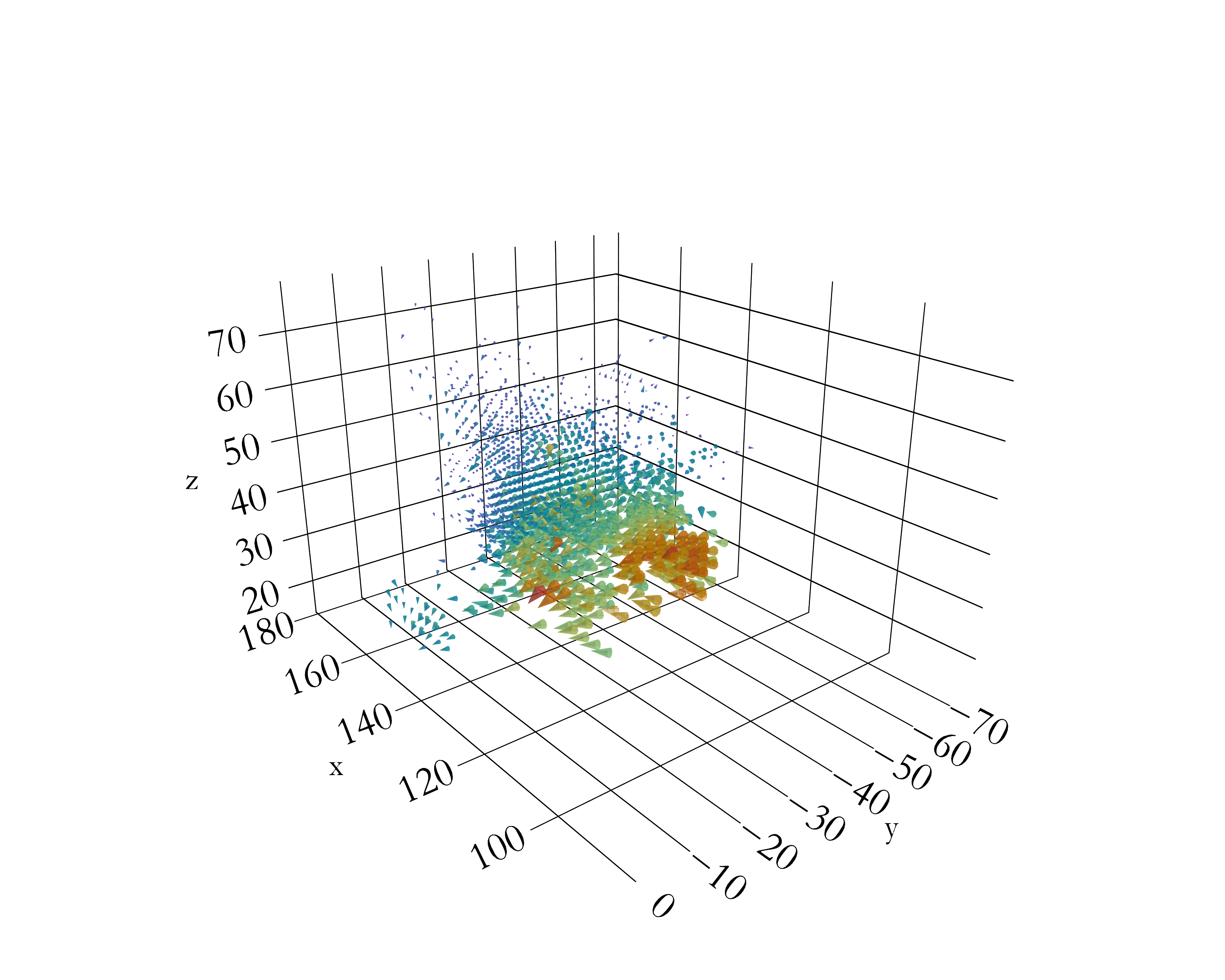}
        \caption{\label{fig:flow_field_add}}       
    \end{subfigure}
    \caption{Posterior mean of the flow fields in the velocity reference frame of \citetalias{2025Grossschedl}. 
    For each plot, we have defined a mask using the density of stars to highlight only parts of the volume that were directly informed by data as described in Sect.~\ref{subsec:vel_field}. 
    Fig.~\ref{fig:flow_field_main} depicts the \texttt{main} field, while Figs.~\ref{fig:flow_field_us} and \ref{fig:flow_field_add} depict the \texttt{secondary} and  $\delta$ field.
    These projections of the vector fields can naturally only provide a first impression and might locally be misleading; for a better depiction, we refer to the 3D interactive version of these plots available \href{https://shutsch.github.io/sco_cen_data.html}{online}.
    \label{fig:flow_fields}}
\end{figure*}

\subsection{Velocity Field} \label{subsec:vel_field}

We show a projection of the main result, the posterior mean of the \texttt{main} and \texttt{secondary} velocity fields, in Fig.~\ref{fig:flow_field_main} in the reference frame defined by \citetalias{2025Grossschedl}.
Additionally, we show the secondary velocity field associated with the \texttt{secondary} sub-selection in  Fig.~\ref{fig:flow_field_us} and the difference of this field to the main field in Fig.~\ref{fig:flow_field_add}.
In these plots and all following plots, we define the border of the \texttt{main} and \texttt{secondary} volume via a mass-weighted kernel density estimate using the \texttt{SigMA} selection of stars at $\rho_\star = 0.005\, \mathrm{pc}^{-3}$.
For reference, we show a contour plot of the stellar density in Appendix~\ref{app:stellar_density}.
Quantitatively, we summarize the minimum and maximum values, their position in the grid, and the root-mean-square (rms) values of the \texttt{main}, \texttt{secondary}, and the $\delta$ fields in Table~\ref{tab:flow_results}.

Visually, the plot in Fig.~\ref{fig:flow_field_main} confirms the expanding velocity pattern of the flow already found in \citetalias{2023Posch, 2025Grossschedl}. 
In combination with the age of the clusters found by \citetalias{2023Ratzenboeck2}, this visual pattern alone already paints a convincing picture of Sco-Cen as an aging star-forming complex that experienced "inside-out" star-formation, where the star-forming gas reservoir was constantly accelerated, a process that has left its imprints in the stellar kinematics.
The main body of Sco-Cen seems to expand from a flow origin at $(x, y, z ) = (106 \pm 38, -70 \pm 27, 27 \pm  15)$\,pc, \citepalias{2025Grossschedl}.
This expansion manifests itself as a bulk motion towards the known secondary centers of star formation in USco and $\sigma$-Cen \citepalias{2023Ratzenboeck2, 2025Grossschedl} and via relatively narrow chains of stars accelerated away from the older parts of Sco-Cen (c.f., Fig.~\ref{fig:orientation} and \citetalias{2025Posch, 2025MiretRoig}). 

The \texttt{secondary} velocity field depicted in Fig.~\ref{fig:flow_field_us} has a higher maximum value than the \texttt{main} field, which, moreover, is placed in the youngest member cluster, the B59 cluster. 
This is consistent with the ongoing feedback-driven momentum input to star-forming clouds. 

The difference between the \texttt{main} and \texttt{secondary} velocity field, $\vec{V}_\delta$ (see Eq. \ref{eq:secondary}), is illustrated in Fig.~\ref{fig:flow_field_add} and is dominated by two distinct features. 
One is a systematic shift in the negative $v_z$ direction and a positive direction in the bulk of USco, which was already noted in \citetalias{2025Grossschedl, 2025Posch}.
The second notable feature is the location maximum of the flow (see Table~\ref{tab:flow_results}), which is centered right within the Lupus~1--4 cluster, and almost fully explains the respective field values in the \texttt{secondary} field.   
This means that this relatively young stellar cluster has a motion completely distinct from the rest of the bulk motion of Sco-Cen,  which, given the age, must indicate a strong momentum impact to the progenitor cloud of  Lupus~1--4 or it being an unrelated structure within the Sco-Cen volume. 
We note that USco and specifically the interface between the Lupus and Ophiuchus molecular clouds have been under the influence of recent supernovae events \citep{2016Breitschwerdt, 2020Neuhauser, 2023BricenoMorales} and ongoing radiative feedback \citep{2025Alves}.
A connection of the morphology and energy density of the flow to these recent events will be pursued in Paper~II. 

\subsection{Power spectra and divergence} \label{subsec:pow_spec}

\begin{figure*}
    \centering
    \begin{subfigure}{0.48\textwidth}
        \includegraphics[width=\textwidth]{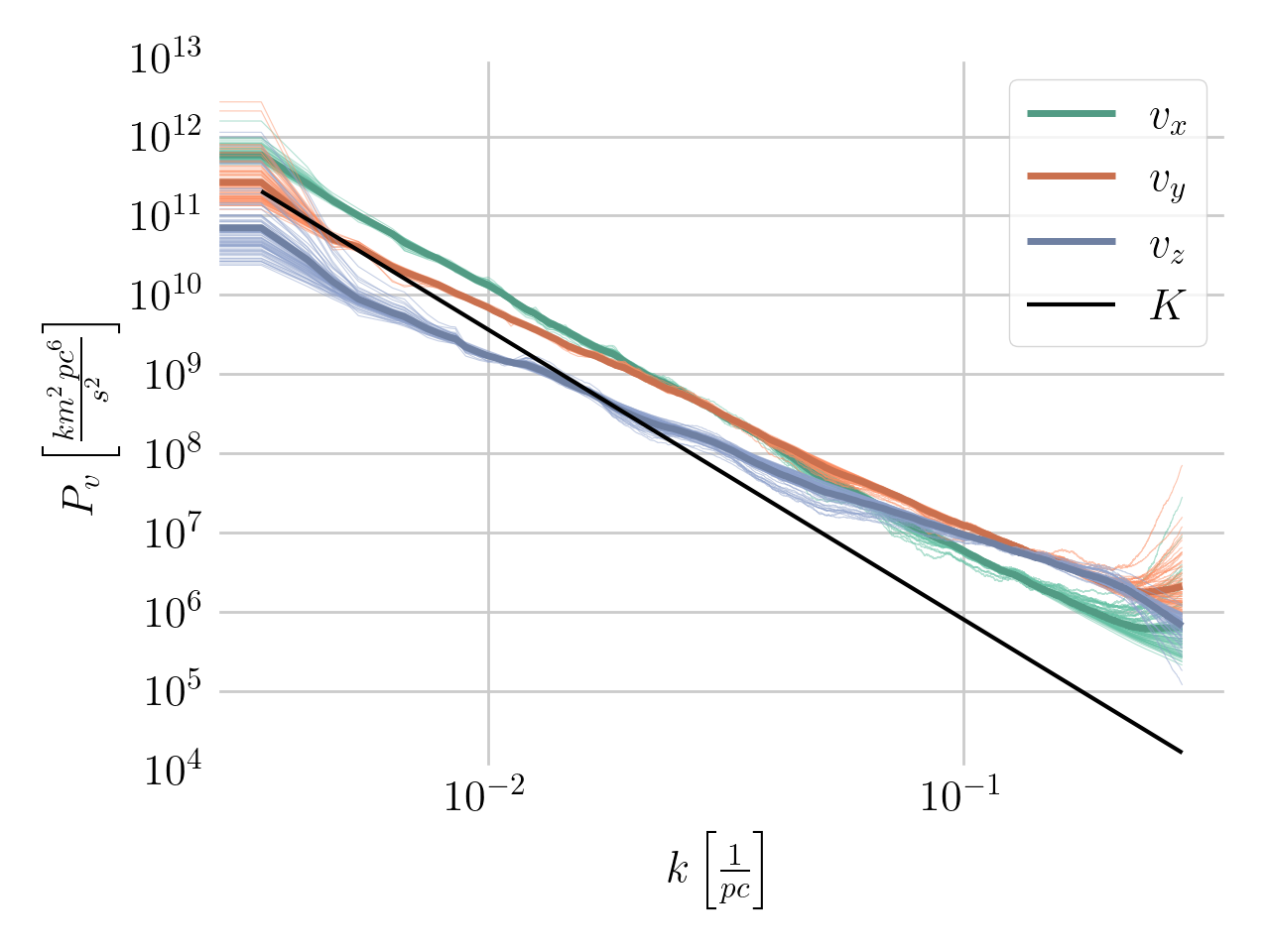}
        \caption{\label{fig:power_main} \texttt{main} field}
    \end{subfigure}
    \begin{subfigure}{0.48\textwidth}
        \includegraphics[width=\textwidth]{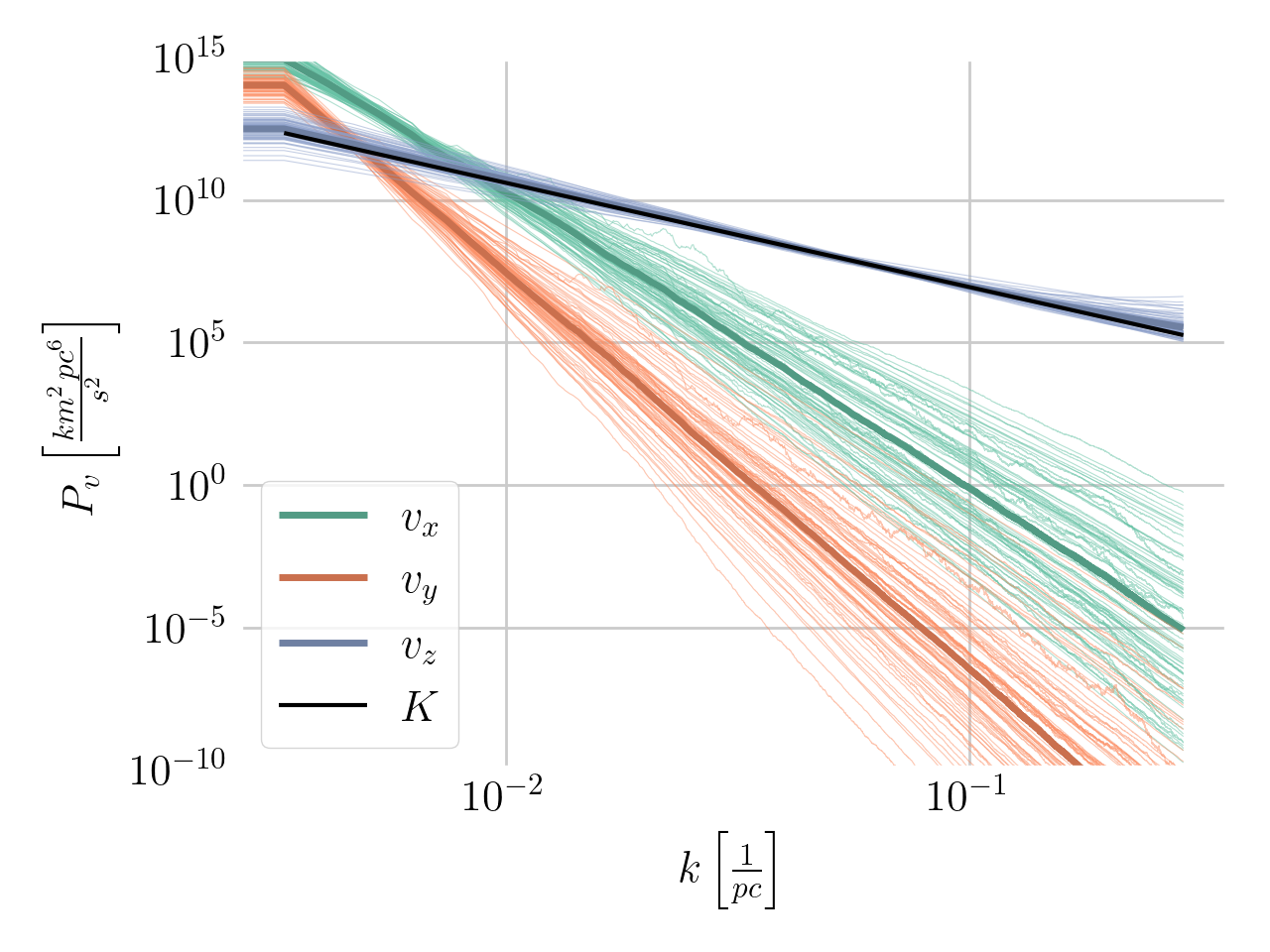}
        \caption{\label{fig:power_us} $\delta$ field}
    \end{subfigure}
    \caption{\label{fig:power}The power spectra of the \texttt{main}  and $\delta$ velocity field components, which are part of the forward model.  
    The black line indicates the Kolmogorov slope $-11/3$, while the posterior samples are indicated as thin colored lines. 
    We have scaled the line indicating the Kolmogorov spectrum to the average of the three component spectra at $k=0.005\,\mathrm{pc}^{-1}$ in each case, for illustrational purposes.}
\end{figure*}

We furthermore calculate several secondary quantities, namely power spectra and derivatives.  
We depict the posterior samples and means of the power spectra of each component of the \texttt{main} and $\delta$ flows in Fig.~\ref{fig:power}, and report the posterior values of the slope parameters in Table~\ref{tab:flow_results}.

In case of the \texttt{main} flow power spectra shown in Fig.~\ref{fig:power_main}, we report that the slopes are consistently higher than the Kolmogorov value of $-11/3$, implying that small scales have significant structure.  
We note that the Kolmogorov slope implies scale invariance, but is not a clear-cut indicator for any physical process, be it random or ordered. 
Interestingly, the Kolmogorov value is found often in ISM-related fields, for instance, as the `Big power law in the sky' \citep{1995Armstrong, 2010Chepurnov, 2024Hutschenreuter} in the diffuse plasma as measured over more than 12 orders of magnitude \citep{2020Ferriere} or Larson's relation \citep{1981Larson}, the dependence of the internal velocity dispersion of clouds on their size.

For the $\delta$ field fitting the systematic difference between the \texttt{main} and \texttt{secondary} fields, we find a very different picture, as shown in Fig.~\ref{fig:power_us}. 
The $z$-component is completely consistent with a self-similar Kolmogorov spectrum, while the $x$- and $y$- fields are completely dominated by the lowest wavenumbers. 
This means that all additional small structure in the difference between the motions of the \texttt{secondary} and \texttt{main}  is driven by a process that accelerates the younger stars downwards, consistent with our discussion in Sect.~\ref{subsec:vel_field}. 

\begin{figure}
    \centering
    \begin{subfigure}{0.85\linewidth}
        \includegraphics[trim=200 0 180 700, clip, width=\linewidth]{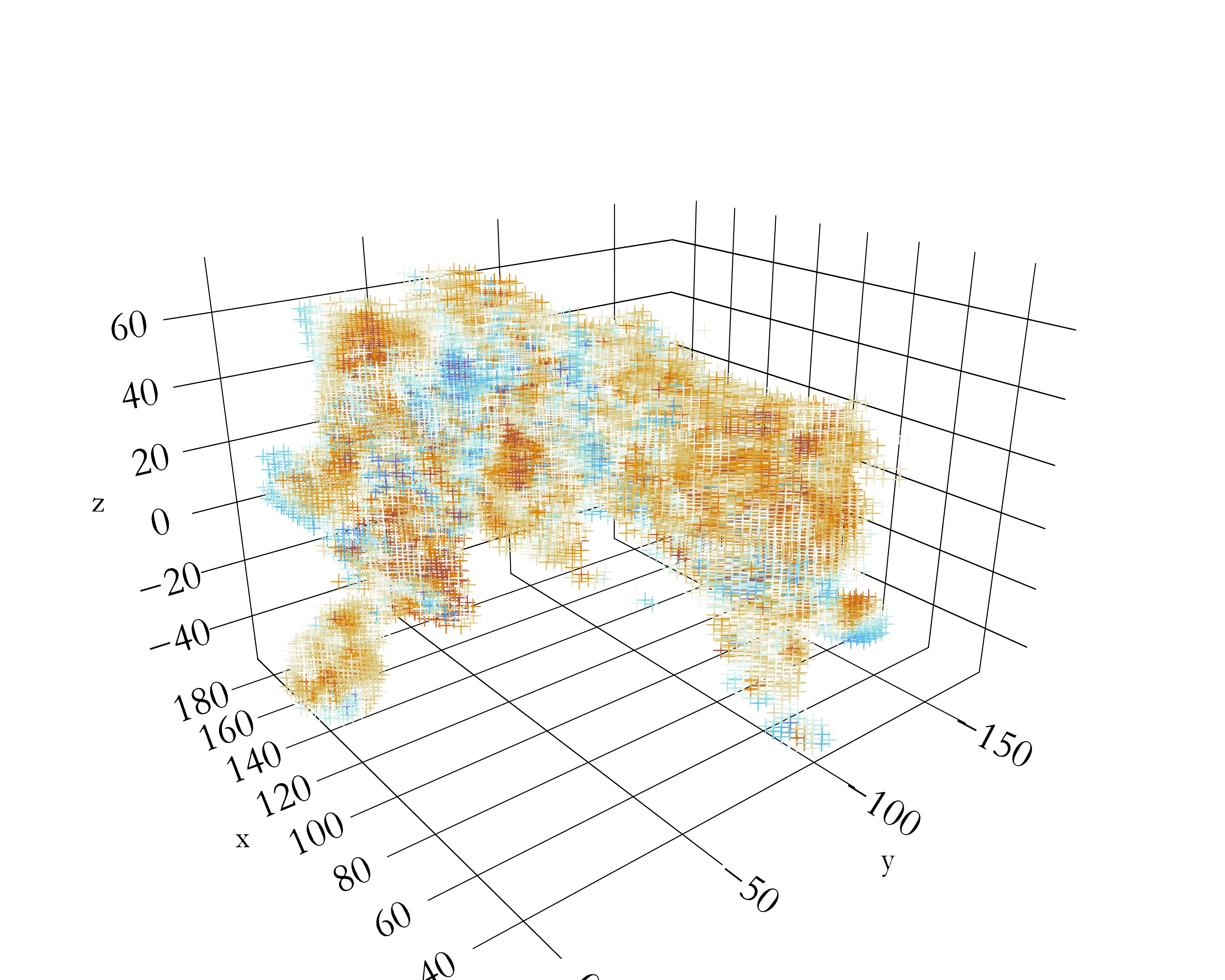}
    \end{subfigure}
    \begin{subfigure}{0.14\linewidth}
        \includegraphics[width=\linewidth]{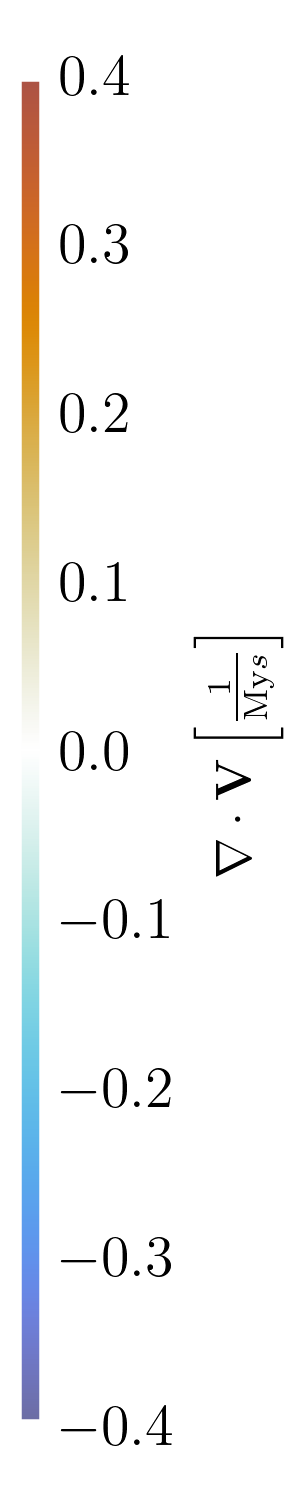}
    \end{subfigure}
    \caption{\label{fig:divergence_field} Posterior mean of the divergence of the \texttt{main} flow field. 
    A 3D interactive version of this plot is available \href{https://shutsch.github.io/sco_cen_data.html}{online}.}
\end{figure}

A 3D vector field also allows the calculation of two first-order differential fields, namely the divergence and the vorticity. 
The divergence of the \texttt{main} field is shown in Fig.~\ref{fig:divergence_field}, while the vorticity of the \texttt{main} field and both quantities for the \texttt{secondary} field are shown in Appendix~\ref{app:additional_derivatives} and online, all calculated via finite differences. 
The divergence indicates the tendency of the flow to contract (if it is negative) or expand (if it is positive). 
The vorticity gives the rate of rotation at a voxel, with positive values indicating right-handed rotation.
Both quantities have the units of one over time, and can be interpreted as local measures of expansion rate and rotational frequency, respectively.  
The divergence plot in Fig.~\ref{fig:divergence_field} shows strong small-scale fluctuations, which are a consequence of the derivatives boosting smaller scales.  
On average, it demonstrates the expanding nature of the Sco-Cen flow field as it is mostly positive throughout the volume.  
The contracting parts are mostly located within and in the direction of USco. 
The expansion rates per voxel have values up to a maximum of about $0.9~\mathrm{Myr}^{-1}$, with the more typical values being around $0.2~\mathrm{Myr}^{-1}$. 

We show the divergence plot of the \texttt{secondary} and $\delta$ fields in Appendix~\ref{app:additional_derivatives}. 
Both show interesting features that can likely be linked to recent feedback events; we again defer this analysis to future work.

\begin{table}[!h]
\begin{small}
\caption{Marginal Posterior means of several characteristic values of the flow component fields and the absolute value fields, for $\vec{V}_\mathtt{main}$, $\vec{V}_\mathtt{secondary}$, and $\vec{V}_\delta$.}
\label{tab:flow_results}
\centering
\resizebox{1\columnwidth}{!}{
\renewcommand{\arraystretch}{1.2}
\begin{tabular}{lrrrr}
\hline 
\hline
\multicolumn{1}{c}{} &
\multicolumn{1}{c}{max} &
\multicolumn{1}{c}{min} &
\multicolumn{1}{c}{r.m.s.} & 
\multicolumn{1}{c}{p.s.~slope} \\
\multicolumn{1}{c}{} &
\multicolumn{1}{c}{km\,s$^{-1}$} &
\multicolumn{1}{c}{km\,s$^{-1}$} &
\multicolumn{1}{c}{km\,s$^{-1}$} & 
\multicolumn{1}{c}{unitless} \\
\hline
$v_{\mathtt{main},x}$ & $3.3 \pm 0.2$ & $-8.2 \pm 0.1$ & $1.97 \pm 0.03$ & $-3.1 \pm 0.2$ \\
$v_{\mathtt{main},y}$ & $8.4 \pm 0.1$ & $-3.4 \pm 0.1 $ & $1.68 \pm 0.03$ & $-2.5 \pm 0.1$ \\
$v_{\mathtt{main},z}$ & $2.6 \pm 0.1$ & $-6.2 \pm 0.1$ & $1.05 \pm 0.01$ & $-2.5 \pm 0.3$ \\
$|\vec{V}_\mathtt{main}|$ & $10.1 \pm 0.1$ & $0.2 \pm 0.1$ & $1.68 \pm 0.02$ & N.A. \\
\hline
$v_{\mathtt{sec},x}$ & $7.3 \pm 0.7$ & $-2.4 \pm 0.4$ & $1.71\pm0.12$ & N.A. \\
$v_{\mathtt{sec},y}$ & $9.4 \pm 0.2$ & $-1.3 \pm 0.2$ & $1.64\pm0.03$ & N.A. \\
$v_{\mathtt{sec},z}$ & $1.4 \pm 0.3$ & $-6.0 \pm 0.2$ & $1.04\pm0.03$ & N.A. \\
$|\vec{V}_\mathtt{sec}|$ & $10.2 \pm 0.2$ & $1.0 \pm 0.2$ & $1.42\pm 0.06$ & N.A. \\
\hline
$v_{\delta,x}$ & $5.8 \pm 0.7$ & $-1.8 \pm 0.7$ & $1.31\pm0.18$ & $-10.3 \pm 1.4$ \\
$v_{\delta,y}$ & $2.5 \pm 0.2$ & $-1.1 \pm 0.2$ & $0.65\pm0.05$ & $-14.0 \pm 1.8$ \\
$v_{\delta,z}$ & $1.2 \pm 0.2$ & $-4.4 \pm 0.2$ & $0.83\pm0.05$ & $-3.7 \pm 0.3$ \\
$|\vec{V}_\delta|$ & $6.3 \pm 0.4$ & $0.4 \pm 0.2$ & $1.26\pm0.14$ & N.A. \\
\hline
\end{tabular}
\renewcommand{\arraystretch}{1}
} 
\tablefoot{
r.m.s.~denotes the root mean square and p.s.~the (unitless) power spectrum slope. 
The p.s. values are only shown if they are part of the forward model, otherwise no value (N.A.) is given.
}
\end{small}
\end{table}

\section{Discussion}
\label{sec:discussion}

\subsection{The driver of expansion}
\label{subsec:expansion}

Inspecting the power spectra of the \texttt{main} field components in Fig.~\ref{fig:power}, we observe that the lines almost coincide at large wavenumbers $k$ (i.e., small scales), with the $x$-component being somewhat damped, but significantly deviating at the largest scales, and with the $z$-component having almost a magnitude less power than the other two, especially compared to $x$. 
This reflects the fact that the large-scale expansion of Sco-Cen is mainly in the Galactic plane, and also hints at Galactic rotation likely not being the main driver of the large-scale structure. 
In this case, one would expect a spherical structure to evolve in a slightly rotated ellipse \citep{1990Palous}, which implies that the $y$-spectrum should dominate over $x$ and $z$ at low $k$ as it points towards Galactic rotation. 
This could be explained with initial conditions, but would require that the proto-Sco-Cen cloud was extremely stratified in the $x$-direction.  

Inspecting the 3D scalar component fields (shown in Appendix~\ref{app:components}) corresponding to the spectra reveals a possible reason for the difference in power. 
The expected pattern from \citet{1990Palous} can be clearly seen in a top-down view of the field, which shows a clear large-scale gradient in velocities that could be readily fit with large $k$ modes. 
In 3D, however, it becomes clear that the $y$-component has a structure that destroys this large-scale pattern to some degree, namely the LCC chain, which has a $y$-velocity opposite to its likely parent structure, the $\sigma$-Cen cluster. 
This indicates that feedback processes can dominate over Galactic rotation even at ages of 20~Myr and scales of up to 100~pc.

At small scales, the observed similarity in power may indicate that the additional power needed to depart from self-similarity stems from a small-scale process, which we conjecture to be stellar feedback.
Here, we lack a good explanation for the somewhat damped spectrum in the $x$-component, as feedback is certainly isotropic on average. 
Since the $x$-direction is the one where RV-data matters the most due to the location of Sco-Cen with respect to the Sun, this might hint at missing small-scale structure in the data due to bad spatial RV sampling and relatively high noise in the RV data (see also the S/N analysis in Appendix~\ref{app:data}), but we cannot corroborate this conjecture.

If we take the picture of the evolution of Sco-Cen in the literature \citepalias{2023Ratzenboeck1, 2023Ratzenboeck2, 2025Posch, 2025Grossschedl} at face value, we can connect the time domain with the phenomenology of the power spectra. 
Each feedback process injects energy at small scales, and the corresponding structure expands out as it ages, which is equivalent to moving to lower $k$ in the power spectrum. 
We note that this proposed explanation is purely a memory effect reflecting the small-scale and localized energy injection, not a true inverse cascade which is observed numerically in, for example, ISM and cosmological magnetic fields \citep{2023Brandenburg}.
This explains the lack of power at larger scales, which are diluted as there is no additional momentum input to offset spreading.
We have to mention that the amplification of smaller scales could, in principle, also be a consequence of an unresolved systematic effect in data, which, in case it is uncorrelated with the flow, would act as white noise in the spectrum and hence amplify smaller scales.
The most likely culprit for this would be unresolved binarity in the RV data.
While we cannot disprove this, we deem it unlikely to dominate the spectra, as this should affect the $x$ and $y$ components more, as these are more often aligned with the radial direction due to the relative location of the Sun to Sco-Cen.

In contrast, the fact that $v_{\delta, z}$ has a Kolmogorov spectrum may indicate that this is the result of a singular event that has injected momentum relatively recently, and that the afore-mentioned spreading effect of feedback-induced structures has not happened yet significantly.
But we note that given the relatively small size of this sub-region compared to the resolution, it is unclear if this is a general result or a consequence of limited resolution.

We reiterate that the prior on the power spectrum is the same for all components, and specifically, the slope has a Gaussian prior centered on $7.3$ with a standard deviation of $3.0$. Since the slope is a log-log quantity, this prior is relatively constrained, hence indicating that this is a data-driven result. 
A caveat in the above discussion is that power spectra are averaged quantities. 
The fact that the \texttt{secondary} field has such a distinct correlation structure and the clear signature of feedback structures overlaying parts that might be driven by Galactic rotation might point towards the need to treat parts of Sco-Cen statistically separately depending on their substructure. 
This will be pursued in future work.

\subsection{Flow substructure and origin}

Figure~\ref{fig:flow_fields} illustrates that the data harbors velocity structure that is not captured by cluster averages. 
Evidence for small scale kinematic structure was already given by \citet{2018Wright}, who interpreted this as indication that the OB association has not gone through a phase of dynamical relaxation.
With reference to the \texttt{SigMA} clustering results, examples for this are significant sub-structure in the (rather extended) location of the  $\sigma$-Cen cluster, velocity patterns within the isolated Cham~1 and 2 clusters, and a notable motion at the side of Sco-Cen facing the Galactic center, mostly traced by stars constituting the  V1062-Sco, UPK\,606, and Centaurus-far clusters. 
In general, the visual impression appears to indicate that these structures have originated from a similar feedback-driven mechanism as the cluster chains. 
The spatial arrangement of the acceleration patterns seen in the stellar flow field potentially indicates the evolution in structure, ranging from bulk motion in proto Sco-Cen, to radial expanding bulk flow, for instance, in USco, to clearly defined cluster chains to smaller low mass extensions, as, for example, traced by stars constituting the L134/183 cluster.  
This might indicate that the morphology of the flow largely depends on the mass of the material that has to be moved by the feedback sources.

A tentative but notable result is the location of the absolute minimum of the velocity field magnitude of the \texttt{main} field at $(x, y, z) =  \left(85\,\mathrm{pc},  -70\,\mathrm{pc},  19\,\mathrm{pc}\right)$, as this may indicate the origin of the flow.
The position is very close but not coinciding (even when considering the error bars) with the geometric center of Sco-Cen as calculated by \citetalias{2025Grossschedl}.
We note that the location of this minimum depends on the choice of reference frame, and is hence far from settled. 
It is very difficult to reconcile the velocity center with any geometrically derived reference frame, as the flow center is located at the edge of stellar density that constitutes the plotting mask.   
It is, moreover, consistent with the on-sky location of the origin of flow of diffuse matter in the ISM, which is very close to the Sun (i.e., several dozens of pc), as inferred by \citet{2024Piecka}. 
This phenomenon may trace the oldest evolution history of Sco-Cen and may hint at anisotropic expansion of the proto-Sco-Cen cloud.
We defer the quantitative analysis of this phenomenon to Paper~II.

\subsection{The evolution of stellar density}
\label{subsec:density_evolution}

The typical divergence value of $0.2~\mathrm{Myr}^{-1}$ can be taken as a characteristic rate of expansion, and we can use it to predict both the future and past density evolution of Sco-Cen to first order approximation using $\rho_t = \rho_0 e^{-\nabla \vec{V}\,t}$. 
The typical stellar density of Sco-Cen is at about 0.1--0.2 stars per pc$^3$, hence, about 3--6 stars per voxel (see Appendix~\ref{app:stellar_density} and Fig.~\ref{fig:density}), while the background field is at about 0.1 stars per pc$^3$. 
This implies that Sco-Cen will have largely diffused as an over-dense structure to the 10\% level in about 11.5--15~Myr and to the 1\% level in about 23--26.5~Myr.   
We note that clusters are typically identified as phase space over-densities~\citep[e.g.,][]{2023Ratzenboeck1, 2023Hunt}, and that the kinematic structure of Sco-Cen might, in principle, allow for tracing it for longer \citep[e.g.,][]{2024Swiggum}.

Going back in time to the likely birth of the Sco-Cen association, 20~Myr ago, we arrive at initial density values of about 5--10  stars per pc$^3$ for the values above. 
We note that in this regime, the projected density values are highly uncertain due to the exponential nature of the approximation; just doubling the divergence value already gives values between 300 and 600 stars per pc$^3$, which interestingly lies in the same order of magnitude as the number of stars per pc$^3$ currently observed, e.g., in the Orion molecular cloud~\citep{1986Herbig, 1998Hillenbrand}.
These numbers are first-order approximations and neither include models for the historical and ongoing star formation nor appreciate the much more complicated spatial structure of the flow. 
A detailed analysis of the history of Sco-Cen will be pursued in follow-up work. 

\subsection{Comparison to Cluster Chains}
\label{subsubsec:compare_chains}
\begin{figure*}
\centering
    \begin{subfigure}[t]{0.31\textwidth}
        \includegraphics[width=\textwidth]{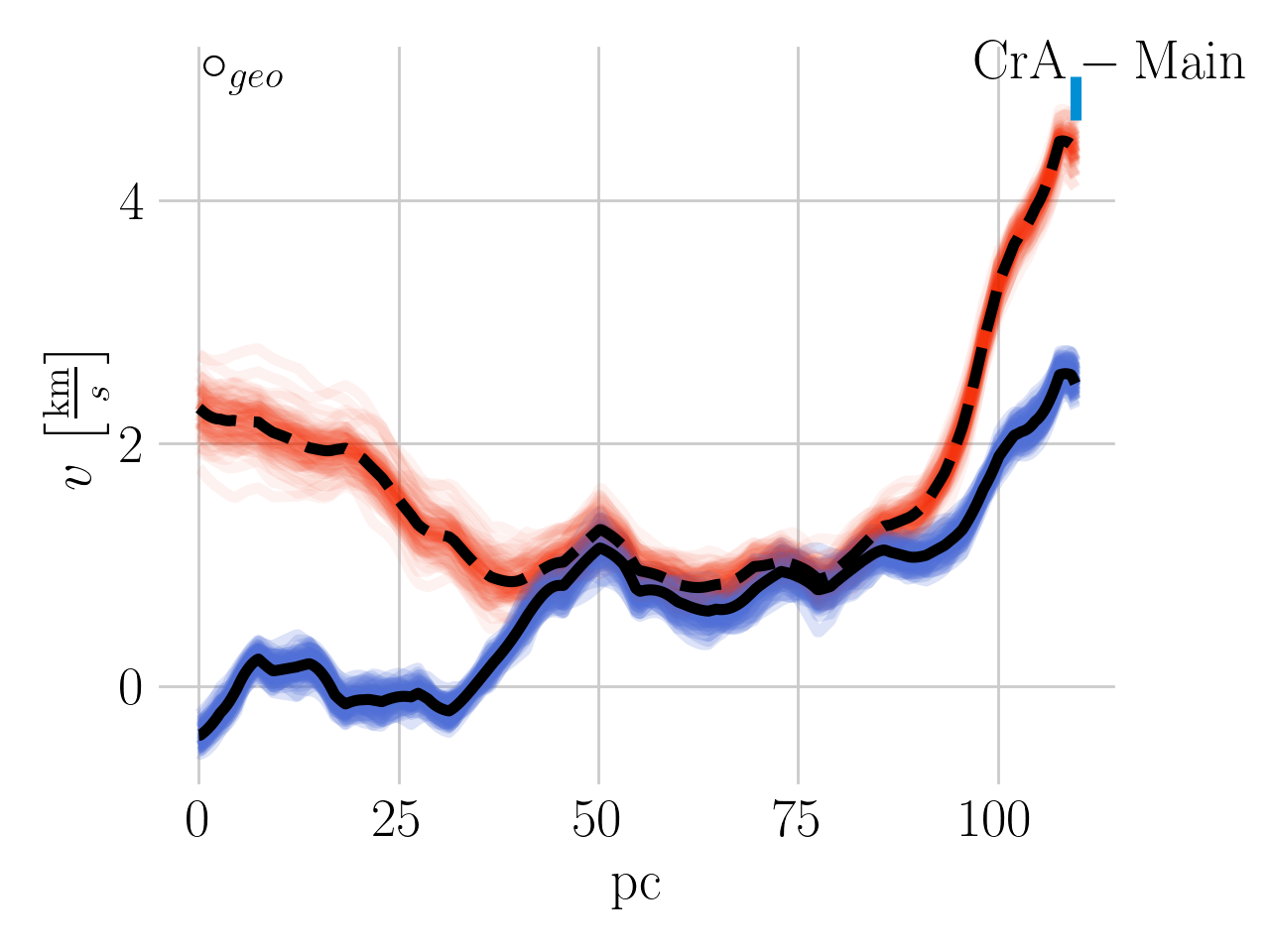}
        \caption{\label{fig:mass_to_cra} CrA-Chain}
    \end{subfigure}
    \begin{subfigure}[t]{0.31\textwidth}
        \includegraphics[width=\textwidth]{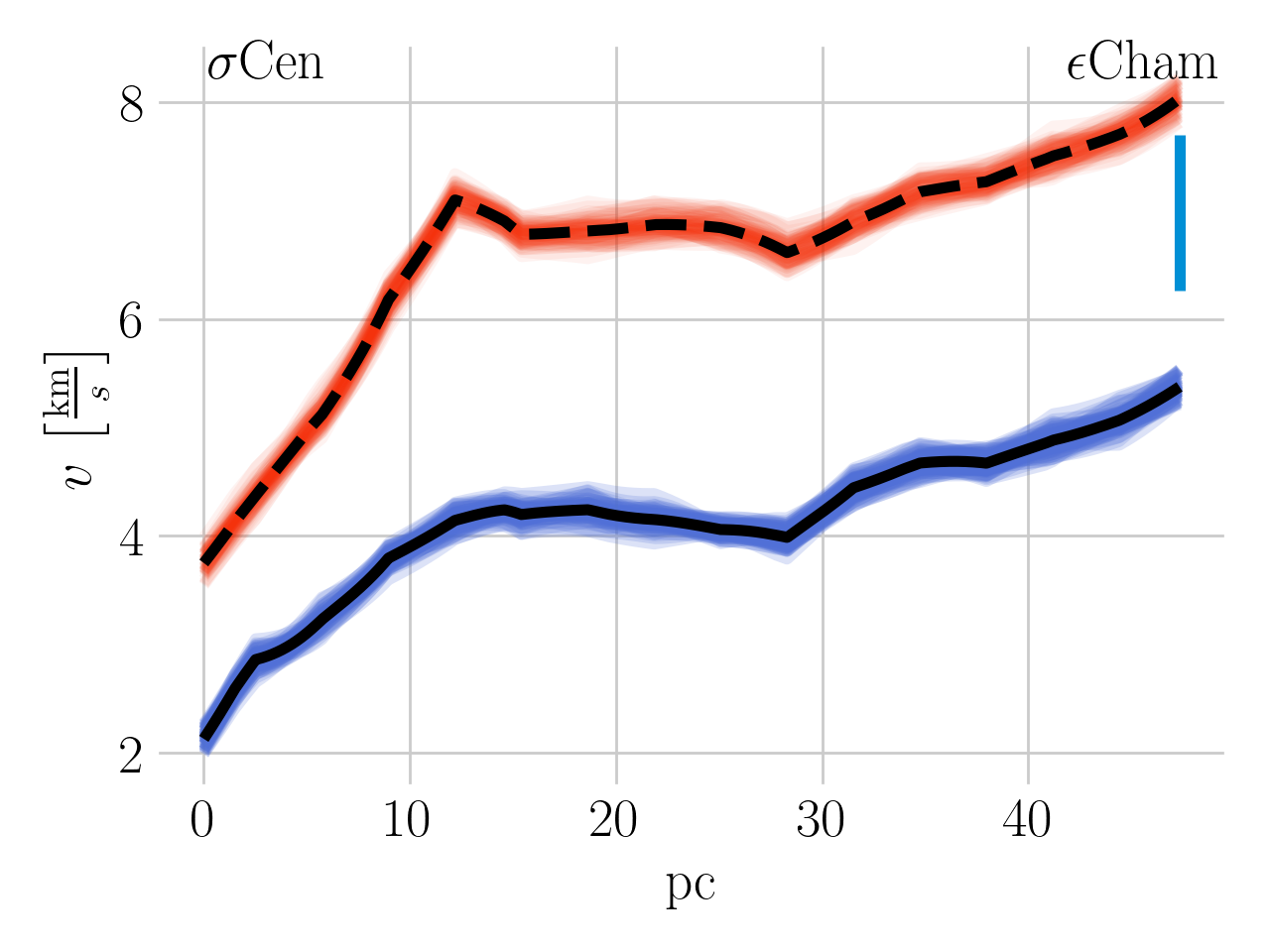}
        \caption{\label{fig:mass_to_epscham} LCC-Chain}
    \end{subfigure}
    \begin{subfigure}[t]{0.31\textwidth}
        \includegraphics[width=\textwidth]{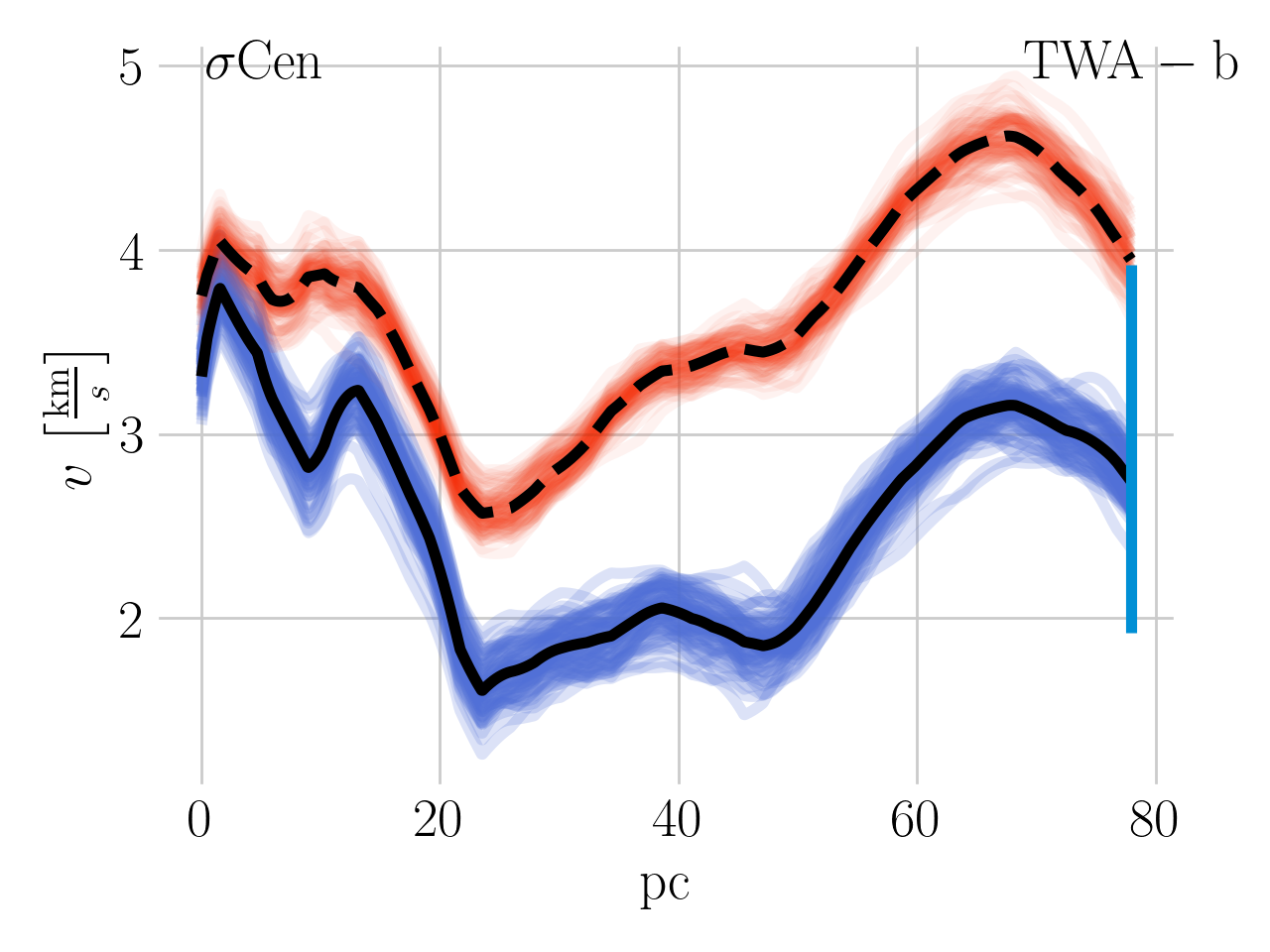}
        \caption{\label{fig:sigcen_to_twa} TWA-chain}
    \end{subfigure}
    \begin{subfigure}[t]{0.2\textwidth}
        \centering
        \includegraphics[width=\textwidth, ]{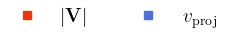}
    \end{subfigure}
    \caption{\label{fig:profiles} Projected velocity profiles of the \texttt{main} velocity field in the reference frame of \citet{2025Posch} along the CrA, LCC, and TWA chains analyzed in \citetalias{2023Posch, 2025Posch} and \citetalias{2025MiretRoig}. 
    We note that in the case of the CrA chain, the starting point is the geometric center of Sco-Cen as defined in \citetalias{2025Grossschedl}, while for the other two, it is the center of the $\sigma$-Cen cluster.
    The projection of the field onto the chain is shown in blue, while the absolute 3D magnitude is depicted in red. 
    All quantities are accompanied by the respective posterior samples in lower opacity to illustrate uncertainties. 
    We have included the values derived in the literature for the respective clusters at the end of the chains on the right of the plots as blue bars.}
\end{figure*}
Cluster chains in Sco-Cen have been recently discovered and are linear alignments of young star clusters with gradients in age, velocity, and mass. 
They are interpreted as signatures of triggered star formation via stellar feedback and thought to represent a major component of Sco-Cen and its star formation history \citepalias{2023Posch, 2025Posch, 2025MiretRoig}.
We make use of these objects to study the systematic differences between clustering and field-based methods.    
Specifically, we analyze the CrA \citepalias{2023Posch, 2025Posch}, LCC \citepalias{2025Posch}, and TWA chains \citepalias{2025MiretRoig}.  
In Fig.~\ref{fig:profiles}, we compare our analysis of the motion along the cluster chains with recently reported cluster velocities from the mentioned literature. 
We show both the projected and absolute values of the velocity field along the line segments that define the chains. All values, including the projections, are reported in the velocity frame of the respective reference publication. 
The comparison of the absolute values of the endpoints of our projections to the literature is not straightforward, as these values derived from clusters are averages over stellar populations, while we only plot the lines until the center of the respective clusters. 
Furthermore, the projection along a single line is an approximation, and the true chains are likely better described by chains of line segments.  
Smaller inconsistencies could also stem from the fact that we have used a slightly more extended RV-data set than \citetalias{2023Posch, 2025Posch, 2025MiretRoig}. 

In all cases, we confirm a positive velocity gradient of the component projected on the chain for most of the respective line segments.
In the case of the CrA chain, the projection reveals that in the first part of the chain, which traverses the main body of Sco-Cen, the velocity field does not align with the linear projection.
At about 25~pc from the start of the chain, the projected field is accelerated and up to about 80~pc is completely aligned with the full vector field, but experiences almost no acceleration. 
In the last 20~pc, the field is strongly accelerated. 
All in all, the CrA chain seems to have been boosted by two singular events, which seem to mirror the clustering results of \citetalias{2023Posch}.
The LCC chain, on the other hand, seems to be more continuously accelerated, with only a small plateau in its central part.  
Interestingly, the projected velocity component seems to have an almost constant offset to the absolute magnitude of the field, indicating that the full structure is moving relative to Sco-Cen independent of the internal acceleration. 
At last, the TWA chain, similarly to CrA, seems to experience acceleration in the beginning, where it is dominated by the internal structure of the \mbox{$\sigma$-Cen} cluster.
After the projection leaves the cluster, the chain sees constant acceleration, interestingly again with an almost constant offset to the absolute magnitude. 
This is similar to the LCC chain, which is almost co-located in $x$ and $y$ position.
There is a small deceleration happening at the end of the chain, which is not reported in \citetalias{2025MiretRoig}, as it likely was lost in the cluster average.
We note that this chain is much less sampled than the CrA and LCC chains, making the results more prone to outliers.  
We will defer a more detailed acceleration analysis to future work, not only along the chains but for the general flow field.

\subsection{Comparison to CO}
\label{subsec:compare_co}

Only small parts of Sco-Cen are still harboring enough dense gas to produce significant, observable CO emission.  
\citet{2009Tobin, 2016Hacar}; and \citet{2021Grossschedl} find that young stars that were recently born and which are still located close to their parent molecular cloud, as traced by CO, have not yet dynamically decoupled from the gas. The line-of-sight motions of these young stars (the RVs) are very similar to the gas RVs, at least in the first several $10^5$ to a few million years after star-formation.
We used the \citet{2001Dame} composite survey for the Ophiuchus and Lupus molecular clouds to derive first moment maps.
To compare the stellar line of sight motions to CO, we used the \citet{2023Edenhofer} dust map to weight and integrate the radial component of the \texttt{secondary} velocity field, as this is the field constrained by the $\rho$-Oph, B59, and Lupus clusters, respectively.
We converted the velocities to the local standard of rest frame (LSR) using the standard solar motion reported in \citet{2011Ando}, 
and the procedure outlined in \citet{2021Grossschedl} and \citetalias{2023Posch}.
This comparison relies on the idea that the dust in the \citet{2023Edenhofer} map traces the same quantity as the CO, which is likely only true to first order approximation. 

\begin{figure}
    \centering
    \includegraphics[width=\linewidth]{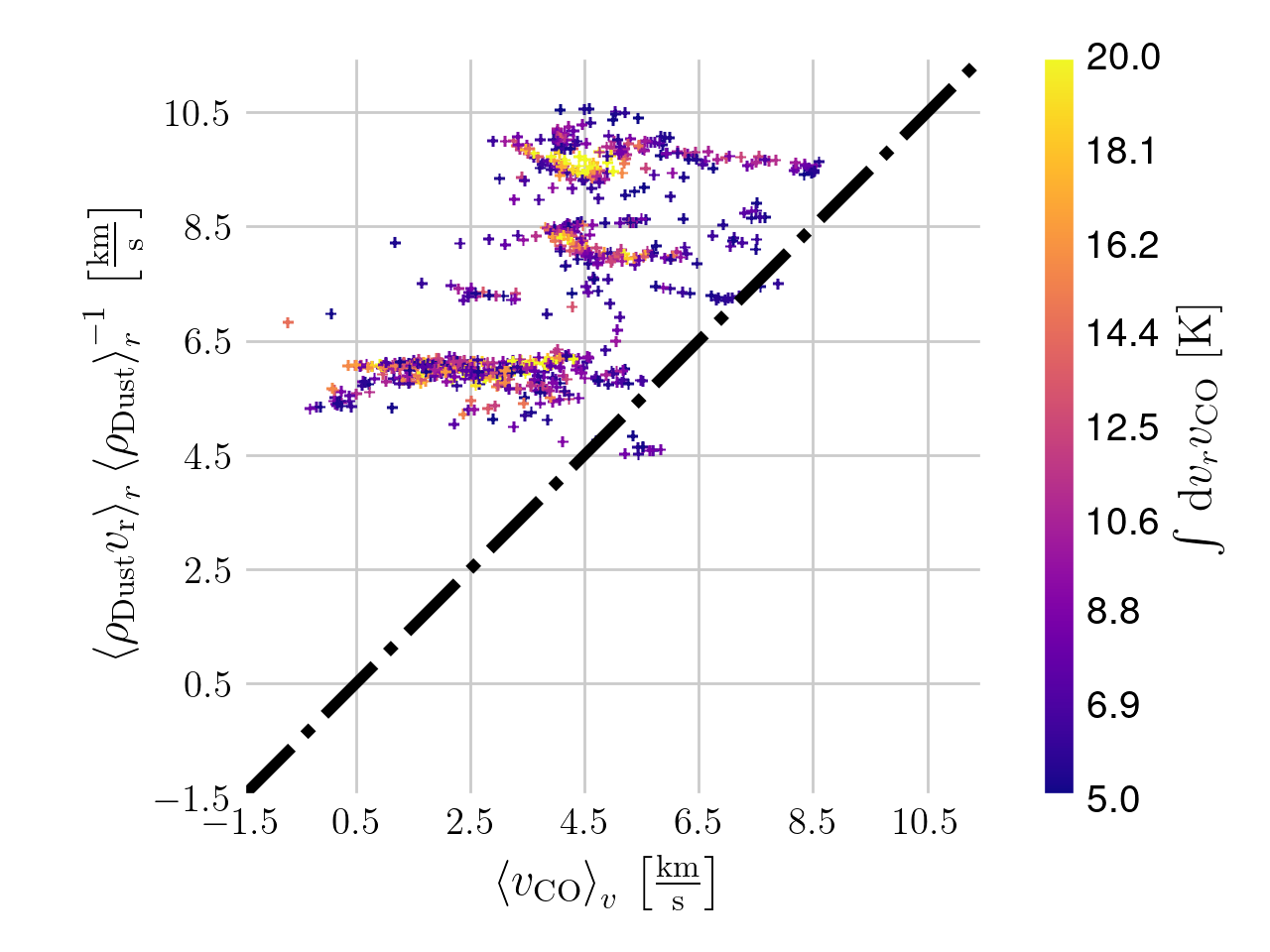}
    \caption{\label{fig:co_scatter}Correlation between the first moment map (i.e., the mean) over all velocity channels and the dust-weighted line-of-sight integral of the RV field of the \texttt{secondary} field in the LSR frame, with integrated CO intensity color coded.}
    \label{fig:co}
\end{figure}

We show a scatter plot between the first moment of the CO map and the dust-weighted integral of the \texttt{secondary} velocity field in the LSR frame in Fig.~\ref{fig:co}.
The plot shows that the CO and velocity field moments show the same large-scale pattern; hence, more negative CO velocities correspond to more negative averaged field moments, albeit with significant scatter.
But there is a clear offset between the two quantities, amounting to several km\,s$^{-1}$, with the CO almost consistently moving to more negative values, i.e., moving slower away from the LSR observer. 
We interpret this offset as ongoing decoupling between the stars and the gas, likely due to ongoing stellar feedback. 
We note that since we lack stars embedded within the molecular clouds (which cannot be observed in \textit{Gaia} due to high extinction), we lack the very young stars in our sample that might bridge this kinematic gap.
Interestingly, the direction of the velocity offset indicates that the source of acceleration cannot come from the young clusters associated with the molecular clouds, as these are located mostly in front of the dense gas, relative to the LSR observer, and should hence accelerate the gas away from us. 
To account for the gas moving more slowly away from the LSR observer than the stars, it would take clusters located behind the gas as seen from the observer.
We note that this is in line with the general trend in this part of Sco-Cen, already discussed in Sect.~\ref{subsec:vel_field} and \ref{subsec:pow_spec}, which similarly sees a general downward motion compared to the main velocity field. 
We hence speculate that the source of feedback that causes this downward acceleration is still ongoing and is located above the \texttt{secondary} population and USco.  

\section{Conclusion}

In this work, we present a method to infer the 3D flow field of stellar OB associations, and illustrate its capabilities by applying it to one of the closest young stellar associations, Sco-Cen. 
We reliably recover both qualitatively and quantitatively several features of the association already identified in the literature, such as the general inside-out acceleration pattern of the association and the cluster-chains connected to it. 
We demonstrate that the velocity field harbors significantly coherent structure below the scale of clusters.  

One strength of a field-level picture is the easy accessibility of statistical descriptors, most importantly, the power spectrum.  
We demonstrate that the correlation structure of Sco-Cen has an excess in small-scale structure relative to a scale-invariant structure. 
This indicates that the observed expansion structure is likely driven at these small scales, which must be feedback processes.
A stretching and acceleration due to Galactic rotation seems to be present, but is likely subdominant. 
Apart from structural insights, the field representation allows us to calculate derivative fields such as divergence and vorticity.
These encode temporal expansion and rotation rates, and we have connected these to the history of Sco-Cen. 

In this work, we mainly focus on the description of the method and a high-level description of the resulting maps, but only scratched the surface of what is possible in the analysis. 
In follow-up work, we will more deeply analyze the connection to ISM by deriving energy and momentum maps for the gas and dust fields in the region. 
We will furthermore analyze the acceleration structure of Sco-Cen in more detail, and deepen the analysis of the fluid characteristics of the field by calculating, e.g., the helicity density.
Moreover, we aim to apply the method to further nearby star-forming regions such as Taurus, Perseus, Orion, and Vela~OB2.  

The largest limitation of our work stems from the RV data, both to the higher observational noise level as well as due to the higher likelihood of systematic biases stemming from, e.g., binaries.
Both future data releases of ongoing surveys as well as newly developed instruments \citep[e.g.][]{2017Majewski, 20124MOST, 2024WEAVE, 2025Brown} will help remedy these issues, and potentially allow for a more detailed and better resolved reconstruction.

\begin{acknowledgements}
We would like to thank the referee for constructive comments.
S.H.~would like to thank Philipp Frank and Alena Rottensteiner for helpful discussions.
Co-funded by the European Union (ERC, ISM-FLOW, 101055318). 
Views and opinions expressed are, however, those of the author(s) only and do not necessarily reflect those of the European Union or the European Research Council. 
Neither the European Union nor the granting authority can be held responsible for them. 
The computational results have been achieved using the Austrian Scientific Computing (ASC) infrastructure.
JG acknowledges funding from the European Union, the Central Bohemian Region, and the Czech Academy of Sciences, as part of the MERIT fellowship (MSCA-COFUND Horizon Europe, Grant agreement 101081195).
The following Python libraries were used in this work: 
\texttt{numpy} \citep{numpy} , \texttt{scipy} \citep{scipy}, \texttt{astropy}
\citep{astropy3}, \texttt{nifty} \citep{niftyre}, \texttt{findiff} \citep{findiff}, \texttt{jax} \citep{jax2018github}, \texttt{jaxbind} \citep{jaxbind}, 
\texttt{cmasher} \citep{2020vanderVelden}, and \texttt{matplotlib} \citep{matplotlib} 
\end{acknowledgements}

\bibliographystyle{aa}
\bibliography{lib}

@ARTICLE{2016Pecaut,
       author = {{Pecaut}, Mark J. and {Mamajek}, Eric E.},
        title = "{The star formation history and accretion-disc fraction among the K-type members of the Scorpius-Centaurus OB association}",
      journal = {\mnras},
         year = 2016,
        month = sep,
       volume = {461},
       number = {1},
        pages = {794-815},
          doi = {10.1093/mnras/stw1300},
archivePrefix = {arXiv},
       eprint = {1605.08789},
 primaryClass = {astro-ph.SR},
       adsurl = {https://ui.adsabs.harvard.edu/abs/2016MNRAS.461..794P}
}

@ARTICLE{2018Queiroz,
       author = {{Queiroz}, A.~B.~A. and {Anders}, F. and {Santiago}, B.~X. and {Chiappini}, C. and {Steinmetz}, M. and {Dal Ponte}, M. and {Stassun}, K.~G. and {da Costa}, L.~N. and {Maia}, M.~A.~G. and {Crestani}, J. and {Beers}, T.~C. and {Fern{\'a}ndez-Trincado}, J.~G. and {Garc{\'\i}a-Hern{\'a}ndez}, D.~A. and {Roman-Lopes}, A. and {Zamora}, O.},
        title = "{StarHorse: a Bayesian tool for determining stellar masses, ages, distances, and extinctions for field stars}",
      journal = {\mnras},
         year = 2018,
        month = may,
       volume = {476},
       number = {2},
        pages = {2556-2583},
          doi = {10.1093/mnras/sty330},
archivePrefix = {arXiv},
       eprint = {1710.09970},
 primaryClass = {astro-ph.IM},
       adsurl = {https://ui.adsabs.harvard.edu/abs/2018MNRAS.476.2556Q}
}

@ARTICLE{2025Armstrong,
       author = {{Armstrong}, Joseph J. and {Tan}, Jonathan C. and {Wright}, Nicholas J. and {Jeffries}, R.~D. and {Kos}, Janez and {Fiorellino}, E. and {Buder}, Sven and {Barrios L{\'o}pez}, D.},
        title = "{Investigating the Upper Scorpius OB association with HERMES ─ I. The spectroscopic sample and 6D kinematics}",
      journal = {\mnras},
         year = 2025,
        month = nov,
       volume = {543},
       number = {3},
        pages = {2349-2373},
          doi = {10.1093/mnras/staf1490},
archivePrefix = {arXiv},
       eprint = {2505.03716},
 primaryClass = {astro-ph.SR},
       adsurl = {https://ui.adsabs.harvard.edu/abs/2025MNRAS.543.2349A}
}

@ARTICLE{2018Wright,
       author = {{Wright}, Nicholas J. and {Mamajek}, Eric E.},
        title = "{The kinematics of the Scorpius-Centaurus OB association from Gaia DR1}",
      journal = {\mnras},
         year = 2018,
        month = may,
       volume = {476},
       number = {1},
        pages = {381-398},
          doi = {10.1093/mnras/sty207},
archivePrefix = {arXiv},
       eprint = {1801.08540},
 primaryClass = {astro-ph.SR},
       adsurl = {https://ui.adsabs.harvard.edu/abs/2018MNRAS.476..381W}
}

@ARTICLE{2025Piecka,
       author = {{Piecka}, M. and {Posch}, L. and {Meingast}, S. and {Hutschenreuter}, S. and {Rottensteiner}, A. and {Alves}, J.},
        title = "{Direct measurement of ISM proper motion with image registration}",
      journal = {\aap},
         year = 2025,
        month = sep,
       volume = {702},
          eid = {L1},
        pages = {L1},
          doi = {10.1051/0004-6361/202556622},
archivePrefix = {arXiv},
       eprint = {2509.04857},
 primaryClass = {astro-ph.GA},
       adsurl = {https://ui.adsabs.harvard.edu/abs/2025A&A...702L...1P}
}

@ARTICLE{2025Brown,
       author = {{Brown}, Anthony G.~A.},
        title = "{Gaia: Ten Years of Surveying the Milky Way and Beyond}",
      journal = {arXiv e-prints},
         year = 2025,
        month = mar,
          eid = {arXiv:2503.01533},
        pages = {arXiv:2503.01533},
          doi = {10.48550/arXiv.2503.01533},
archivePrefix = {arXiv},
       eprint = {2503.01533},
 primaryClass = {astro-ph.GA},
       adsurl = {https://ui.adsabs.harvard.edu/abs/2025arXiv250301533B}
}

@article{1998Hillenbrand,
doi = {10.1086/305076},
url = {https://dx.doi.org/10.1086/305076},
year = {1998},
month = {jan},
publisher = {},
volume = {492},
number = {2},
pages = {540},
author = {Hillenbrand, Lynne A. and Hartmann, Lee W.},
title = {A Preliminary Study of the Orion Nebula Cluster Structure and Dynamics},
journal = {The Astrophysical Journal}
}

@ARTICLE{1986Herbig,
       author = {{Herbig}, G.~H. and {Terndrup}, D.~M.},
        title = "{The Trapezium Cluster of the Orion Nebula}",
      journal = {\apj},
         year = 1986,
        month = aug,
       volume = {307},
        pages = {609},
          doi = {10.1086/164447},
       adsurl = {https://ui.adsabs.harvard.edu/abs/1986ApJ...307..609H}
}

@software{jax2018github,
  author = {James Bradbury and Roy Frostig and Peter Hawkins and Matthew James Johnson and Chris Leary and Dougal Maclaurin and George Necula and Adam Paszke and Jake Vander{P}las and Skye Wanderman-{M}ilne and Qiao Zhang},
  title = {{JAX}: composable transformations of {P}ython+{N}um{P}y programs},
  url = {http://github.com/jax-ml/jax},
  version = {0.3.13},
  year = {2018},
}

@ARTICLE{astropy3,
       author = {{Astropy Collaboration} and {Price-Whelan}, Adrian M. and {Lim}, Pey Lian and {Earl}, Nicholas and {Starkman}, Nathaniel and {Bradley}, Larry and {Shupe}, David L. and {Patil}, Aarya A. and {Corrales}, Lia and {Brasseur}, C.~E. and {N{\"o}the}, Maximilian and {Donath}, Axel and {Tollerud}, Erik and {Morris}, Brett M. and {Ginsburg}, Adam and {Vaher}, Eero and {Weaver}, Benjamin A. and {Tocknell}, James and {Jamieson}, William and {van Kerkwijk}, Marten H. and {Robitaille}, Thomas P. and {Merry}, Bruce and {Bachetti}, Matteo and {G{\"u}nther}, H. Moritz and {Aldcroft}, Thomas L. and {Alvarado-Montes}, Jaime A. and {Archibald}, Anne M. and {B{\'o}di}, Attila and {Bapat}, Shreyas and {Barentsen}, Geert and {Baz{\'a}n}, Juanjo and {Biswas}, Manish and {Boquien}, M{\'e}d{\'e}ric and {Burke}, D.~J. and {Cara}, Daria and {Cara}, Mihai and {Conroy}, Kyle E. and {Conseil}, Simon and {Craig}, Matthew W. and {Cross}, Robert M. and {Cruz}, Kelle L. and {D'Eugenio}, Francesco and {Dencheva}, Nadia and {Devillepoix}, Hadrien A.~R. and {Dietrich}, J{\"o}rg P. and {Eigenbrot}, Arthur Davis and {Erben}, Thomas and {Ferreira}, Leonardo and {Foreman-Mackey}, Daniel and {Fox}, Ryan and {Freij}, Nabil and {Garg}, Suyog and {Geda}, Robel and {Glattly}, Lauren and {Gondhalekar}, Yash and {Gordon}, Karl D. and {Grant}, David and {Greenfield}, Perry and {Groener}, Austen M. and {Guest}, Steve and {Gurovich}, Sebastian and {Handberg}, Rasmus and {Hart}, Akeem and {Hatfield-Dodds}, Zac and {Homeier}, Derek and {Hosseinzadeh}, Griffin and {Jenness}, Tim and {Jones}, Craig K. and {Joseph}, Prajwel and {Kalmbach}, J. Bryce and {Karamehmetoglu}, Emir and {Ka{\l}uszy{\'n}ski}, Miko{\l}aj and {Kelley}, Michael S.~P. and {Kern}, Nicholas and {Kerzendorf}, Wolfgang E. and {Koch}, Eric W. and {Kulumani}, Shankar and {Lee}, Antony and {Ly}, Chun and {Ma}, Zhiyuan and {MacBride}, Conor and {Maljaars}, Jakob M. and {Muna}, Demitri and {Murphy}, N.~A. and {Norman}, Henrik and {O'Steen}, Richard and {Oman}, Kyle A. and {Pacifici}, Camilla and {Pascual}, Sergio and {Pascual-Granado}, J. and {Patil}, Rohit R. and {Perren}, Gabriel I. and {Pickering}, Timothy E. and {Rastogi}, Tanuj and {Roulston}, Benjamin R. and {Ryan}, Daniel F. and {Rykoff}, Eli S. and {Sabater}, Jose and {Sakurikar}, Parikshit and {Salgado}, Jes{\'u}s and {Sanghi}, Aniket and {Saunders}, Nicholas and {Savchenko}, Volodymyr and {Schwardt}, Ludwig and {Seifert-Eckert}, Michael and {Shih}, Albert Y. and {Jain}, Anany Shrey and {Shukla}, Gyanendra and {Sick}, Jonathan and {Simpson}, Chris and {Singanamalla}, Sudheesh and {Singer}, Leo P. and {Singhal}, Jaladh and {Sinha}, Manodeep and {Sip{\H{o}}cz}, Brigitta M. and {Spitler}, Lee R. and {Stansby}, David and {Streicher}, Ole and {{\v{S}}umak}, Jani and {Swinbank}, John D. and {Taranu}, Dan S. and {Tewary}, Nikita and {Tremblay}, Grant R. and {de Val-Borro}, Miguel and {Van Kooten}, Samuel J. and {Vasovi{\'c}}, Zlatan and {Verma}, Shresth and {de Miranda Cardoso}, Jos{\'e} Vin{\'\i}cius and {Williams}, Peter K.~G. and {Wilson}, Tom J. and {Winkel}, Benjamin and {Wood-Vasey}, W.~M. and {Xue}, Rui and {Yoachim}, Peter and {Zhang}, Chen and {Zonca}, Andrea and {Astropy Project Contributors}},
        title = "{The Astropy Project: Sustaining and Growing a Community-oriented Open-source Project and the Latest Major Release (v5.0) of the Core Package}",
      journal = {\apj},
         year = 2022,
        month = aug,
       volume = {935},
       number = {2},
          eid = {167},
        pages = {167},
          doi = {10.3847/1538-4357/ac7c74},
archivePrefix = {arXiv},
       eprint = {2206.14220},
 primaryClass = {astro-ph.IM},
       adsurl = {https://ui.adsabs.harvard.edu/abs/2022ApJ...935..167A}
}

@ARTICLE{scipy,
  author  = {Virtanen, Pauli and Gommers, Ralf and Oliphant, Travis E. and
            Haberland, Matt and Reddy, Tyler and Cournapeau, David and
            Burovski, Evgeni and Peterson, Pearu and Weckesser, Warren and
            Bright, Jonathan and {van der Walt}, St{\'e}fan J. and
            Brett, Matthew and Wilson, Joshua and Millman, K. Jarrod and
            Mayorov, Nikolay and Nelson, Andrew R. J. and Jones, Eric and
            Kern, Robert and Larson, Eric and Carey, C J and
            Polat, {\.I}lhan and Feng, Yu and Moore, Eric W. and
            {VanderPlas}, Jake and Laxalde, Denis and Perktold, Josef and
            Cimrman, Robert and Henriksen, Ian and Quintero, E. A. and
            Harris, Charles R. and Archibald, Anne M. and
            Ribeiro, Ant{\^o}nio H. and Pedregosa, Fabian and
            {van Mulbregt}, Paul and {SciPy 1.0 Contributors}},
  title   = {{{SciPy} 1.0: Fundamental Algorithms for Scientific
            Computing in Python}},
  journal = {Nature Methods},
  year    = {2020},
  volume  = {17},
  pages   = {261--272},
  adsurl  = {https://rdcu.be/b08Wh},
  doi     = {10.1038/s41592-019-0686-2},
}

@Article{numpy,
 title         = {Array programming with {NumPy}},
 author        = {Charles R. Harris and K. Jarrod Millman and St{\'{e}}fan J.
                 van der Walt and Ralf Gommers and Pauli Virtanen and David
                 Cournapeau and Eric Wieser and Julian Taylor and Sebastian
                 Berg and Nathaniel J. Smith and Robert Kern and Matti Picus
                 and Stephan Hoyer and Marten H. van Kerkwijk and Matthew
                 Brett and Allan Haldane and Jaime Fern{\'{a}}ndez del
                 R{\'{i}}o and Mark Wiebe and Pearu Peterson and Pierre
                 G{\'{e}}rard-Marchant and Kevin Sheppard and Tyler Reddy and
                 Warren Weckesser and Hameer Abbasi and Christoph Gohlke and
                 Travis E. Oliphant},
 year          = {2020},
 month         = sep,
 journal       = {Nature},
 volume        = {585},
 number        = {7825},
 pages         = {357--362},
 doi           = {10.1038/s41586-020-2649-2},
 publisher     = {Springer Science and Business Media {LLC}},
 url           = {https://doi.org/10.1038/s41586-020-2649-2}
}

@Article{matplotlib,
  Author    = {Hunter, J. D.},
  Title     = {Matplotlib: A 2D graphics environment},
  Journal   = {Computing in Science \& Engineering},
  Volume    = {9},
  Number    = {3},
  Pages     = {90--95},
  publisher = {IEEE COMPUTER SOC},
  doi       = {10.1109/MCSE.2007.55},
  year      = 2007
}

@article{jaxbind,
    title = {JAXbind: Bind any function to JAX},
    author = {Jakob Roth and Martin Reinecke and Gordian Edenhofer},
    year = {2024},
    journal = {Journal of Open Source Software},
    publisher = {The Open Journal},
    volume = {9},
    number = {98},
    pages = {6532},
    doi = {10.21105/joss.06532},
    url = {https://doi.org/10.21105/joss.06532},
}

@ARTICLE{2020vanderVelden,
       author = {{van der Velden}, Ellert},
        title = "{CMasher: Scientific colormaps for making accessible, informative and 'cmashing' plots}",
      journal = {The Journal of Open Source Software},
         year = 2020,
        month = feb,
       volume = {5},
       number = {46},
          eid = {2004},
        pages = {2004},
          doi = {10.21105/joss.02004},
archivePrefix = {arXiv},
       eprint = {2003.01069},
 primaryClass = {eess.IV},
       adsurl = {https://ui.adsabs.harvard.edu/abs/2020JOSS....5.2004V}
}

@article{niftyre,
  title     = {Re-Envisioning Numerical Information Field Theory (NIFTy.re): A Library for Gaussian Processes and Variational Inference},
  author    = {Gordian Edenhofer and Philipp Frank and Jakob Roth and Reimar H. Leike and Massin Guerdi and Lukas I. Scheel-Platz and Matteo Guardiani and Vincent Eberle and Margret Westerkamp and Torsten A. Enßlin},
  year      = {2024},
  journal   = {Journal of Open Source Software},
  publisher = {The Open Journal},
  volume    = {9},
  number    = {98},
  pages     = {6593},
  doi       = {10.21105/joss.06593},
  url       = {https://doi.org/10.21105/joss.06593},
}

@misc{findiff,
  title = {{findiff} Software Package},
  author = {M. Baer},
  url = {https://github.com/maroba/findiff},
  key = {findiff},
  note = {\url{https://github.com/maroba/findiff}},
  year = {2018}
}

@ARTICLE{2020Neuhauser,
       author = {{Neuh{\"a}user}, R. and {Gie{\ss}ler}, F. and {Hambaryan}, V.~V.},
        title = "{A nearby recent supernova that ejected the runaway star {\ensuremath{\zeta}} Oph, the pulsar PSR B1706-16, and $^{60}$Fe found on Earth}",
      journal = {\mnras},
         year = 2020,
        month = oct,
       volume = {498},
       number = {1},
        pages = {899-917},
          doi = {10.1093/mnras/stz2629},
archivePrefix = {arXiv},
       eprint = {1909.06850},
 primaryClass = {astro-ph.HE},
       adsurl = {https://ui.adsabs.harvard.edu/abs/2020MNRAS.498..899N}
}

@ARTICLE{2023BricenoMorales,
       author = {{Brice{\~n}o-Morales}, Geovanny and {Chanam{\'e}}, Julio},
        title = "{Substructure, supernovae, and a time-resolved star formation history for Upper Scorpius}",
      journal = {\mnras},
         year = 2023,
        month = jun,
       volume = {522},
       number = {1},
        pages = {1288-1309},
          doi = {10.1093/mnras/stad608},
archivePrefix = {arXiv},
       eprint = {2205.01735},
 primaryClass = {astro-ph.GA},
       adsurl = {https://ui.adsabs.harvard.edu/abs/2023MNRAS.522.1288B}
}

@ARTICLE{1990Palous,
       author = {{Palous}, J. and {Franco}, J. and {Tenorio-Tagle}, G.},
        title = "{The evolution of superstructures expanding in differentially rotatingdisks.}",
      journal = {\aap},
         year = 1990,
        month = jan,
       volume = {227},
        pages = {175-182},
       adsurl = {https://ui.adsabs.harvard.edu/abs/1990A&A...227..175P}
}

@ARTICLE{1981Larson,
       author = {{Larson}, R.~B.},
        title = "{Turbulence and star formation in molecular clouds.}",
      journal = {\mnras},
         year = 1981,
        month = mar,
       volume = {194},
        pages = {809-826},
          doi = {10.1093/mnras/194.4.809},
       adsurl = {https://ui.adsabs.harvard.edu/abs/1981MNRAS.194..809L}
}

@ARTICLE{2023Brandenburg,
       author = {{Brandenburg}, Axel and {Ntormousi}, Evangelia},
        title = "{Galactic Dynamos}",
      journal = {\araa},
         year = 2023,
        month = aug,
       volume = {61},
        pages = {561-606},
          doi = {10.1146/annurev-astro-071221-052807},
archivePrefix = {arXiv},
       eprint = {2211.03476},
 primaryClass = {astro-ph.GA},
       adsurl = {https://ui.adsabs.harvard.edu/abs/2023ARA&A..61..561B}
}

@ARTICLE{2024WEAVE,
       author = {{Jin}, Shoko and {Trager}, Scott C. and {Dalton}, Gavin B. and {Aguerri}, J. Alfonso L. and {Drew}, J.~E. and {Falc{\'o}n-Barroso}, Jes{\'u}s and {G{\"a}nsicke}, Boris T. and {Hill}, Vanessa and {Iovino}, Angela and {Pieri}, Matthew M. and {Poggianti}, Bianca M. and {Smith}, D.~J.~B. and {Vallenari}, Antonella and {Abrams}, Don Carlos and {Aguado}, David S. and {Antoja}, Teresa and {Arag{\'o}n-Salamanca}, Alfonso and {Ascasibar}, Yago and {Babusiaux}, Carine and {Balcells}, Marc and {Barrena}, R. and {Battaglia}, Giuseppina and {Belokurov}, Vasily and {Bensby}, Thomas and {Bonifacio}, Piercarlo and {Bragaglia}, Angela and {Carrasco}, Esperanza and {Carrera}, Ricardo and {Cornwell}, Daniel J. and {Dom{\'\i}nguez-Palmero}, Lilian and {Duncan}, Kenneth J. and {Famaey}, Benoit and {Fari{\~n}a}, Cecilia and {Gonzalez}, Oscar A. and {Guest}, Steve and {Hatch}, Nina A. and {Hess}, Kelley M. and {Hoskin}, Matthew J. and {Irwin}, Mike and {Knapen}, Johan H. and {Koposov}, Sergey E. and {Kuchner}, Ulrike and {Laigle}, Clotilde and {Lewis}, Jim and {Longhetti}, Marcella and {Lucatello}, Sara and {M{\'e}ndez-Abreu}, Jairo and {Mercurio}, Amata and {Molaeinezhad}, Alireza and {Mongui{\'o}}, Maria and {Morrison}, Sean and {Murphy}, David N.~A. and {Peralta de Arriba}, Luis and {P{\'e}rez}, Isabel and {P{\'e}rez-R{\`a}fols}, Ignasi and {Pic{\'o}}, Sergio and {Raddi}, Roberto and {Romero-G{\'o}mez}, Merc{\`e} and {Royer}, Fr{\'e}d{\'e}ric and {Siebert}, Arnaud and {Seabroke}, George M. and {Som}, Debopam and {Terrett}, David and {Thomas}, Guillaume and {Wesson}, Roger and {Worley}, C. Clare and {Alfaro}, Emilio J. and {Allende Prieto}, Carlos and {Alonso-Santiago}, Javier and {Amos}, Nicholas J. and {Ashley}, Richard P. and {Balaguer-N{\'u}{\~n}ez}, Lola and {Balbinot}, Eduardo and {Bellazzini}, Michele and {Benn}, Chris R. and {Berlanas}, Sara R. and {Bernard}, Edouard J. and {Best}, Philip and {Bettoni}, Daniela and {Bianco}, Andrea and {Bishop}, Georgia and {Blomqvist}, Michael and {Boeche}, Corrado and {Bolzonella}, Micol and {Bonoli}, Silvia and {Bosma}, Albert and {Britavskiy}, Nikolay and {Busarello}, Gianni and {Caffau}, Elisabetta and {Cantat-Gaudin}, Tristan and {Castro-Ginard}, Alfred and {Couto}, Guilherme and {Carbajo-Hijarrubia}, Juan and {Carter}, David and {Casamiquela}, Laia and {Conrado}, Ana M. and {Corcho-Caballero}, Pablo and {Costantin}, Luca and {Deason}, Alis and {de Burgos}, Abel and {De Grandi}, Sabrina and {Di Matteo}, Paola and {Dom{\'\i}nguez-G{\'o}mez}, Jes{\'u}s and {Dorda}, Ricardo and {Drake}, Alyssa and {Dutta}, Rajeshwari and {Erkal}, Denis and {Feltzing}, Sofia and {Ferr{\'e}-Mateu}, Anna and {Feuillet}, Diane and {Figueras}, Francesca and {Fossati}, Matteo and {Franciosini}, Elena and {Frasca}, Antonio and {Fumagalli}, Michele and {Gallazzi}, Anna and {Garc{\'\i}a-Benito}, Rub{\'e}n and {Gentile Fusillo}, Nicola and {Gebran}, Marwan and {Gilbert}, James and {Gledhill}, T.~M. and {Gonz{\'a}lez Delgado}, Rosa M. and {Greimel}, Robert and {Guarcello}, Mario Giuseppe and {Guerra}, Jose and {Gullieuszik}, Marco and {Haines}, Christopher P. and {Hardcastle}, Martin J. and {Harris}, Amy and {Haywood}, Misha and {Helmi}, Amina and {Hernandez}, Nauzet and {Herrero}, Artemio and {Hughes}, Sarah and {Ir{\v{s}}i{\v{c}}}, Vid and {Jablonka}, Pascale and {Jarvis}, Matt J. and {Jordi}, Carme and {Kondapally}, Rohit and {Kordopatis}, Georges and {Krogager}, Jens-Kristian and {La Barbera}, Francesco and {Lam}, Man I. and {Larsen}, S{\o}ren S. and {Lemasle}, Bertrand and {Lewis}, Ian J. and {Lhom{\'e}}, Emilie and {Lind}, Karin and {Lodi}, Marcello and {Longobardi}, Alessia and {Lonoce}, Ilaria and {Magrini}, Laura and {Ma{\'\i}z Apell{\'a}niz}, Jes{\'u}s and {Marchal}, Olivier and {Marco}, Amparo and {Martin}, Nicolas F. and {Matsuno}, Tadafumi and {Maurogordato}, Sophie and {Merluzzi}, Paola and {Miralda-Escud{\'e}}, Jordi and {Molinari}, Emilio and {Monari}, Giacomo and {Morelli}, Lorenzo and {Mottram}, Christopher J. and {Naylor}, Tim and {Negueruela}, Ignacio and {O{\~n}orbe}, Jose and {Pancino}, Elena and {Peirani}, S{\'e}bastien and {Peletier}, Reynier F. and {Pozzetti}, Lucia and {Rainer}, Monica and {Ramos}, Pau and {Read}, Shaun C. and {Rossi}, Elena Maria and {R{\"o}ttgering}, Huub J.~A. and {Rubi{\~n}o-Mart{\'\i}n}, Jose Alberto and {Sabater}, Jose and {San Juan}, Jos{\'e} and {Sanna}, Nicoletta and {Schallig}, Ellen and {Schiavon}, Ricardo P. and {Schultheis}, Mathias and {Serra}, Paolo and {Shimwell}, Timothy W. and {Sim{\'o}n-D{\'\i}az}, Sergio and {Smith}, Russell J. and {Sordo}, Rosanna and {Sorini}, Daniele and {Soubiran}, Caroline and {Starkenburg}, Else and {Steele}, Iain A. and {Stott}, John and {Stuik}, Remko and {Tolstoy}, Eline and {Tortora}, Crescenzo and {Tsantaki}, Maria and {Van der Swaelmen}, Mathieu and {van Weeren}, Reinout J. and {Vergani}, Daniela},
        title = "{The wide-field, multiplexed, spectroscopic facility WEAVE: Survey design, overview, and simulated implementation}",
      journal = {\mnras},
         year = 2024,
        month = may,
       volume = {530},
       number = {3},
        pages = {2688-2730},
          doi = {10.1093/mnras/stad557},
archivePrefix = {arXiv},
       eprint = {2212.03981},
 primaryClass = {astro-ph.IM},
       adsurl = {https://ui.adsabs.harvard.edu/abs/2024MNRAS.530.2688J}
}

@INPROCEEDINGS{20124MOST,
       author = {{de Jong}, Roelof S. and {Bellido-Tirado}, Olga and {Chiappini}, Cristina and {Depagne}, {\'E}ric and {Haynes}, Roger and {Johl}, Diana and {Schnurr}, Olivier and {Schwope}, Axel and {Walcher}, Jakob and {Dionies}, Frank and {Haynes}, Dionne and {Kelz}, Andreas and {Kitaura}, Francisco S. and {Lamer}, Georg and {Minchev}, Ivan and {M{\"u}ller}, Volker and {Nuza}, Sebasti{\'a}n. E. and {Olaya}, Jean-Christophe and {Piffl}, Tilmann and {Popow}, Emil and {Steinmetz}, Matthias and {Ural}, Ugur and {Williams}, Mary and {Winkler}, Roland and {Wisotzki}, Lutz and {Ansorge}, Wolfgang R. and {Banerji}, Manda and {Gonzalez Solares}, Eduardo and {Irwin}, Mike and {Kennicutt}, Robert C. and {King}, Dave and {McMahon}, Richard G. and {Koposov}, Sergey and {Parry}, Ian R. and {Sun}, David and {Walton}, Nicholas A. and {Finger}, Gert and {Iwert}, Olaf and {Krumpe}, Mirko and {Lizon}, Jean-Louis and {Vincenzo}, Mainieri and {Amans}, Jean-Philippe and {Bonifacio}, Piercarlo and {Cohen}, Mathieu and {Francois}, Patrick and {Jagourel}, Pascal and {Mignot}, Shan B. and {Royer}, Fr{\'e}d{\'e}ric and {Sartoretti}, Paola and {Bender}, Ralf and {Grupp}, Frank and {Hess}, Hans-Joachim and {Lang-Bardl}, Florian and {Muschielok}, Bernard and {B{\"o}hringer}, Hans and {Boller}, Thomas and {Bongiorno}, Angela and {Brusa}, Marcella and {Dwelly}, Tom and {Merloni}, Andrea and {Nandra}, Kirpal and {Salvato}, Mara and {Pragt}, Johannes H. and {Navarro}, Ram{\'o}n and {Gerlofsma}, Gerrit and {Roelfsema}, Ronald and {Dalton}, Gavin B. and {Middleton}, Kevin F. and {Tosh}, Ian A. and {Boeche}, Corrado and {Caffau}, Elisabetta and {Christlieb}, Norbert and {Grebel}, Eva K. and {Hansen}, Camilla and {Koch}, Andreas and {Ludwig}, Hans-G. and {Quirrenbach}, Andreas and {Sbordone}, Luca and {Seifert}, Walter and {Thimm}, Guido and {Trifonov}, Trifon and {Helmi}, Amina and {Trager}, Scott C. and {Feltzing}, Sofia and {Korn}, Andreas and {Boland}, Wilfried},
        title = "{4MOST: 4-metre multi-object spectroscopic telescope}",
    booktitle = {Ground-based and Airborne Instrumentation for Astronomy IV},
         year = 2012,
       editor = {{McLean}, Ian S. and {Ramsay}, Suzanne K. and {Takami}, Hideki},
       series = {Society of Photo-Optical Instrumentation Engineers (SPIE) Conference Series},
       volume = {8446},
        month = sep,
          eid = {84460T},
        pages = {84460T},
          doi = {10.1117/12.926239},
archivePrefix = {arXiv},
       eprint = {1206.6885},
 primaryClass = {astro-ph.IM},
       adsurl = {https://ui.adsabs.harvard.edu/abs/2012SPIE.8446E..0TD}
}

@ARTICLE{2025Alves,
       author = {{Alves}, Jo{\~a}o and {Lombardi}, Marco and {Lada}, Charles J.},
        title = "{HP2 Survey: V. Ophiuchus: Filament formation in a dispersing cloud complex}",
      journal = {\aap},
         year = 2025,
        month = may,
       volume = {697},
          eid = {A208},
        pages = {A208},
          doi = {10.1051/0004-6361/202452881},
archivePrefix = {arXiv},
       eprint = {2501.13931},
 primaryClass = {astro-ph.GA},
       adsurl = {https://ui.adsabs.harvard.edu/abs/2025A&A...697A.208A}
}

@ARTICLE{2023Hunt,
       author = {{Hunt}, Emily L. and {Reffert}, Sabine},
        title = "{Improving the open cluster census. II. An all-sky cluster catalogue with Gaia DR3}",
      journal = {\aap},
         year = 2023,
        month = may,
       volume = {673},
          eid = {A114},
        pages = {A114},
          doi = {10.1051/0004-6361/202346285},
archivePrefix = {arXiv},
       eprint = {2303.13424},
 primaryClass = {astro-ph.GA},
       adsurl = {https://ui.adsabs.harvard.edu/abs/2023A&A...673A.114H}
}

@INPROCEEDINGS{2023Chevance,
       author = {{Chevance}, M. and {Krumholz}, M.~R. and {McLeod}, A.~F. and {Ostriker}, E.~C. and {Rosolowsky}, E.~W. and {Sternberg}, A.},
        title = "{The Life and Times of Giant Molecular Clouds}",
    booktitle = {Protostars and Planets VII},
         year = 2023,
       editor = {{Inutsuka}, S. and {Aikawa}, Y. and {Muto}, T. and {Tomida}, K. and {Tamura}, M.},
       series = {Astronomical Society of the Pacific Conference Series},
       volume = {534},
        month = jul,
        pages = {1},
          doi = {10.48550/arXiv.2203.09570},
archivePrefix = {arXiv},
       eprint = {2203.09570},
 primaryClass = {astro-ph.GA},
       adsurl = {https://ui.adsabs.harvard.edu/abs/2023ASPC..534....1C}
}

@ARTICLE{2025Arunima,
       author = {{Arunima}, Arunima and {Krumholz}, Mark R. and {Ireland}, Michael J. and {Zhang}, Chuhan and {Hu}, Zipeng},
        title = "{Fundamental limits to orbit reconstruction due to non-conservation of stellar actions in a Milky Way-like simulation}",
      journal = {\mnras},
         year = 2025,
        month = oct,
       volume = {543},
       number = {1},
        pages = {358-374},
          doi = {10.1093/mnras/staf1515},
archivePrefix = {arXiv},
       eprint = {2503.00373},
 primaryClass = {astro-ph.GA},
       adsurl = {https://ui.adsabs.harvard.edu/abs/2025MNRAS.543..358A}
}

@article{2007McKee,
   author = "McKee, Christopher F. and Ostriker, Eve C.",
   title = "Theory of Star Formation", 
   journal= "Annual Review of Astronomy and Astrophysics",
   year = "2007",
   volume = "45",
   number = "Volume 45, 2007",
   pages = "565-687",
   doi = "https://doi.org/10.1146/annurev.astro.45.051806.110602",
   url = "https://www.annualreviews.org/content/journals/10.1146/annurev.astro.45.051806.110602",
   publisher = "Annual Reviews",
   issn = "1545-4282",
   type = "Journal Article",
  }

@ARTICLE{2009Tobin,
       author = {{Tobin}, John J. and {Hartmann}, Lee and {Furesz}, Gabor and {Mateo}, Mario and {Megeath}, S. Tom},
        title = "{Kinematics of the Orion Nebula Cluster: Velocity Substructure and Spectroscopic Binaries}",
      journal = {\apj},
         year = 2009,
        month = jun,
       volume = {697},
       number = {2},
        pages = {1103-1118},
          doi = {10.1088/0004-637X/697/2/1103},
archivePrefix = {arXiv},
       eprint = {0903.2775},
 primaryClass = {astro-ph.SR},
       adsurl = {https://ui.adsabs.harvard.edu/abs/2009ApJ...697.1103T}
}

@ARTICLE{2016Hacar,
       author = {{Hacar}, A. and {Alves}, J. and {Forbrich}, J. and {Meingast}, S. and {Kubiak}, K. and {Gro{\ss}schedl}, J.},
        title = "{APOGEE strings: A fossil record of the gas kinematic structure}",
      journal = {\aap},
         year = 2016,
        month = may,
       volume = {589},
          eid = {A80},
        pages = {A80},
          doi = {10.1051/0004-6361/201527805},
archivePrefix = {arXiv},
       eprint = {1602.01854},
 primaryClass = {astro-ph.GA},
       adsurl = {https://ui.adsabs.harvard.edu/abs/2016A&A...589A..80H}
}

@ARTICLE{1999ATruelove,
       author = {{Truelove}, J. Kelly and {McKee}, Christopher F.},
        title = "{Evolution of Nonradiative Supernova Remnants}",
      journal = {\apjs},
         year = 1999,
        month = feb,
       volume = {120},
       number = {2},
        pages = {299-326},
          doi = {10.1086/313176},
       adsurl = {https://ui.adsabs.harvard.edu/abs/1999ApJS..120..299T}
}

@ARTICLE{2021Grossschedl,
       author = {{Gro{\ss}schedl}, Josefa E. and {Alves}, Jo{\~a}o and {Meingast}, Stefan and {Herbst-Kiss}, Gabor},
        title = "{3D dynamics of the Orion cloud complex. Discovery of coherent radial gas motions at the 100-pc scale}",
      journal = {\aap},
         year = 2021,
        month = mar,
       volume = {647},
          eid = {A91},
        pages = {A91},
          doi = {10.1051/0004-6361/202038913},
archivePrefix = {arXiv},
       eprint = {2007.07254},
 primaryClass = {astro-ph.SR},
       adsurl = {https://ui.adsabs.harvard.edu/abs/2021A&A...647A..91G}
}

@INPROCEEDINGS{2023Zucker,
       author = {{Zucker}, C. and {Alves}, J. and {Goodman}, A. and {Meingast}, S. and {Galli}, P.},
        title = "{The Solar Neighborhood in the Age of Gaia}",
    booktitle = {Protostars and Planets VII},
         year = 2023,
       editor = {{Inutsuka}, S. and {Aikawa}, Y. and {Muto}, T. and {Tomida}, K. and {Tamura}, M.},
       series = {Astronomical Society of the Pacific Conference Series},
       volume = {534},
        month = jul,
        pages = {43},
          doi = {10.48550/arXiv.2212.00067},
archivePrefix = {arXiv},
       eprint = {2212.00067},
 primaryClass = {astro-ph.GA},
       adsurl = {https://ui.adsabs.harvard.edu/abs/2023ASPC..534...43Z}
}

@ARTICLE{2022Rybizki,
       author = {{Rybizki}, Jan and {Green}, Gregory M. and {Rix}, Hans-Walter and {El-Badry}, Kareem and {Demleitner}, Markus and {Zari}, Eleonora and {Udalski}, Andrzej and {Smart}, Richard L. and {Gould}, Andrew},
        title = "{A classifier for spurious astrometric solutions in Gaia eDR3}",
      journal = {\mnras},
         year = 2022,
        month = feb,
       volume = {510},
       number = {2},
        pages = {2597-2616},
          doi = {10.1093/mnras/stab3588},
archivePrefix = {arXiv},
       eprint = {2101.11641},
 primaryClass = {astro-ph.IM},
       adsurl = {https://ui.adsabs.harvard.edu/abs/2022MNRAS.510.2597R}
}

@ARTICLE{2021Forbes,
       author = {{Forbes}, John C. and {Alves}, Jo{\~a}o and {Lin}, Douglas N.~C.},
        title = "{A Solar System formation analogue in the Ophiuchus star-forming complex}",
      journal = {Nature Astronomy},
         year = 2021,
        month = oct,
       volume = {5},
        pages = {1009-1016},
          doi = {10.1038/s41550-021-01442-9},
archivePrefix = {arXiv},
       eprint = {2108.09326},
 primaryClass = {astro-ph.EP},
       adsurl = {https://ui.adsabs.harvard.edu/abs/2021NatAs...5.1009F}
}

@ARTICLE{2010Poeppel,
       author = {{P{\"o}ppel}, W.~G.~L. and {Bajaja}, E. and {Arnal}, E.~M. and {Morras}, R.},
        title = "{The interstellar medium surrounding the Scorpius-Centaurus association revisited}",
      journal = {\aap},
         year = 2010,
        month = mar,
       volume = {512},
          eid = {A83},
        pages = {A83},
          doi = {10.1051/0004-6361/200811290},
       adsurl = {https://ui.adsabs.harvard.edu/abs/2010A&A...512A..83P}
}

@ARTICLE{2007AMakarovB,
       author = {{Makarov}, Valeri V.},
        title = "{Signatures of Dynamical Star Formation in the Ophiuchus Association of Pre-Main-Sequence Stars}",
      journal = {\apj},
         year = 2007,
        month = dec,
       volume = {670},
       number = {2},
        pages = {1225-1233},
          doi = {10.1086/522669},
       adsurl = {https://ui.adsabs.harvard.edu/abs/2007ApJ...670.1225M}
}

@ARTICLE{2007MakarovA,
       author = {{Makarov}, Valeri V.},
        title = "{Unraveling the Origins of Nearby Young Stars}",
      journal = {\apjs},
         year = 2007,
        month = mar,
       volume = {169},
       number = {1},
        pages = {105-119},
          doi = {10.1086/509887},
       adsurl = {https://ui.adsabs.harvard.edu/abs/2007ApJS..169..105M}
}

@ARTICLE{1999deBruijne,
       author = {{de Bruijne}, Jos H.~J.},
        title = "{Structure and colour-magnitude diagrams of Scorpius OB2 based on kinematic modelling of Hipparcos data}",
      journal = {\mnras},
         year = 1999,
        month = dec,
       volume = {310},
       number = {3},
        pages = {585-617},
          doi = {10.1046/j.1365-8711.1999.02953.x},
       adsurl = {https://ui.adsabs.harvard.edu/abs/1999MNRAS.310..585D}
}

@ARTICLE{1992deGeus,
       author = {{de Geus}, E.~J.},
        title = "{Interactions of stars and interstellar matter in Scorpio Centaurus.}",
      journal = {\aap},
         year = 1992,
        month = aug,
       volume = {262},
        pages = {258-270},
       adsurl = {https://ui.adsabs.harvard.edu/abs/1992A&A...262..258D}
}

@ARTICLE{1989deGeus,
       author = {{de Geus}, E.~J. and {de Zeeuw}, P.~T. and {Lub}, J.},
        title = "{Physical parameters of stars in the Scorpio-Centaurus OB association.}",
      journal = {\aap},
         year = 1989,
        month = jun,
       volume = {216},
        pages = {44-61},
       adsurl = {https://ui.adsabs.harvard.edu/abs/1989A&A...216...44D}
}

@ARTICLE{1946Blaauw,
       author = {{Blaauw}, A.},
        title = "{A Study of the Scorpio-Centaurus Cluster}",
      journal = {Publications of the Kapteyn Astronomical Laboratory Groningen},
         year = 1946,
        month = jan,
       volume = {52},
        pages = {1-132},
       adsurl = {https://ui.adsabs.harvard.edu/abs/1946PGro...52....1B}
}

@ARTICLE{1914Kapteyn,
       author = {{Kapteyn}, J.~C.},
        title = "{On the individual parallaxes of the brighter galactic helium stars in the southern hemisphere, together with considerations on the parallax of stars in general.}",
      journal = {\apj},
         year = 1914,
        month = jul,
       volume = {40},
        pages = {43-126},
          doi = {10.1086/142098},
       adsurl = {https://ui.adsabs.harvard.edu/abs/1914ApJ....40...43K}
}

@ARTICLE{2006Fuchs,
       author = {{Fuchs}, B. and {Breitschwerdt}, D. and {de Avillez}, M.~A. and {Dettbarn}, C. and {Flynn}, C.},
        title = "{The search for the origin of the Local Bubble redivivus}",
      journal = {\mnras},
         year = 2006,
        month = dec,
       volume = {373},
       number = {3},
        pages = {993-1003},
          doi = {10.1111/j.1365-2966.2006.11044.x},
archivePrefix = {arXiv},
       eprint = {astro-ph/0609227},
 primaryClass = {astro-ph},
       adsurl = {https://ui.adsabs.harvard.edu/abs/2006MNRAS.373..993F}
}

@ARTICLE{2016Breitschwerdt,
       author = {{Breitschwerdt}, D. and {Feige}, J. and {Schulreich}, M.~M. and {Avillez}, M.~A. De. and {Dettbarn}, C. and {Fuchs}, B.},
        title = "{The locations of recent supernovae near the Sun from modelling $^{60}$Fe transport}",
      journal = {\nat},
         year = 2016,
        month = apr,
       volume = {532},
       number = {7597},
        pages = {73-76},
          doi = {10.1038/nature17424},
       adsurl = {https://ui.adsabs.harvard.edu/abs/2016Natur.532...73B}
}

@ARTICLE{2001Dame,
       author = {{Dame}, T.~M. and {Hartmann}, Dap and {Thaddeus}, P.},
        title = "{The Milky Way in Molecular Clouds: A New Complete CO Survey}",
      journal = {\apj},
         year = 2001,
        month = feb,
       volume = {547},
       number = {2},
        pages = {792-813},
          doi = {10.1086/318388},
archivePrefix = {arXiv},
       eprint = {astro-ph/0009217},
 primaryClass = {astro-ph},
       adsurl = {https://ui.adsabs.harvard.edu/abs/2001ApJ...547..792D}
}

@ARTICLE{2010Chepurnov,
       author = {{Chepurnov}, A. and {Lazarian}, A.},
        title = "{Extending the Big Power Law in the Sky with Turbulence Spectra from Wisconsin H{\ensuremath{\alpha}} Mapper Data}",
      journal = {\apj},
         year = 2010,
        month = feb,
       volume = {710},
       number = {1},
        pages = {853-858},
          doi = {10.1088/0004-637X/710/1/853},
archivePrefix = {arXiv},
       eprint = {0905.4413},
 primaryClass = {astro-ph.GA},
       adsurl = {https://ui.adsabs.harvard.edu/abs/2010ApJ...710..853C}
}

@ARTICLE{1995Armstrong,
       author = {{Armstrong}, J.~W. and {Rickett}, B.~J. and {Spangler}, S.~R.},
        title = "{Electron Density Power Spectrum in the Local Interstellar Medium}",
      journal = {\apj},
         year = 1995,
        month = apr,
       volume = {443},
        pages = {209},
          doi = {10.1086/175515},
       adsurl = {https://ui.adsabs.harvard.edu/abs/1995ApJ...443..209A}
}

@ARTICLE{2020Ferriere,
       author = {{Ferri{\`e}re}, K.},
        title = "{Plasma turbulence in the interstellar medium}",
      journal = {Plasma Physics and Controlled Fusion},
         year = 2020,
        month = jan,
       volume = {62},
       number = {1},
        pages = {014014},
          doi = {10.1088/1361-6587/ab49eb},
archivePrefix = {arXiv},
       eprint = {1912.08237},
 primaryClass = {astro-ph.GA},
       adsurl = {https://ui.adsabs.harvard.edu/abs/2020PPCF...62a4014F}
}

@ARTICLE{2025Posch,
       author = {{Posch}, Laura and {Alves}, Jo{\~a}o and {Miret-Roig}, N{\'u}ria and {Ratzenb{\"o}ck}, Sebastian and {Gro{\ss}schedl}, Josefa and {Meingast}, Stefan and {Swiggum}, Cameren and {Konietzka}, Ralf},
        title = "{The physical properties of cluster chains}",
      journal = {\aap},
         year = 2025,
        month = jan,
       volume = {693},
          eid = {A175},
        pages = {A175},
          doi = {10.1051/0004-6361/202451312},
archivePrefix = {arXiv},
       eprint = {2410.18080},
 primaryClass = {astro-ph.GA},
       adsurl = {https://ui.adsabs.harvard.edu/abs/2025A&A...693A.175P}
}

@ARTICLE{2025MiretRoig,
       author = {{Miret-Roig}, N. and {Alves}, J. and {Ratzenb{\"o}ck}, S. and {Galli}, P.~A.~B. and {Bouy}, H. and {Figueras}, F. and {Gro{\ss}schedl}, J. and {Meingast}, S. and {Posch}, L. and {Rottensteiner}, A. and {Swiggum}, C. and {Wagner}, N.},
        title = "{The TW Hydrae Association is a cluster chain of Sco-Cen}",
      journal = {\aap},
         year = 2025,
        month = feb,
       volume = {694},
          eid = {A60},
        pages = {A60},
          doi = {10.1051/0004-6361/202452558},
archivePrefix = {arXiv},
       eprint = {2501.11716},
 primaryClass = {astro-ph.SR},
       adsurl = {https://ui.adsabs.harvard.edu/abs/2025A&A...694A..60M}
}

@ARTICLE{2025Grossschedl,
       author = {{Gro{\ss}schedl}, Josefa E. and {Alves}, Jo{\~a}o and {Ratzenb{\"o}ck}, Sebastian and {Miret-Roig}, N{\'u}ria and {Hacar}, Alvaro and {Hutschenreuter}, Sebastian and {Posch}, Laura},
        title = "{The evolution of velocity dispersion in the Sco-Cen association}",
      journal = {\aap},
         year = {subm.},
          eid = {arXiv:2509.19487},
        pages = {arXiv:2509.19487},
          doi = {10.48550/arXiv.2509.19487},
archivePrefix = {arXiv},
       eprint = {2509.19487},
 primaryClass = {astro-ph.GA},
       adsurl = {https://ui.adsabs.harvard.edu/abs/2025arXiv250919487G}
}

@ARTICLE{2020Leike,
       author = {{Leike}, R.~H. and {Glatzle}, M. and {En{\ss}lin}, T.~A.},
        title = "{Resolving nearby dust clouds}",
      journal = {\aap},
         year = 2020,
        month = jul,
       volume = {639},
          eid = {A138},
        pages = {A138},
          doi = {10.1051/0004-6361/202038169},
archivePrefix = {arXiv},
       eprint = {2004.06732},
 primaryClass = {astro-ph.GA},
       adsurl = {https://ui.adsabs.harvard.edu/abs/2020A&A...639A.138L}
}

@ARTICLE{2024Westerkamp,
       author = {{Westerkamp}, M. and {Eberle}, V. and {Guardiani}, M. and {Frank}, P. and {Scheel-Platz}, L. and {Arras}, P. and {Knollm{\"u}ller}, J. and {Stadler}, J. and {En{\ss}lin}, T.},
        title = "{The first spatio-spectral Bayesian imaging of SN1006 in X-rays}",
      journal = {\aap},
         year = 2024,
        month = apr,
       volume = {684},
          eid = {A155},
        pages = {A155},
          doi = {10.1051/0004-6361/202347750},
archivePrefix = {arXiv},
       eprint = {2308.09176},
 primaryClass = {astro-ph.HE},
       adsurl = {https://ui.adsabs.harvard.edu/abs/2024A&A...684A.155W}
}

@ARTICLE{2023Scheel-Platz,
       author = {{Scheel-Platz}, L.~I. and {Knollm{\"u}ller}, J. and {Arras}, P. and {Frank}, P. and {Reinecke}, M. and {J{\"u}stel}, D. and {En{\ss}lin}, T.~A.},
        title = "{Multicomponent imaging of the Fermi gamma-ray sky in the spatio-spectral domain}",
      journal = {\aap},
         year = 2023,
        month = dec,
       volume = {680},
          eid = {A2},
        pages = {A2},
          doi = {10.1051/0004-6361/202243819},
archivePrefix = {arXiv},
       eprint = {2204.09360},
 primaryClass = {astro-ph.HE},
       adsurl = {https://ui.adsabs.harvard.edu/abs/2023A&A...680A...2S}
}

@ARTICLE{2025Soeding,
       author = {{S{\"o}ding}, Laurin and {Edenhofer}, Gordian and {En{\ss}lin}, Torsten A. and {Frank}, Philipp and {Kissmann}, Ralf and {Phan}, Vo Hong Minh and {Ram{\'\i}rez}, Andr{\'e}s and {Zandinejad}, Hanieh and {Mertsch}, Philipp},
        title = "{Spatially coherent 3D distributions of HI and CO in the Milky Way}",
      journal = {\aap},
         year = 2025,
        month = jan,
       volume = {693},
          eid = {A139},
        pages = {A139},
          doi = {10.1051/0004-6361/202451361},
archivePrefix = {arXiv},
       eprint = {2407.02859},
 primaryClass = {astro-ph.GA},
       adsurl = {https://ui.adsabs.harvard.edu/abs/2025A&A...693A.139S}
}

@ARTICLE{2024Castro-Ginard,
       author = {{Castro-Ginard}, Alfred and {Penoyre}, Zephyr and {Casey}, Andrew R. and {Brown}, Anthony G.~A. and {Belokurov}, Vasily and {Cantat-Gaudin}, Tristan and {Drimmel}, Ronald and {Fouesneau}, Morgan and {Khanna}, Shourya and {Kurbatov}, Evgeny P. and {Price-Whelan}, Adrian M. and {Rix}, Hans-Walter and {Smart}, Richard L.},
        title = "{Gaia DR3 detectability of unresolved binary systems}",
      journal = {\aap},
         year = 2024,
        month = aug,
       volume = {688},
          eid = {A1},
        pages = {A1},
          doi = {10.1051/0004-6361/202450172},
archivePrefix = {arXiv},
       eprint = {2404.14127},
 primaryClass = {astro-ph.GA},
       adsurl = {https://ui.adsabs.harvard.edu/abs/2024A&A...688A...1C}
}

@ARTICLE{2025Stiskalek,
       author = {{Stiskalek}, Richard and {Desmond}, Harry and {Devriendt}, Julien and {Slyz}, Adrianne and {Lavaux}, Guilhem and {Hudson}, Michael J. and {Bartlett}, Deaglan J. and {Courtois}, H{\'e}l{\`e}ne M.},
        title = "{The Velocity Field Olympics: Assessing velocity field reconstructions with direct distance tracers}",
      journal = {\mnras},
         year = 2025,
        month = nov,
          doi = {10.1093/mnras/staf1960},
archivePrefix = {arXiv},
       eprint = {2502.00121},
 primaryClass = {astro-ph.CO},
       adsurl = {https://ui.adsabs.harvard.edu/abs/2025MNRAS.tmp.1852S}
}

@ARTICLE{1964Blaauw,
       author = {{Blaauw}, Adriaan},
        title = "{The O Associations in the Solar Neighborhood}",
      journal = {\araa},
         year = 1964,
        month = jan,
       volume = {2},
        pages = {213},
          doi = {10.1146/annurev.aa.02.090164.001241},
       adsurl = {https://ui.adsabs.harvard.edu/abs/1964ARA&A...2..213B}
}

@ARTICLE{2024QuintanaArxiv,
       author = {{Quintana}, Alexis L.},
        title = "{OB associations: from stellar to galactic scales}",
      journal = {arXiv e-prints},
         year = 2024,
        month = dec,
          eid = {arXiv:2412.10769},
        pages = {arXiv:2412.10769},
          doi = {10.48550/arXiv.2412.10769},
archivePrefix = {arXiv},
       eprint = {2412.10769},
 primaryClass = {astro-ph.GA},
       adsurl = {https://ui.adsabs.harvard.edu/abs/2024arXiv241210769Q}
}

@ARTICLE{2017Kunder,
       author = {{Kunder}, Andrea and {Kordopatis}, Georges and {Steinmetz}, Matthias and {Zwitter}, Toma{\v{z}} and {McMillan}, Paul J. and {Casagrande}, Luca and {Enke}, Harry and {Wojno}, Jennifer and {Valentini}, Marica and {Chiappini}, Cristina and {Matijevi{\v{c}}}, Gal and {Siviero}, Alessandro and {de Laverny}, Patrick and {Recio-Blanco}, Alejandra and {Bijaoui}, Albert and {Wyse}, Rosemary F.~G. and {Binney}, James and {Grebel}, E.~K. and {Helmi}, Amina and {Jofre}, Paula and {Antoja}, Teresa and {Gilmore}, Gerard and {Siebert}, Arnaud and {Famaey}, Benoit and {Bienaym{\'e}}, Olivier and {Gibson}, Brad K. and {Freeman}, Kenneth C. and {Navarro}, Julio F. and {Munari}, Ulisse and {Seabroke}, George and {Anguiano}, Borja and {{\v{Z}}erjal}, Maru{\v{s}}a and {Minchev}, Ivan and {Reid}, Warren and {Bland-Hawthorn}, Joss and {Kos}, Janez and {Sharma}, Sanjib and {Watson}, Fred and {Parker}, Quentin A. and {Scholz}, Ralf-Dieter and {Burton}, Donna and {Cass}, Paul and {Hartley}, Malcolm and {Fiegert}, Kristin and {Stupar}, Milorad and {Ritter}, Andreas and {Hawkins}, Keith and {Gerhard}, Ortwin and {Chaplin}, W.~J. and {Davies}, G.~R. and {Elsworth}, Y.~P. and {Lund}, M.~N. and {Miglio}, A. and {Mosser}, B.},
        title = "{The Radial Velocity Experiment (RAVE): Fifth Data Release}",
      journal = {\aj},
         year = 2017,
        month = feb,
       volume = {153},
       number = {2},
          eid = {75},
        pages = {75},
          doi = {10.3847/1538-3881/153/2/75},
archivePrefix = {arXiv},
       eprint = {1609.03210},
 primaryClass = {astro-ph.SR},
       adsurl = {https://ui.adsabs.harvard.edu/abs/2017AJ....153...75K}
}

@ARTICLE{2020aSteinmetz,
       author = {{Steinmetz}, Matthias and {Matijevi{\v{c}}}, Gal and {Enke}, Harry and {Zwitter}, Toma{\v{z}} and {Guiglion}, Guillaume and {McMillan}, Paul J. and {Kordopatis}, Georges and {Valentini}, Marica and {Chiappini}, Cristina and {Casagrande}, Luca and {Wojno}, Jennifer and {Anguiano}, Borja and {Bienaym{\'e}}, Olivier and {Bijaoui}, Albert and {Binney}, James and {Burton}, Donna and {Cass}, Paul and {de Laverny}, Patrick and {Fiegert}, Kristin and {Freeman}, Kenneth and {Fulbright}, Jon P. and {Gibson}, Brad K. and {Gilmore}, Gerard and {Grebel}, Eva K. and {Helmi}, Amina and {Kunder}, Andrea and {Munari}, Ulisse and {Navarro}, Julio F. and {Parker}, Quentin and {Ruchti}, Gregory R. and {Recio-Blanco}, Alejandra and {Reid}, Warren and {Seabroke}, George M. and {Siviero}, Alessandro and {Siebert}, Arnaud and {Stupar}, Milorad and {Watson}, Fred and {Williams}, Mary E.~K. and {Wyse}, Rosemary F.~G. and {Anders}, Friedrich and {Antoja}, Teresa and {Birko}, Danijela and {Bland-Hawthorn}, Joss and {Bossini}, Diego and {Garc{\'\i}a}, Rafael A. and {Carrillo}, Ismael and {Chaplin}, William J. and {Elsworth}, Yvonne and {Famaey}, Benoit and {Gerhard}, Ortwin and {Jofre}, Paula and {Just}, Andreas and {Mathur}, Savita and {Miglio}, Andrea and {Minchev}, Ivan and {Monari}, Giacomo and {Mosser}, Benoit and {Ritter}, Andreas and {Rodrigues}, Thaise S. and {Scholz}, Ralf-Dieter and {Sharma}, Sanjib and {Sysoliatina}, Kseniia and {RAVE Collaboration}},
        title = "{The Sixth Data Release of the Radial Velocity Experiment (RAVE). I. Survey Description, Spectra, and Radial Velocities}",
      journal = {\aj},
         year = {2020a},
        month = aug,
       volume = {160},
       number = {2},
          eid = {82},
        pages = {82},
          doi = {10.3847/1538-3881/ab9ab9},
archivePrefix = {arXiv},
       eprint = {2002.04377},
 primaryClass = {astro-ph.SR},
       adsurl = {https://ui.adsabs.harvard.edu/abs/2020AJ....160...82S}
}

@ARTICLE{2022Jackson,
       author = {{Jackson}, R.~J. and {Jeffries}, R.~D. and {Wright}, N.~J. and {Randich}, S. and {Sacco}, G. and {Bragaglia}, A. and {Hourihane}, A. and {Tognelli}, E. and {Degl'Innocenti}, S. and {Prada Moroni}, P.~G. and {Gilmore}, G. and {Bensby}, T. and {Pancino}, E. and {Smiljanic}, R. and {Bergemann}, M. and {Carraro}, G. and {Franciosini}, E. and {Gonneau}, A. and {Jofr{\'e}}, P. and {Lewis}, J. and {Magrini}, L. and {Morbidelli}, L. and {Prisinzano}, L. and {Worley}, C. and {Zaggia}, S. and {Tautvai{\v{s}}iene}, G. and {Guti{\'e}rrez Albarr{\'a}n}, M.~L. and {Montes}, D. and {Jim{\'e}nez-Esteban}, F.},
        title = "{The Gaia-ESO Survey: Membership probabilities for stars in 63 open and 7 globular clusters from 3D kinematics}",
      journal = {\mnras},
         year = 2022,
        month = jan,
       volume = {509},
       number = {2},
        pages = {1664-1680},
          doi = {10.1093/mnras/stab3032},
archivePrefix = {arXiv},
       eprint = {2110.10477},
 primaryClass = {astro-ph.SR},
       adsurl = {https://ui.adsabs.harvard.edu/abs/2022MNRAS.509.1664J}
}

@ARTICLE{2017Sacco,
       author = {{Sacco}, G.~G. and {Spina}, L. and {Randich}, S. and {Palla}, F. and {Parker}, R.~J. and {Jeffries}, R.~D. and {Jackson}, R. and {Meyer}, M.~R. and {Mapelli}, M. and {Lanzafame}, A.~C. and {Bonito}, R. and {Damiani}, F. and {Franciosini}, E. and {Frasca}, A. and {Klutsch}, A. and {Prisinzano}, L. and {Tognelli}, E. and {Degl'Innocenti}, S. and {Prada Moroni}, P.~G. and {Alfaro}, E.~J. and {Micela}, G. and {Prusti}, T. and {Barrado}, D. and {Biazzo}, K. and {Bouy}, H. and {Bravi}, L. and {Lopez-Santiago}, J. and {Wright}, N.~J. and {Bayo}, A. and {Gilmore}, G. and {Bragaglia}, A. and {Flaccomio}, E. and {Koposov}, S.~E. and {Pancino}, E. and {Casey}, A.~R. and {Costado}, M.~T. and {Donati}, P. and {Hourihane}, A. and {Jofr{\'e}}, P. and {Lardo}, C. and {Lewis}, J. and {Magrini}, L. and {Monaco}, L. and {Morbidelli}, L. and {Sousa}, S.~G. and {Worley}, C.~C. and {Zaggia}, S.},
        title = "{The Gaia-ESO Survey: Structural and dynamical properties of the young cluster Chamaeleon I}",
      journal = {\aap},
         year = 2017,
        month = may,
       volume = {601},
          eid = {A97},
        pages = {A97},
          doi = {10.1051/0004-6361/201629698},
archivePrefix = {arXiv},
       eprint = {1701.03741},
 primaryClass = {astro-ph.SR},
       adsurl = {https://ui.adsabs.harvard.edu/abs/2017A&A...601A..97S}
}

@ARTICLE{2012Gilmore,
       author = {{Gilmore}, G. and {Randich}, S. and {Asplund}, M. and {Binney}, J. and {Bonifacio}, P. and {Drew}, J. and {Feltzing}, S. and {Ferguson}, A. and {Jeffries}, R. and {Micela}, G. and {Negueruela}, I. and {Prusti}, T. and {Rix}, H. -W. and {Vallenari}, A. and {Alfaro}, E. and {Allende-Prieto}, C. and {Babusiaux}, C. and {Bensby}, T. and {Blomme}, R. and {Bragaglia}, A. and {Flaccomio}, E. and {Fran{\c{c}}ois}, P. and {Irwin}, M. and {Koposov}, S. and {Korn}, A. and {Lanzafame}, A. and {Pancino}, E. and {Paunzen}, E. and {Recio-Blanco}, A. and {Sacco},  G. and {et al.} and {Gaia-ESO Survey Team}},
        title = "{The Gaia-ESO Public Spectroscopic Survey}",
      journal = {The Messenger},
         year = 2012,
        month = mar,
       volume = {147},
        pages = {25-31},
       adsurl = {https://ui.adsabs.harvard.edu/abs/2012Msngr.147...25G}
}

@ARTICLE{2021Buder,
       author = {{Buder}, Sven and {Sharma}, Sanjib and {Kos}, Janez and {Amarsi}, Anish M. and {Nordlander}, Thomas and {Lind}, Karin and {Martell}, Sarah L. and {Asplund}, Martin and {Bland-Hawthorn}, Joss and {Casey}, Andrew R. and {de Silva}, Gayandhi M. and {D'Orazi}, Valentina and {Freeman}, Ken C. and {Hayden}, Michael R. and {Lewis}, Geraint F. and {Lin}, Jane and {Schlesinger}, Katharine J. and {Simpson}, Jeffrey D. and {Stello}, Dennis and {Zucker}, Daniel B. and {Zwitter}, Toma{\v{z}} and {Beeson}, Kevin L. and {Buck}, Tobias and {Casagrande}, Luca and {Clark}, Jake T. and {{\v{C}}otar}, Klemen and {da Costa}, Gary S. and {de Grijs}, Richard and {Feuillet}, Diane and {Horner}, Jonathan and {Kafle}, Prajwal R. and {Khanna}, Shourya and {Kobayashi}, Chiaki and {Liu}, Fan and {Montet}, Benjamin T. and {Nandakumar}, Govind and {Nataf}, David M. and {Ness}, Melissa K. and {Spina}, Lorenzo and {Tepper-Garc{\'\i}a}, Thor and {Ting}, Yuan-Sen and {Traven}, Gregor and {Vogrin{\v{c}}i{\v{c}}}, Rok and {Wittenmyer}, Robert A. and {Wyse}, Rosemary F.~G. and {{\v{Z}}erjal}, Maru{\v{s}}a and {Galah Collaboration}},
        title = "{The GALAH+ survey: Third data release}",
      journal = {\mnras},
         year = 2021,
        month = sep,
       volume = {506},
       number = {1},
        pages = {150-201},
          doi = {10.1093/mnras/stab1242},
archivePrefix = {arXiv},
       eprint = {2011.02505},
 primaryClass = {astro-ph.GA},
       adsurl = {https://ui.adsabs.harvard.edu/abs/2021MNRAS.506..150B}
}

@ARTICLE{2015DeSilva,
       author = {{De Silva}, G.~M. and {Freeman}, K.~C. and {Bland-Hawthorn}, J. and {Martell}, S. and {de Boer}, E. Wylie and {Asplund}, M. and {Keller}, S. and {Sharma}, S. and {Zucker}, D.~B. and {Zwitter}, T. and {Anguiano}, B. and {Bacigalupo}, C. and {Bayliss}, D. and {Beavis}, M.~A. and {Bergemann}, M. and {Campbell}, S. and {Cannon}, R. and {Carollo}, D. and {Casagrande}, L. and {Casey}, A.~R. and {Da Costa}, G. and {D'Orazi}, V. and {Dotter}, A. and {Duong}, L. and {Heger}, A. and {Ireland}, M.~J. and {Kafle}, P.~R. and {Kos}, J. and {Lattanzio}, J. and {Lewis}, G.~F. and {Lin}, J. and {Lind}, K. and {Munari}, U. and {Nataf}, D.~M. and {O'Toole}, S. and {Parker}, Q. and {Reid}, W. and {Schlesinger}, K.~J. and {Sheinis}, A. and {Simpson}, J.~D. and {Stello}, D. and {Ting}, Y. -S. and {Traven}, G. and {Watson}, F. and {Wittenmyer}, R. and {Yong}, D. and {{\v{Z}}erjal}, M.},
        title = "{The GALAH survey: scientific motivation}",
      journal = {\mnras},
         year = 2015,
        month = may,
       volume = {449},
       number = {3},
        pages = {2604-2617},
          doi = {10.1093/mnras/stv327},
archivePrefix = {arXiv},
       eprint = {1502.04767},
 primaryClass = {astro-ph.GA},
       adsurl = {https://ui.adsabs.harvard.edu/abs/2015MNRAS.449.2604D}
}

@ARTICLE{2021Santana,
       author = {{Santana}, Felipe A. and {Beaton}, Rachael L. and {Covey}, Kevin R. and {O'Connell}, Julia E. and {Longa-Pe{\~n}a}, Pen{\'e}lope and {Cohen}, Roger and {Fern{\'a}ndez-Trincado}, Jos{\'e} G. and {Hayes}, Christian R. and {Zasowski}, Gail and {Sobeck}, Jennifer S. and {Majewski}, Steven R. and {Chojnowski}, S.~D. and {De Lee}, Nathan and {Oelkers}, Ryan J. and {Stringfellow}, Guy S. and {Almeida}, Andr{\'e}s and {Anguiano}, Borja and {Donor}, John and {Frinchaboy}, Peter M. and {Hasselquist}, Sten and {Johnson}, Jennifer A. and {Kollmeier}, Juna A. and {Nidever}, David L. and {Price-Whelan}, Adrian M. and {Rojas-Arriagada}, Alvaro and {Schultheis}, Mathias and {Shetrone}, Matthew and {Simon}, Joshua D. and {Aerts}, Conny and {Borissova}, Jura and {Drout}, Maria R. and {Geisler}, Doug and {Law}, C.~Y. and {Medina}, Nicolas and {Minniti}, Dante and {Monachesi}, Antonela and {Mu{\~n}oz}, Ricardo R. and {Poleski}, Rados{\l}aw and {Roman-Lopes}, Alexandre and {Schlaufman}, Kevin C. and {Stutz}, Amelia M. and {Teske}, Johanna and {Tkachenko}, Andrew and {Van Saders}, Jennifer L. and {Weinberger}, Alycia J. and {Zoccali}, Manuela},
        title = "{Final Targeting Strategy for the SDSS-IV APOGEE-2S Survey}",
      journal = {\aj},
         year = 2021,
        month = dec,
       volume = {162},
       number = {6},
          eid = {303},
        pages = {303},
          doi = {10.3847/1538-3881/ac2cbc},
archivePrefix = {arXiv},
       eprint = {2108.11908},
 primaryClass = {astro-ph.GA},
       adsurl = {https://ui.adsabs.harvard.edu/abs/2021AJ....162..303S}
}

@ARTICLE{2011Ando,
       author = {{Ando}, Kazuma and {Nagayama}, Takumi and {Omodaka}, Toshihiro and {Handa}, Toshihiro and {Imai}, Hiroshi and {Nakagawa}, Akiharu and {Nakanishi}, Hiroyuki and {Honma}, Mareki and {Kobayashi}, Hideyuki and {Miyaji}, Takeshi},
        title = "{Astrometry of Galactic Star-Forming Region ON2N with VERA: Estimation of the Galactic Constants}",
      journal = {\pasj},
         year = 2011,
        month = feb,
       volume = {63},
        pages = {45},
          doi = {10.1093/pasj/63.1.45},
archivePrefix = {arXiv},
       eprint = {1012.5715},
 primaryClass = {astro-ph.GA},
       adsurl = {https://ui.adsabs.harvard.edu/abs/2011PASJ...63...45A}
}

@ARTICLE{2017Majewski,
       author = {{Majewski}, Steven R. and {Schiavon}, Ricardo P. and {Frinchaboy}, Peter M. and {Allende Prieto}, Carlos and {Barkhouser}, Robert and {Bizyaev}, Dmitry and {Blank}, Basil and {Brunner}, Sophia and {Burton}, Adam and {Carrera}, Ricardo and {Chojnowski}, S. Drew and {Cunha}, K{\'a}tia and {Epstein}, Courtney and {Fitzgerald}, Greg and {Garc{\'\i}a P{\'e}rez}, Ana E. and {Hearty}, Fred R. and {Henderson}, Chuck and {Holtzman}, Jon A. and {Johnson}, Jennifer A. and {Lam}, Charles R. and {Lawler}, James E. and {Maseman}, Paul and {M{\'e}sz{\'a}ros}, Szabolcs and {Nelson}, Matthew and {Nguyen}, Duy Coung and {Nidever}, David L. and {Pinsonneault}, Marc and {Shetrone}, Matthew and {Smee}, Stephen and {Smith}, Verne V. and {Stolberg}, Todd and {Skrutskie}, Michael F. and {Walker}, Eric and {Wilson}, John C. and {Zasowski}, Gail and {Anders}, Friedrich and {Basu}, Sarbani and {Beland}, Stephane and {Blanton}, Michael R. and {Bovy}, Jo and {Brownstein}, Joel R. and {Carlberg}, Joleen and {Chaplin}, William and {Chiappini}, Cristina and {Eisenstein}, Daniel J. and {Elsworth}, Yvonne and {Feuillet}, Diane and {Fleming}, Scott W. and {Galbraith-Frew}, Jessica and {Garc{\'\i}a}, Rafael A. and {Garc{\'\i}a-Hern{\'a}ndez}, D. An{\'\i}bal and {Gillespie}, Bruce A. and {Girardi}, L{\'e}o and {Gunn}, James E. and {Hasselquist}, Sten and {Hayden}, Michael R. and {Hekker}, Saskia and {Ivans}, Inese and {Kinemuchi}, Karen and {Klaene}, Mark and {Mahadevan}, Suvrath and {Mathur}, Savita and {Mosser}, Beno{\^\i}t and {Muna}, Demitri and {Munn}, Jeffrey A. and {Nichol}, Robert C. and {O'Connell}, Robert W. and {Parejko}, John K. and {Robin}, A.~C. and {Rocha-Pinto}, Helio and {Schultheis}, Matthias and {Serenelli}, Aldo M. and {Shane}, Neville and {Silva Aguirre}, Victor and {Sobeck}, Jennifer S. and {Thompson}, Benjamin and {Troup}, Nicholas W. and {Weinberg}, David H. and {Zamora}, Olga},
        title = "{The Apache Point Observatory Galactic Evolution Experiment (APOGEE)}",
      journal = {\aj},
         year = 2017,
        month = sep,
       volume = {154},
       number = {3},
          eid = {94},
        pages = {94},
          doi = {10.3847/1538-3881/aa784d},
archivePrefix = {arXiv},
       eprint = {1509.05420},
 primaryClass = {astro-ph.IM},
       adsurl = {https://ui.adsabs.harvard.edu/abs/2017AJ....154...94M}
}

@ARTICLE{2013Murphy,
       author = {{Murphy}, Simon J. and {Lawson}, Warrick A. and {Bessell}, Michael S.},
        title = "{Re-examining the membership and origin of the ɛ Cha association}",
      journal = {\mnras},
         year = 2013,
        month = oct,
       volume = {435},
       number = {2},
        pages = {1325-1349},
          doi = {10.1093/mnras/stt1375},
archivePrefix = {arXiv},
       eprint = {1305.4177},
 primaryClass = {astro-ph.SR},
       adsurl = {https://ui.adsabs.harvard.edu/abs/2013MNRAS.435.1325M}
}

@ARTICLE{2006Torres,
       author = {{Torres}, C.~A.~O. and {Quast}, G.~R. and {da Silva}, L. and {de La Reza}, R. and {Melo}, C.~H.~F. and {Sterzik}, M.},
        title = "{Search for associations containing young stars (SACY). I. Sample and searching method}",
      journal = {\aap},
         year = 2006,
        month = dec,
       volume = {460},
       number = {3},
        pages = {695-708},
          doi = {10.1051/0004-6361:20065602},
archivePrefix = {arXiv},
       eprint = {astro-ph/0609258},
 primaryClass = {astro-ph},
       adsurl = {https://ui.adsabs.harvard.edu/abs/2006A&A...460..695T}
}

@ARTICLE{2001Joergens,
       author = {{Joergens}, V. and {Guenther}, E.},
        title = "{UVES spectra of young brown dwarfs in Cha I: Radial and rotational velocities}",
      journal = {\aap},
         year = 2001,
        month = nov,
       volume = {379},
        pages = {L9-L12},
          doi = {10.1051/0004-6361:20011337},
archivePrefix = {arXiv},
       eprint = {astro-ph/0110175},
 primaryClass = {astro-ph},
       adsurl = {https://ui.adsabs.harvard.edu/abs/2001A&A...379L...9J}
}

@ARTICLE{2012Nguyen,
       author = {{Nguyen}, Duy Cuong and {Brandeker}, Alexis and {van Kerkwijk}, Marten H. and {Jayawardhana}, Ray},
        title = "{Close Companions to Young Stars. I. A Large Spectroscopic Survey in Chamaeleon I and Taurus-Auriga}",
      journal = {\apj},
         year = 2012,
        month = feb,
       volume = {745},
       number = {2},
          eid = {119},
        pages = {119},
          doi = {10.1088/0004-637X/745/2/119},
archivePrefix = {arXiv},
       eprint = {1112.0002},
 primaryClass = {astro-ph.SR},
       adsurl = {https://ui.adsabs.harvard.edu/abs/2012ApJ...745..119N}
}

@ARTICLE{2006Jilinski,
       author = {{Jilinski}, E. and {Daflon}, S. and {Cunha}, K. and {de La Reza}, R.},
        title = "{Radial velocity measurements of B stars in the Scorpius-Centaurus association}",
      journal = {\aap},
         year = 2006,
        month = mar,
       volume = {448},
       number = {3},
        pages = {1001-1006},
          doi = {10.1051/0004-6361:20041614},
archivePrefix = {arXiv},
       eprint = {astro-ph/0601643},
 primaryClass = {astro-ph},
       adsurl = {https://ui.adsabs.harvard.edu/abs/2006A&A...448.1001J}
}

@ARTICLE{2013Galli,
       author = {{Galli}, P.~A.~B. and {Bertout}, C. and {Teixeira}, R. and {Ducourant}, C.},
        title = "{A kinematic study and membership analysis of the Lupus star-forming region}",
      journal = {\aap},
         year = 2013,
        month = oct,
       volume = {558},
          eid = {A77},
        pages = {A77},
          doi = {10.1051/0004-6361/201220704},
archivePrefix = {arXiv},
       eprint = {1309.7799},
 primaryClass = {astro-ph.GA},
       adsurl = {https://ui.adsabs.harvard.edu/abs/2013A&A...558A..77G}
}

@ARTICLE{2011Chen,
       author = {{Chen}, Christine H. and {Mamajek}, Eric E. and {Bitner}, Martin A. and {Pecaut}, Mark and {Su}, Kate Y.~L. and {Weinberger}, Alycia J.},
        title = "{A Magellan MIKE and Spitzer MIPS Study of 1.5-1.0 M $_{sun}$ Stars in Scorpius-Centaurus}",
      journal = {\apj},
         year = 2011,
        month = sep,
       volume = {738},
       number = {2},
          eid = {122},
        pages = {122},
          doi = {10.1088/0004-637X/738/2/122},
       adsurl = {https://ui.adsabs.harvard.edu/abs/2011ApJ...738..122C}
}

@ARTICLE{2006James,
       author = {{James}, D.~J. and {Melo}, C. and {Santos}, N.~C. and {Bouvier}, J.},
        title = "{Fundamental properties of pre-main sequence stars in young, southern star forming regions: metallicities}",
      journal = {\aap},
         year = 2006,
        month = feb,
       volume = {446},
       number = {3},
        pages = {971-983},
          doi = {10.1051/0004-6361:20053900},
archivePrefix = {arXiv},
       eprint = {astro-ph/0510596},
 primaryClass = {astro-ph},
       adsurl = {https://ui.adsabs.harvard.edu/abs/2006A&A...446..971J}
}

@ARTICLE{2022Abdurrouf,
       author = {{Abdurro'uf} and {Accetta}, Katherine and {Aerts}, Conny and {Silva Aguirre}, V{\'\i}ctor and {Ahumada}, Romina and {Ajgaonkar}, Nikhil and {Filiz Ak}, N. and {Alam}, Shadab and {Allende Prieto}, Carlos and {Almeida}, Andr{\'e}s and {Anders}, Friedrich and {Anderson}, Scott F. and {Andrews}, Brett H. and {Anguiano}, Borja and {Aquino-Ort{\'\i}z}, Erik and {Arag{\'o}n-Salamanca}, Alfonso and {Argudo-Fern{\'a}ndez}, Maria and {Ata}, Metin and {Aubert}, Marie and {Avila-Reese}, Vladimir and {Badenes}, Carles and {Barb{\'a}}, Rodolfo H. and {Barger}, Kat and {Barrera-Ballesteros}, Jorge K. and {Beaton}, Rachael L. and {Beers}, Timothy C. and {Belfiore}, Francesco and {Bender}, Chad F. and {Bernardi}, Mariangela and {Bershady}, Matthew A. and {Beutler}, Florian and {Bidin}, Christian Moni and {Bird}, Jonathan C. and {Bizyaev}, Dmitry and {Blanc}, Guillermo A. and {Blanton}, Michael R. and {Boardman}, Nicholas Fraser and {Bolton}, Adam S. and {Boquien}, M{\'e}d{\'e}ric and {Borissova}, Jura and {Bovy}, Jo and {Brandt}, W.~N. and {Brown}, Jordan and {Brownstein}, Joel R. and {Brusa}, Marcella and {Buchner}, Johannes and {Bundy}, Kevin and {Burchett}, Joseph N. and {Bureau}, Martin and {Burgasser}, Adam and {Cabang}, Tuesday K. and {Campbell}, Stephanie and {Cappellari}, Michele and {Carlberg}, Joleen K. and {Wanderley}, F{\'a}bio Carneiro and {Carrera}, Ricardo and {Cash}, Jennifer and {Chen}, Yan-Ping and {Chen}, Wei-Huai and {Cherinka}, Brian and {Chiappini}, Cristina and {Choi}, Peter Doohyun and {Chojnowski}, S. Drew and {Chung}, Haeun and {Clerc}, Nicolas and {Cohen}, Roger E. and {Comerford}, Julia M. and {Comparat}, Johan and {da Costa}, Luiz and {Covey}, Kevin and {Crane}, Jeffrey D. and {Cruz-Gonzalez}, Irene and {Culhane}, Connor and {Cunha}, Katia and {Dai}, Y. Sophia and {Damke}, Guillermo and {Darling}, Jeremy and {Davidson}, James W., Jr. and {Davies}, Roger and {Dawson}, Kyle and {De Lee}, Nathan and {Diamond-Stanic}, Aleksandar M. and {Cano-D{\'\i}az}, Mariana and {S{\'a}nchez}, Helena Dom{\'\i}nguez and {Donor}, John and {Duckworth}, Chris and {Dwelly}, Tom and {Eisenstein}, Daniel J. and {Elsworth}, Yvonne P. and {Emsellem}, Eric and {Eracleous}, Mike and {Escoffier}, Stephanie and {Fan}, Xiaohui and {Farr}, Emily and {Feng}, Shuai and {Fern{\'a}ndez-Trincado}, Jos{\'e} G. and {Feuillet}, Diane and {Filipp}, Andreas and {Fillingham}, Sean P. and {Frinchaboy}, Peter M. and {Fromenteau}, Sebastien and {Galbany}, Llu{\'\i}s and {Garc{\'\i}a}, Rafael A. and {Garc{\'\i}a-Hern{\'a}ndez}, D.~A. and {Ge}, Junqiang and {Geisler}, Doug and {Gelfand}, Joseph and {G{\'e}ron}, Tobias and {Gibson}, Benjamin J. and {Goddy}, Julian and {Godoy-Rivera}, Diego and {Grabowski}, Kathleen and {Green}, Paul J. and {Greener}, Michael and {Grier}, Catherine J. and {Griffith}, Emily and {Guo}, Hong and {Guy}, Julien and {Hadjara}, Massinissa and {Harding}, Paul and {Hasselquist}, Sten and {Hayes}, Christian R. and {Hearty}, Fred and {Hern{\'a}ndez}, Jes{\'u}s and {Hill}, Lewis and {Hogg}, David W. and {Holtzman}, Jon A. and {Horta}, Danny and {Hsieh}, Bau-Ching and {Hsu}, Chin-Hao and {Hsu}, Yun-Hsin and {Huber}, Daniel and {Huertas-Company}, Marc and {Hutchinson}, Brian and {Hwang}, Ho Seong and {Ibarra-Medel}, H{\'e}ctor J. and {Chitham}, Jacob Ider and {Ilha}, Gabriele S. and {Imig}, Julie and {Jaekle}, Will and {Jayasinghe}, Tharindu and {Ji}, Xihan and {Johnson}, Jennifer A. and {Jones}, Amy and {J{\"o}nsson}, Henrik and {Katkov}, Ivan and {Khalatyan}, Arman, Dr. and {Kinemuchi}, Karen and {Kisku}, Shobhit and {Knapen}, Johan H. and {Kneib}, Jean-Paul and {Kollmeier}, Juna A. and {Kong}, Miranda and {Kounkel}, Marina and {Kreckel}, Kathryn and {Krishnarao}, Dhanesh and {Lacerna}, Ivan and {Lane}, Richard R. and {Langgin}, Rachel and {Lavender}, Ramon and {Law}, David R. and {Lazarz}, Daniel and {Leung}, Henry W. and {Leung}, Ho-Hin and {Lewis}, Hannah M. and {Li}, Cheng and {Li}, Ran and {Lian}, Jianhui and {Liang}, Fu-Heng and {Lin}, Lihwai and {Lin}, Yen-Ting and {Lin}, Sicheng and {Lintott}, Chris and {Long}, Dan and {Longa-Pe{\~n}a}, Pen{\'e}lope and {L{\'o}pez-Cob{\'a}}, Carlos and {Lu}, Shengdong and {Lundgren}, Britt F. and {Luo}, Yuanze and {Mackereth}, J. Ted and {de la Macorra}, Axel and {Mahadevan}, Suvrath and {Majewski}, Steven R. and {Manchado}, Arturo and {Mandeville}, Travis and {Maraston}, Claudia and {Margalef-Bentabol}, Berta and {Masseron}, Thomas and {Masters}, Karen L. and {Mathur}, Savita and {McDermid}, Richard M. and {Mckay}, Myles and {Merloni}, Andrea and {Merrifield}, Michael and {Meszaros}, Szabolcs and {Miglio}, Andrea and {Di Mille}, Francesco and {Minniti}, Dante and {Minsley}, Rebecca and {Monachesi}, Antonela and {Moon}, Jeongin and {Mosser}, Benoit and {Mulchaey}, John and {Muna}, Demitri and {Mu{\~n}oz}, Ricardo R. and {Myers}, Adam D. and {Myers}, Natalie and {Nadathur}, Seshadri and {Nair}, Preethi and {Nandra}, Kirpal and {Neumann}, Justus and {Newman}, Jeffrey A. and {Nidever}, David L. and {Nikakhtar}, Farnik and {Nitschelm}, Christian and {O'Connell}, Julia E. and {Garma-Oehmichen}, Luis and {Luan Souza de Oliveira}, Gabriel and {Olney}, Richard and {Oravetz}, Daniel and {Ortigoza-Urdaneta}, Mario and {Osorio}, Yeisson and {Otter}, Justin and {Pace}, Zachary J. and {Padilla}, Nelson and {Pan}, Kaike and {Pan}, Hsi-An and {Parikh}, Taniya and {Parker}, James and {Peirani}, Sebastien and {Pe{\~n}a Ram{\'\i}rez}, Karla and {Penny}, Samantha and {Percival}, Will J. and {Perez-Fournon}, Ismael and {Pinsonneault}, Marc and {Poidevin}, Fr{\'e}d{\'e}rick and {Poovelil}, Vijith Jacob and {Price-Whelan}, Adrian M. and {B{\'a}rbara de Andrade Queiroz}, Anna and {Raddick}, M. Jordan and {Ray}, Amy and {Rembold}, Sandro Barboza and {Riddle}, Nicole and {Riffel}, Rogemar A. and {Riffel}, Rog{\'e}rio and {Rix}, Hans-Walter and {Robin}, Annie C. and {Rodr{\'\i}guez-Puebla}, Aldo and {Roman-Lopes}, Alexandre and {Rom{\'a}n-Z{\'u}{\~n}iga}, Carlos and {Rose}, Benjamin and {Ross}, Ashley J. and {Rossi}, Graziano and {Rubin}, Kate H.~R. and {Salvato}, Mara and {S{\'a}nchez}, Seb{\'a}stian F. and {S{\'a}nchez-Gallego}, Jos{\'e} R. and {Sanderson}, Robyn and {Santana Rojas}, Felipe Antonio and {Sarceno}, Edgar and {Sarmiento}, Regina and {Sayres}, Conor and {Sazonova}, Elizaveta and {Schaefer}, Adam L. and {Schiavon}, Ricardo and {Schlegel}, David J. and {Schneider}, Donald P. and {Schultheis}, Mathias and {Schwope}, Axel and {Serenelli}, Aldo and {Serna}, Javier and {Shao}, Zhengyi and {Shapiro}, Griffin and {Sharma}, Anubhav and {Shen}, Yue and {Shetrone}, Matthew and {Shu}, Yiping and {Simon}, Joshua D. and {Skrutskie}, M.~F. and {Smethurst}, Rebecca and {Smith}, Verne and {Sobeck}, Jennifer and {Spoo}, Taylor and {Sprague}, Dani and {Stark}, David V. and {Stassun}, Keivan G. and {Steinmetz}, Matthias and {Stello}, Dennis and {Stone-Martinez}, Alexander and {Storchi-Bergmann}, Thaisa and {Stringfellow}, Guy S. and {Stutz}, Amelia and {Su}, Yung-Chau and {Taghizadeh-Popp}, Manuchehr and {Talbot}, Michael S. and {Tayar}, Jamie and {Telles}, Eduardo and {Teske}, Johanna and {Thakar}, Ani and {Theissen}, Christopher and {Tkachenko}, Andrew and {Thomas}, Daniel and {Tojeiro}, Rita and {Hernandez Toledo}, Hector and {Troup}, Nicholas W. and {Trump}, Jonathan R. and {Trussler}, James and {Turner}, Jacqueline and {Tuttle}, Sarah and {Unda-Sanzana}, Eduardo and {V{\'a}zquez-Mata}, Jos{\'e} Antonio and {Valentini}, Marica and {Valenzuela}, Octavio and {Vargas-Gonz{\'a}lez}, Jaime and {Vargas-Maga{\~n}a}, Mariana and {Alfaro}, Pablo Vera and {Villanova}, Sandro and {Vincenzo}, Fiorenzo and {Wake}, David and {Warfield}, Jack T. and {Washington}, Jessica Diane and {Weaver}, Benjamin Alan and {Weijmans}, Anne-Marie and {Weinberg}, David H. and {Weiss}, Achim and {Westfall}, Kyle B. and {Wild}, Vivienne and {Wilde}, Matthew C. and {Wilson}, John C. and {Wilson}, Robert F. and {Wilson}, Mikayla and {Wolf}, Julien and {Wood-Vasey}, W.~M. and {Yan}, Renbin and {Zamora}, Olga and {Zasowski}, Gail and {Zhang}, Kai and {Zhao}, Cheng and {Zheng}, Zheng and {Zheng}, Zheng and {Zhu}, Kai},
        title = "{The Seventeenth Data Release of the Sloan Digital Sky Surveys: Complete Release of MaNGA, MaStar, and APOGEE-2 Data}",
      journal = {\apjs},
         year = {2022},
        month = apr,
       volume = {259},
       number = {2},
          eid = {35},
        pages = {35},
          doi = {10.3847/1538-4365/ac4414},
archivePrefix = {arXiv},
       eprint = {2112.02026},
 primaryClass = {astro-ph.GA},
       adsurl = {https://ui.adsabs.harvard.edu/abs/2022ApJS..259...35A}
}

@ARTICLE{2023Fang,
       author = {{Fang}, Min and {Pascucci}, Ilaria and {Edwards}, Suzan and {Gorti}, Uma and {Hillenbrand}, Lynne A. and {Carpenter}, John M.},
        title = "{A High-resolution Optical Survey of Upper Sco: Evidence for Coevolution of Accretion and Disk Winds}",
      journal = {\apj},
         year = 2023,
        month = mar,
       volume = {945},
       number = {2},
          eid = {112},
        pages = {112},
          doi = {10.3847/1538-4357/acb2c9},
archivePrefix = {arXiv},
       eprint = {2301.09240},
 primaryClass = {astro-ph.SR},
       adsurl = {https://ui.adsabs.harvard.edu/abs/2023ApJ...945..112F}
}

@ARTICLE{1999Wichmann,
       author = {{Wichmann}, R. and {Covino}, E. and {Alcal{\'a}}, J.~M. and {Krautter}, J. and {Allain}, S. and {Hauschildt}, P.~H.},
        title = "{High-resolution spectroscopy of ROSAT-discovered weak-line T Tauri stars near Lupus}",
      journal = {\mnras},
         year = 1999,
        month = aug,
       volume = {307},
       number = {4},
        pages = {909-918},
          doi = {10.1046/j.1365-8711.1999.02666.x},
       adsurl = {https://ui.adsabs.harvard.edu/abs/1999MNRAS.307..909W}
}

@ARTICLE{2017Frasca,
       author = {{Frasca}, A. and {Biazzo}, K. and {Alcal{\'a}}, J.~M. and {Manara}, C.~F. and {Stelzer}, B. and {Covino}, E. and {Antoniucci}, S.},
        title = "{X-shooter spectroscopy of young stellar objects in Lupus. Atmospheric parameters, membership, and activity diagnostics}",
      journal = {\aap},
         year = 2017,
        month = jun,
       volume = {602},
          eid = {A33},
        pages = {A33},
          doi = {10.1051/0004-6361/201630108},
archivePrefix = {arXiv},
       eprint = {1703.01251},
 primaryClass = {astro-ph.SR},
       adsurl = {https://ui.adsabs.harvard.edu/abs/2017A&A...602A..33F}
}

@ARTICLE{2006Gontcharov,
       author = {{Gontcharov}, G.~A.},
        title = "{Pulkovo Compilation of Radial Velocities for 35 495 Hipparcos stars in a common system}",
      journal = {Astronomy Letters},
         year = 2006,
        month = nov,
       volume = {32},
       number = {11},
        pages = {759-771},
          doi = {10.1134/S1063773706110065},
archivePrefix = {arXiv},
       eprint = {1606.08053},
 primaryClass = {astro-ph.SR},
       adsurl = {https://ui.adsabs.harvard.edu/abs/2006AstL...32..759G}
}

@ARTICLE{2007Guenther,
       author = {{Guenther}, E.~W. and {Esposito}, M. and {Mundt}, R. and {Covino}, E. and {Alcal{\'a}}, J.~M. and {Cusano}, F. and {Stecklum}, B.},
        title = "{Pre-main sequence spectroscopic binaries suitable for VLTI observations}",
      journal = {\aap},
         year = 2007,
        month = jun,
       volume = {467},
       number = {3},
        pages = {1147-1155},
          doi = {10.1051/0004-6361:20065686},
archivePrefix = {arXiv},
       eprint = {astro-ph/0702268},
 primaryClass = {astro-ph},
       adsurl = {https://ui.adsabs.harvard.edu/abs/2007A&A...467.1147G}
}

@ARTICLE{2012Dahm,
       author = {{Dahm}, S.~E. and {Slesnick}, Catherine L. and {White}, R.~J.},
        title = "{A Correlation between Circumstellar Disks and Rotation in the Upper Scorpius OB Association}",
      journal = {\apj},
         year = 2012,
        month = jan,
       volume = {745},
       number = {1},
          eid = {56},
        pages = {56},
          doi = {10.1088/0004-637X/745/1/56},
archivePrefix = {arXiv},
       eprint = {1110.0536},
 primaryClass = {astro-ph.SR},
       adsurl = {https://ui.adsabs.harvard.edu/abs/2012ApJ...745...56D}
}

@ARTICLE{2012Biazzo,
       author = {{Biazzo}, K. and {Alcal{\'a}}, J.~M. and {Covino}, E. and {Frasca}, A. and {Getman}, F. and {Spezzi}, L.},
        title = "{The Chamaeleon II low-mass star-forming region: radial velocities, elemental abundances, and accretion properties {\ensuremath{\star}}}",
      journal = {\aap},
         year = 2012,
        month = nov,
       volume = {547},
          eid = {A104},
        pages = {A104},
          doi = {10.1051/0004-6361/201219680},
archivePrefix = {arXiv},
       eprint = {1209.5316},
 primaryClass = {astro-ph.SR},
       adsurl = {https://ui.adsabs.harvard.edu/abs/2012A&A...547A.104B}
}

@ARTICLE{2023Katz,
       author = {{Katz}, D. and {Sartoretti}, P. and {Guerrier}, A. and {Panuzzo}, P. and {Seabroke}, G.~M. and {Th{\'e}venin}, F. and {Cropper}, M. and {Benson}, K. and {Blomme}, R. and {Haigron}, R. and {Marchal}, O. and {Smith}, M. and {Baker}, S. and {Chemin}, L. and {Damerdji}, Y. and {David}, M. and {Dolding}, C. and {Fr{\'e}mat}, Y. and {Gosset}, E. and {Jan{\ss}en}, K. and {Jasniewicz}, G. and {Lobel}, A. and {Plum}, G. and {Samaras}, N. and {Snaith}, O. and {Soubiran}, C. and {Vanel}, O. and {Zwitter}, T. and {Antoja}, T. and {Arenou}, F. and {Babusiaux}, C. and {Brouillet}, N. and {Caffau}, E. and {Di Matteo}, P. and {Fabre}, C. and {Fabricius}, C. and {Fragkoudi}, F. and {Haywood}, M. and {Huckle}, H.~E. and {Hottier}, C. and {Lasne}, Y. and {Leclerc}, N. and {Mastrobuono-Battisti}, A. and {Royer}, F. and {Teyssier}, D. and {Zorec}, J. and {Crifo}, F. and {Jean-Antoine Piccolo}, A. and {Turon}, C. and {Viala}, Y.},
        title = "{Gaia Data Release 3. Properties and validation of the radial velocities}",
      journal = {\aap},
         year = 2023,
        month = jun,
       volume = {674},
          eid = {A5},
        pages = {A5},
          doi = {10.1051/0004-6361/202244220},
archivePrefix = {arXiv},
       eprint = {2206.05902},
 primaryClass = {astro-ph.GA},
       adsurl = {https://ui.adsabs.harvard.edu/abs/2023A&A...674A...5K}
}

@ARTICLE{2024Edenhofer,
       author = {{Edenhofer}, Gordian and {Alves}, Jo{\~a}o and {Zucker}, Catherine and {Posch}, Laura and {En{\ss}lin}, Torsten A.},
        title = "{The ``C'': The large Chameleon-Musca-Coalsack cloud}",
      journal = {\aap},
         year = 2024,
        month = jul,
       volume = {687},
          eid = {L9},
        pages = {L9},
          doi = {10.1051/0004-6361/202450374},
archivePrefix = {arXiv},
       eprint = {2404.09592},
 primaryClass = {astro-ph.GA},
       adsurl = {https://ui.adsabs.harvard.edu/abs/2024A&A...687L...9E}
}

@ARTICLE{2022Zucker,
       author = {{Zucker}, Catherine and {Goodman}, Alyssa A. and {Alves}, Jo{\~a}o and {Bialy}, Shmuel and {Foley}, Michael and {Speagle}, Joshua S. and {Gro{\^I}{\texttwosuperior}schedl}, Josefa and {Finkbeiner}, Douglas P. and {Burkert}, Andreas and {Khimey}, Diana and {Swiggum}, Cameren},
        title = "{Star formation near the Sun is driven by expansion of the Local Bubble}",
      journal = {\nat},
         year = 2022,
        month = jan,
       volume = {601},
       number = {7893},
        pages = {334-337},
          doi = {10.1038/s41586-021-04286-5},
archivePrefix = {arXiv},
       eprint = {2201.05124},
 primaryClass = {astro-ph.GA},
       adsurl = {https://ui.adsabs.harvard.edu/abs/2022Natur.601..334Z}
}

@ARTICLE{2024Swiggum,
       author = {{Swiggum}, Cameren and {Alves}, Jo{\~a}o and {Benjamin}, Robert and {Ratzenb{\"o}ck}, Sebastian and {Miret-Roig}, N{\'u}ria and {Gro{\ss}schedl}, Josefa and {Meingast}, Stefan and {Goodman}, Alyssa and {Konietzka}, Ralf and {Zucker}, Catherine and {Hunt}, Emily L. and {Reffert}, Sabine},
        title = "{Most nearby young star clusters formed in three massive complexes}",
      journal = {\nat},
         year = 2024,
        month = jul,
       volume = {631},
       number = {8019},
        pages = {49-53},
          doi = {10.1038/s41586-024-07496-9},
archivePrefix = {arXiv},
       eprint = {2406.06510},
 primaryClass = {astro-ph.GA},
       adsurl = {https://ui.adsabs.harvard.edu/abs/2024Natur.631...49S}
}

@ARTICLE{2012Oppermann,
   author = {{Oppermann}, N. and {Junklewitz}, H. and {Robbers}, G. and {Bell}, M.~R. and
	{En{\ss}lin}, T.~A. and {Bonafede}, A. and {Braun}, R. and {Brown}, J.~C. and
	{Clarke}, T.~E. and {Feain}, I.~J. and {Gaensler}, B.~M. and
	{Hammond}, A. and {Harvey-Smith}, L. and {Heald}, G. and {Johnston-Hollitt}, M. and
	{Klein}, U. and {Kronberg}, P.~P. and {Mao}, S.~A. and {McClure-Griffiths}, N.~M. and
	{O'Sullivan}, S.~P. and {Pratley}, L. and {Robishaw}, T. and
	{Roy}, S. and {Schnitzeler}, D.~H.~F.~M. and {Sotomayor-Beltran}, C. and
	{Stevens}, J. and {Stil}, J.~M. and {Sunstrum}, C. and {Tanna}, A. and
	{Taylor}, A.~R. and {Van Eck}, C.~L.},
    title = "{An improved map of the Galactic Faraday sky}",
  journal = {\aap},
archivePrefix = "arXiv",
   eprint = {1111.6186},
 primaryClass = "astro-ph.GA",
     year = 2012,
    month = jun,
   volume = 542,
      eid = {A93},
    pages = {A93},
      doi = {10.1051/0004-6361/201118526},
   adsurl = {http://adsabs.harvard.edu/abs/2012A26A...542A..93O}
}

@ARTICLE{2019Damiani,
       author = {{Damiani}, F. and {Prisinzano}, L. and {Pillitteri}, I. and {Micela}, G. and {Sciortino}, S.},
        title = "{Stellar population of Sco OB2 revealed by Gaia DR2 data}",
      journal = {\aap},
         year = 2019,
        month = mar,
       volume = {623},
          eid = {A112},
        pages = {A112},
          doi = {10.1051/0004-6361/201833994},
archivePrefix = {arXiv},
       eprint = {1807.11884},
 primaryClass = {astro-ph.SR},
       adsurl = {https://ui.adsabs.harvard.edu/abs/2019A&A...623A.112D}
}

@ARTICLE{2018Krause,
       author = {{Krause}, Martin G.~H. and {Burkert}, Andreas and {Diehl}, Roland and {Fierlinger}, Katharina and {Gaczkowski}, Benjamin and {Kroell}, Daniel and {Ngoumou}, Judith and {Roccatagliata}, Veronica and {Siegert}, Thomas and {Preibisch}, Thomas},
        title = "{Surround and Squash: the impact of superbubbles on the interstellar medium in Scorpius-Centaurus OB2}",
      journal = {\aap},
         year = 2018,
        month = nov,
       volume = {619},
          eid = {A120},
        pages = {A120},
          doi = {10.1051/0004-6361/201732416},
archivePrefix = {arXiv},
       eprint = {1808.04788},
 primaryClass = {astro-ph.GA},
       adsurl = {https://ui.adsabs.harvard.edu/abs/2018A&A...619A.120K}
}

@ARTICLE{2018Robitaille,
       author = {{Robitaille}, J. -F. and {Scaife}, A.~M.~M. and {Carretti}, E. and {Haverkorn}, M. and {Crocker}, R.~M. and {Kesteven}, M.~J. and {Poppi}, S. and {Staveley-Smith}, L.},
        title = "{Interstellar magnetic cannon targeting the Galactic halo. A young bubble at the origin of the Ophiuchus and Lupus molecular complexes}",
      journal = {\aap},
         year = 2018,
        month = sep,
       volume = {617},
          eid = {A101},
        pages = {A101},
          doi = {10.1051/0004-6361/201833358},
archivePrefix = {arXiv},
       eprint = {1807.04054},
 primaryClass = {astro-ph.GA},
       adsurl = {https://ui.adsabs.harvard.edu/abs/2018A&A...617A.101R}
}

@ARTICLE{1989Loren,
       author = {{Loren}, Robert B.},
        title = "{The Cobwebs of Ophiuchus. I. Strands of 13CO: The Mass Distribution}",
      journal = {\apj},
         year = 1989,
        month = mar,
       volume = {338},
        pages = {902},
          doi = {10.1086/167244},
       adsurl = {https://ui.adsabs.harvard.edu/abs/1989ApJ...338..902L}
}

@ARTICLE{1993Harju,
       author = {{Harju}, J. and {Haikala}, L.~K. and {Mattila}, K. and {Mauersberger}, R. and {Booth}, R.~S. and {Nordh}, H.~L.},
        title = "{Large scale structure of the R Coronae Australis cloud core.}",
      journal = {\aap},
         year = 1993,
        month = nov,
       volume = {278},
        pages = {569-583},
       adsurl = {https://ui.adsabs.harvard.edu/abs/1993A&A...278..569H}
}

@ARTICLE{2021Zucker,
       author = {{Zucker}, Catherine and {Goodman}, Alyssa and {Alves}, Jo{\~a}o and {Bialy}, Shmuel and {Koch}, Eric W. and {Speagle}, Joshua S. and {Foley}, Michael M. and {Finkbeiner}, Douglas and {Leike}, Reimar and {En{\ss}lin}, Torsten and {Peek}, Joshua E.~G. and {Edenhofer}, Gordian},
        title = "{On the Three-dimensional Structure of Local Molecular Clouds}",
      journal = {\apj},
         year = 2021,
        month = sep,
       volume = {919},
       number = {1},
          eid = {35},
        pages = {35},
          doi = {10.3847/1538-4357/ac1f96},
archivePrefix = {arXiv},
       eprint = {2109.09765},
 primaryClass = {astro-ph.GA},
       adsurl = {https://ui.adsabs.harvard.edu/abs/2021ApJ...919...35Z}
}

@ARTICLE{2008Nehme,
       author = {{Nehm{\'e}}, C. and {Gry}, C. and {Boulanger}, F. and {Le Bourlot}, J. and {Pineau Des For{\^e}ts}, G. and {Falgarone}, E.},
        title = "{Multi-wavelength observations of a nearby multi-phase interstellar cloud}",
      journal = {\aap},
         year = 2008,
        month = may,
       volume = {483},
       number = {2},
        pages = {471-484},
          doi = {10.1051/0004-6361:20078373},
archivePrefix = {arXiv},
       eprint = {0802.4059},
 primaryClass = {astro-ph},
       adsurl = {https://ui.adsabs.harvard.edu/abs/2008A&A...483..471N}
}

@ARTICLE{1995Frisch,
       author = {{Frisch}, Priscilla C.},
        title = "{Characteristics of Nearby Interstellar Matter}",
      journal = {\ssr},
         year = 1995,
        month = may,
       volume = {72},
       number = {3-4},
        pages = {499-592},
          doi = {10.1007/BF00749006},
       adsurl = {https://ui.adsabs.harvard.edu/abs/1995SSRv...72..499F}
}

@ARTICLE{2008Redfield,
       author = {{Redfield}, Seth and {Linsky}, Jeffrey L.},
        title = "{The Structure of the Local Interstellar Medium. IV. Dynamics, Morphology, Physical Properties, and Implications of Cloud-Cloud Interactions}",
      journal = {\apj},
         year = 2008,
        month = jan,
       volume = {673},
       number = {1},
        pages = {283-314},
          doi = {10.1086/524002},
archivePrefix = {arXiv},
       eprint = {0804.1802},
 primaryClass = {astro-ph},
       adsurl = {https://ui.adsabs.harvard.edu/abs/2008ApJ...673..283R}
}

@ARTICLE{2024Piecka,
       author = {{Piecka}, M. and {Hutschenreuter}, S. and {Alves}, J.},
        title = "{Towards a complete picture of the Sco-Cen outflow}",
      journal = {\aap},
         year = 2024,
        month = sep,
       volume = {689},
          eid = {A84},
        pages = {A84},
          doi = {10.1051/0004-6361/202450936},
archivePrefix = {arXiv},
       eprint = {2407.13226},
 primaryClass = {astro-ph.GA},
       adsurl = {https://ui.adsabs.harvard.edu/abs/2024A&A...689A..84P}
}

@ARTICLE{2019Ensslin,
       author = {{En{\ss}lin}, Torsten A.},
        title = "{Information Theory for Fields}",
      journal = {Annalen der Physik},
         year = 2019,
        month = mar,
       volume = {531},
       number = {3},
        pages = {1800127},
          doi = {10.1002/andp.201800127},
archivePrefix = {arXiv},
       eprint = {1804.03350},
 primaryClass = {astro-ph.CO},
       adsurl = {https://ui.adsabs.harvard.edu/abs/2019AnP...53100127E}
}

@ARTICLE{2023Posch,
       author = {{Posch}, Laura and {Miret-Roig}, N{\'u}ria and {Alves}, Jo{\~a}o and {Ratzenb{\"o}ck}, Sebastian and {Gro{\ss}schedl}, Josefa and {Meingast}, Stefan and {Zucker}, Catherine and {Burkert}, Andreas},
        title = "{The Corona Australis star formation complex is accelerating away from the Galactic plane}",
      journal = {\aap},
         year = 2023,
        month = nov,
       volume = {679},
          eid = {L10},
        pages = {L10},
          doi = {10.1051/0004-6361/202347186},
archivePrefix = {arXiv},
       eprint = {2310.14373},
 primaryClass = {astro-ph.GA},
       adsurl = {https://ui.adsabs.harvard.edu/abs/2023A&A...679L..10P}
}

@ARTICLE{2022Miret-Roig,
       author = {{Miret-Roig}, N. and {Galli}, P.~A.~B. and {Olivares}, J. and {Bouy}, H. and {Alves}, J. and {Barrado}, D.},
        title = "{The star formation history of Upper Scorpius and Ophiuchus. A 7D picture: positions, kinematics, and dynamical traceback ages}",
      journal = {\aap},
         year = 2022,
        month = nov,
       volume = {667},
          eid = {A163},
        pages = {A163},
          doi = {10.1051/0004-6361/202244709},
archivePrefix = {arXiv},
       eprint = {2209.12938},
 primaryClass = {astro-ph.GA},
       adsurl = {https://ui.adsabs.harvard.edu/abs/2022A&A...667A.163M}
}

@ARTICLE{2023Ratzenboeck1,
       author = {{Ratzenb{\"o}ck}, Sebastian and {Gro{\ss}schedl}, Josefa E. and {M{\"o}ller}, Torsten and {Alves}, Jo{\~a}o and {Bomze}, Immanuel and {Meingast}, Stefan},
        title = "{Significance mode analysis (SigMA) for hierarchical structures. An application to the Sco-Cen OB association}",
      journal = {\aap},
         year = {2023a},
        month = sep,
       volume = {677},
          eid = {A59},
        pages = {A59},
          doi = {10.1051/0004-6361/202243690},
archivePrefix = {arXiv},
       eprint = {2211.14225},
 primaryClass = {astro-ph.GA},
       adsurl = {https://ui.adsabs.harvard.edu/abs/2023A&A...677A..59R}
}

@ARTICLE{2023Ratzenboeck2,
       author = {{Ratzenb{\"o}ck}, Sebastian and {Gro{\ss}schedl}, Josefa E. and {Alves}, Jo{\~a}o and {Miret-Roig}, N{\'u}ria and {Bomze}, Immanuel and {Forbes}, John and {Goodman}, Alyssa and {Hacar}, {\'A}lvaro and {Lin}, Doug and {Meingast}, Stefan and {M{\"o}ller}, Torsten and {Piecka}, Martin and {Posch}, Laura and {Rottensteiner}, Alena and {Swiggum}, Cameren and {Zucker}, Catherine},
        title = "{The star formation history of the Sco-Cen association. Coherent star formation patterns in space and time}",
      journal = {\aap},
         year = {2023b},
        month = oct,
       volume = {678},
          eid = {A71},
        pages = {A71},
          doi = {10.1051/0004-6361/202346901},
archivePrefix = {arXiv},
       eprint = {2302.07853},
 primaryClass = {astro-ph.SR},
       adsurl = {https://ui.adsabs.harvard.edu/abs/2023A&A...678A..71R}
}

@ARTICLE{2024Hutschenreuter,
       author = {{Hutschenreuter}, Sebastian and {Haverkorn}, Marijke and {Frank}, Philipp and {Raycheva}, Nergis C. and {En{\ss}lin}, Torsten A.},
        title = "{Disentangling the Faraday rotation sky}",
      journal = {\aap},
         year = 2024,
        month = oct,
       volume = {690},
          eid = {A314},
        pages = {A314},
          doi = {10.1051/0004-6361/202346740},
archivePrefix = {arXiv},
       eprint = {2304.12350},
 primaryClass = {astro-ph.GA},
       adsurl = {https://ui.adsabs.harvard.edu/abs/2024A&A...690A.314H}
}

@ARTICLE{2022Hutschenreuter,
       author = {{Hutschenreuter}, S. and {Anderson}, C.~S. and {Betti}, S. and {Bower}, G.~C. and {Brown}, J. -A. and {Br{\"u}ggen}, M. and {Carretti}, E. and {Clarke}, T. and {Clegg}, A. and {Costa}, A. and {Croft}, S. and {Van Eck}, C. and {Gaensler}, B.~M. and {de Gasperin}, F. and {Haverkorn}, M. and {Heald}, G. and {Hull}, C.~L.~H. and {Inoue}, M. and {Johnston-Hollitt}, M. and {Kaczmarek}, J. and {Law}, C. and {Ma}, Y.~K. and {MacMahon}, D. and {Mao}, S.~A. and {Riseley}, C. and {Roy}, S. and {Shanahan}, R. and {Shimwell}, T. and {Stil}, J. and {Sobey}, C. and {O'Sullivan}, S.~P. and {Tasse}, C. and {Vacca}, V. and {Vernstrom}, T. and {Williams}, P.~K.~G. and {Wright}, M. and {En{\ss}lin}, T.~A.},
        title = "{The Galactic Faraday rotation sky 2020}",
      journal = {\aap},
         year = 2022,
        month = jan,
       volume = {657},
          eid = {A43},
        pages = {A43},
          doi = {10.1051/0004-6361/202140486},
archivePrefix = {arXiv},
       eprint = {2102.01709},
 primaryClass = {astro-ph.GA},
       adsurl = {https://ui.adsabs.harvard.edu/abs/2022A&A...657A..43H}
}

@misc{2023Edenhofer,
      title={A Parsec-Scale Galactic 3D Dust Map out to 1.25 kpc from the Sun}, 
      author={Gordian Edenhofer and Catherine Zucker and Philipp Frank and Andrew K. Saydjari and Joshua S. Speagle and Douglas Finkbeiner and Torsten Enßlin},
      year={2023},
      eprint={2308.01295},
      archivePrefix={arXiv},
      primaryClass={astro-ph.GA}
}

@ARTICLE{2022Arras,
       author = {{Arras}, Philipp and {Frank}, Philipp and {Haim}, Philipp and {Knollm{\"u}ller}, Jakob and {Leike}, Reimar and {Reinecke}, Martin and {En{\ss}lin}, Torsten},
        title = "{Variable structures in M87* from space, time and frequency resolved interferometry}",
      journal = {Nature Astronomy},
         year = 2022,
        month = jan,
       volume = {6},
        pages = {259-269},
          doi = {10.1038/s41550-021-01548-0},
archivePrefix = {arXiv},
       eprint = {2002.05218},
 primaryClass = {astro-ph.IM},
       adsurl = {https://ui.adsabs.harvard.edu/abs/2022NatAs...6..259A}
}

@ARTICLE{2021Frank,
       author = {{Frank}, Philipp and {Leike}, Reimar and {En{\ss}lin}, Torsten A.},
        title = "{Geometric Variational Inference}",
      journal = {Entropy},
         year = 2021,
        month = jul,
       volume = {23},
       number = {7},
        pages = {853},
          doi = {10.3390/e23070853},
archivePrefix = {arXiv},
       eprint = {2105.10470},
 primaryClass = {stat.ME},
       adsurl = {https://ui.adsabs.harvard.edu/abs/2021Entrp..23..853F}
}

@ARTICLE{2019Leike,
       author = {{Leike}, R.~H. and {En{\ss}lin}, T.~A.},
        title = "{Charting nearby dust clouds using Gaia data only}",
      journal = {\aap},
         year = 2019,
        month = nov,
       volume = {631},
          eid = {A32},
        pages = {A32},
          doi = {10.1051/0004-6361/201935093},
archivePrefix = {arXiv},
       eprint = {1901.05971},
 primaryClass = {astro-ph.GA},
       adsurl = {https://ui.adsabs.harvard.edu/abs/2019A&A...631A..32L}
}

@article{luhman_census_2022,
    title = {A {Census} of the {Stellar} {Populations} in the {Sco}-{Cen} {Complex}*},
    volume = {163},
    url = {http://dx.doi.org/10.3847/1538-3881/ac35e2},
    doi = {10.3847/1538-3881/ac35e2},
    number = {1},
    journal = {The Astronomical Journal},
    author = {Luhman, K L},
    year = {2022},
    pages = {24},
}

\appendix

\section{Further details on the data}
\label{app:data}

\begin{figure*}
    \centering
\begin{minipage}[c]{0.48\linewidth}
    \centering
\includegraphics[width=\linewidth]{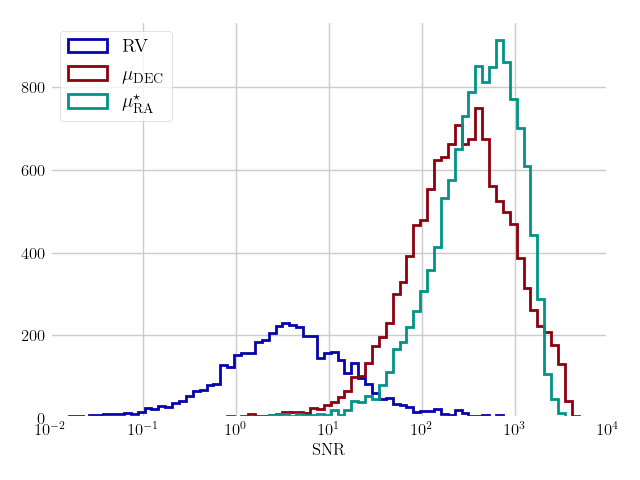}
\caption{Illustrating the quality of the used data sets, as traced by the S/N, i.e., the ratio of data value over observational errors.}
\label{fig:data_S/N}
\end{minipage}
\hfill
\begin{minipage}[c]{0.48\linewidth}
\includegraphics[width=\linewidth]{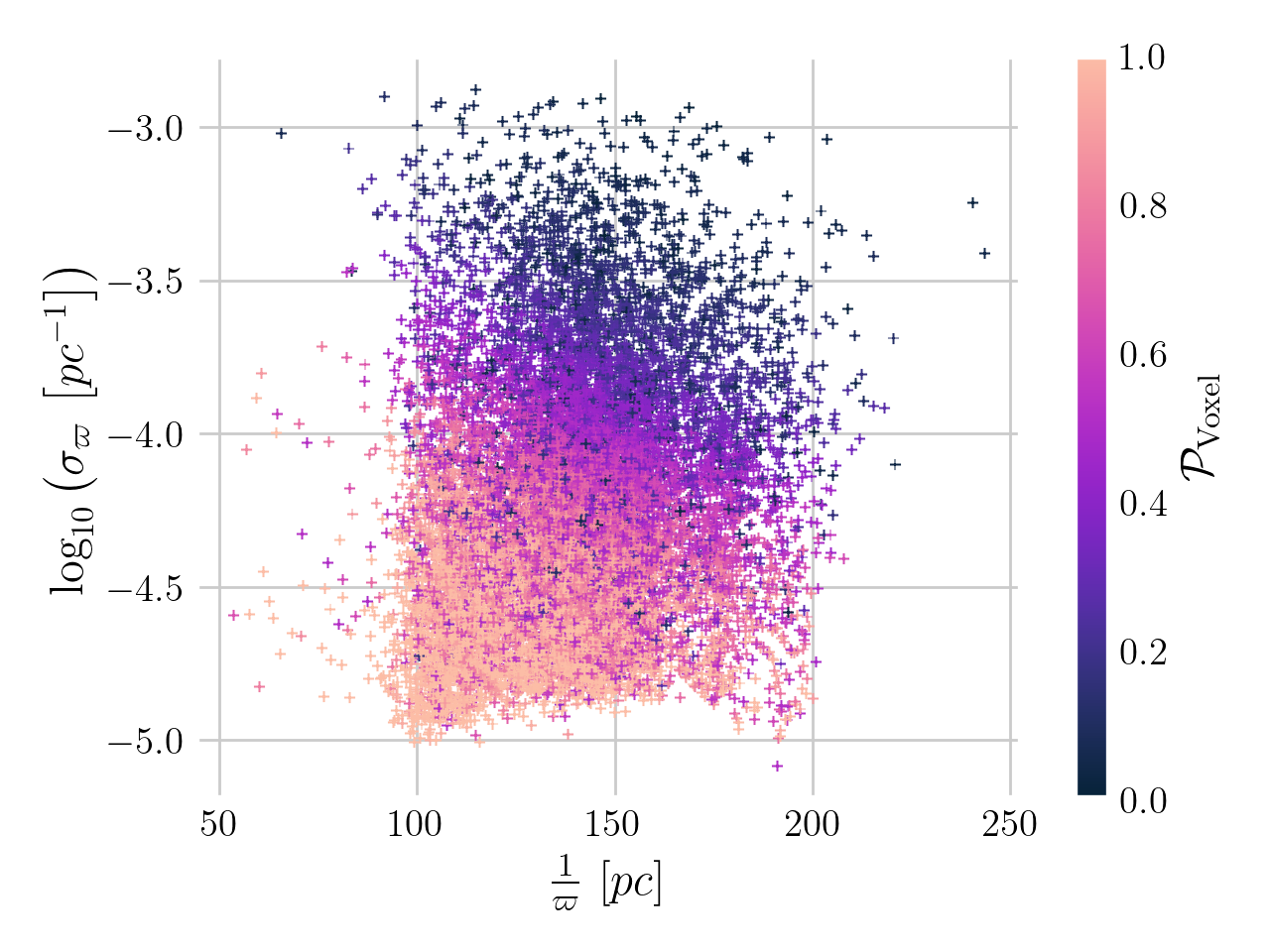}
\caption{Scatter plot of observed distance as derived from $1/\varpi$ vs. the errors on $\varpi$. 
Color-coded is the likelihood of the stars to actually be in the voxel where they are found, according to our grid definition and the $\varpi$ noise statistics.
} 
\label{fig:data_distance_error}
\end{minipage}%
\end{figure*}

\begin{table}
\centering
\begin{small}
\caption{Overview of the clusters in the \texttt{main} selection.}
\label{tab:summary_main_selection}
\renewcommand{\arraystretch}{1.4}
\begin{tabular}{lrr}
\hline \hline
\texttt{SigMA} name & $N_{\rm stars}$ & age [Myr] \\
\hline
Antares & 502 & $12.7^{+0.4}_{-1.3}$ \\
$\rho$ Sco & 240 & $13.7^{+1.3}_{-0.6}$ \\
Scorpio-Body & 373 & $14.7^{+0.8}_{-0.7}$ \\
US-foreground & 276 & $19.1^{+2.4}_{-1.3}$ \\
V1062-Sco & 1029 & $15.0^{+0.9}_{-1.4}$ \\
$\mu$ Sco & 54 & $17.2^{+0.9}_{-2.4}$ \\
Libra-South & 71 & $20.0^{+2.5}_{-2.2}$ \\
$\eta$ Lup & 769 & $15.3^{+0.6}_{-0.3}$ \\
$\phi$ Lup & 1114 & $16.9^{+0.9}_{-0.6}$ \\
$e$ Lup & 516 & $20.9^{+0.7}_{+0.8}$ \\
UPK606 & 131 & $13.4^{+1.0}_{-0.7}$ \\
$\rho$ Lup & 246 & $14.4^{+0.4}_{-0.9}$ \\
$\nu$ Cen & 1737 & $15.7^{+0.3}_{-0.9}$ \\
$\sigma$ Cen & 1805 & $15.5^{+0.6}_{-0.5}$ \\
Acrux & 394 & $11.2^{+1.0}_{-1.0}$ \\
Musca-foreground & 95 & $10.2^{+1.0}_{-0.7}$ \\
$\epsilon$ Cham & 39 & $8.8^{+0.6}_{-0.4}$ \\
$\eta$ Cham & 30 & $9.4^{+1.4}_{-0.9}$ \\
Pipe-North & 42 & $15.9^{+1.6}_{-2.1}$ \\
$\theta$ Oph & 98 & $15.4^{+0.8}_{-1.9}$ \\
CrA-Main & 96 & $8.5^{+2.0}_{-2.4}$ \\
CrA-North & 351 & $11.6^{+0.5}_{-0.6}$ \\
Scorpio-Sting & 132 & $14.5^{+0.6}_{-0.6}$ \\
Centaurus-Far & 99 & $8.5^{+1.1}_{-1.3}$ \\
Chamaeleon-1 & 192 & $3.8^{+1.9}_{-0.9}$ \\
Chamaeleon-2 & 54 & $2.8^{+0.7}_{-0.9}$ \\
L134/L183 & 24 & $9.6^{+1.7}_{-2.2}$ \\
\hline
\end{tabular}
\end{small}
\end{table}

\begin{table}[!h]
\centering
\begin{small}
\caption{Overview of the clusters in the \texttt{secondary} selection.}
\label{tab:summary_secondary_selection}
\renewcommand{\arraystretch}{1.4}
\begin{tabular}{lcr}
\hline 
\hline
\texttt{SigMA} name & $N_{\rm stars}$ & age [Myr] \\
\hline
B59 & 32 & 3.4$^{+3.1}_{-0.9}$ \\ 
$\beta$-Sco & 285  & 7.6$^{+0.8}_{-0.7}$ \\   
$\delta$-Sco & 691 &  9.8$^{+1.2}_{-1.4}$ \\   
$\nu$-Sco & 150 & 5.8$^{+1.8}_{-0.5}$ \\   
$\sigma$-Sco & 544 & 10.0$^{+1.0}_{-0.5}$ \\   
$\rho$-Oph/L1688 & 535 & 3.8$^{+0.4}_{-0.4}$ \\   
Lupus-1--4 & 226 & 6.0$^{+0.6}_{-0.9}$ \\   
L134/L138 & 24 & 9.6$^{+1.7}_{-2.2}$ \\ 
\hline
\end{tabular}
\renewcommand{\arraystretch}{1}
\end{small}
\end{table}

\begin{figure}
    \centering
    \begin{subfigure}{0.84\linewidth}
        \includegraphics[trim=200 50 350 730, clip, width=\linewidth]{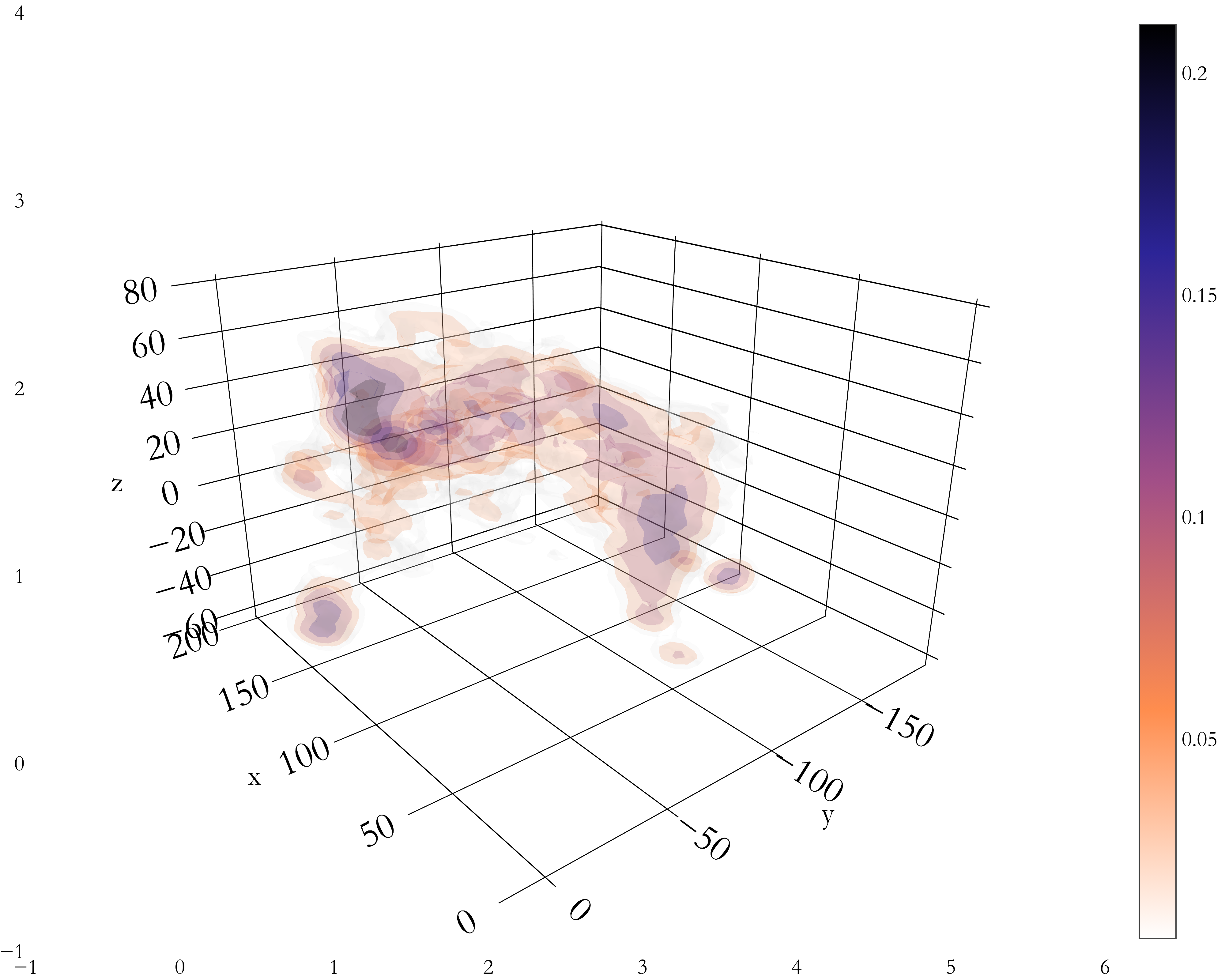}
    \end{subfigure}
    \begin{subfigure}{0.15\linewidth}
        \includegraphics[width=\linewidth]{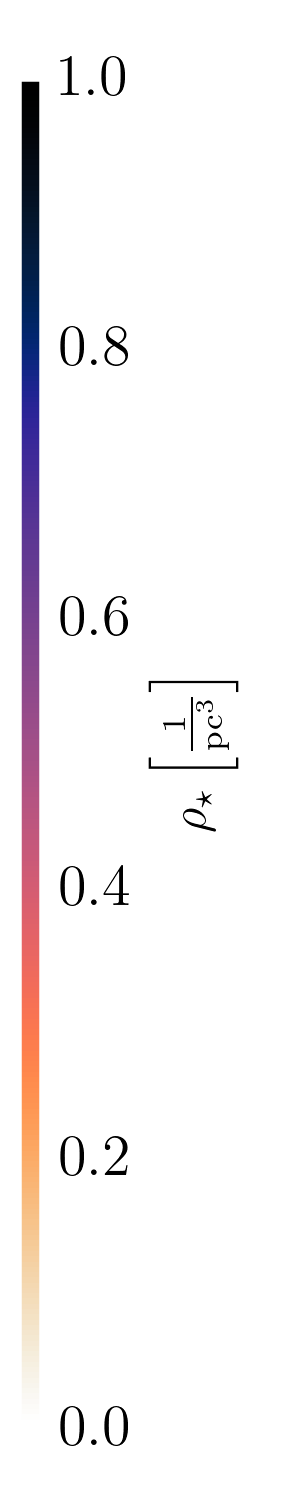}
    \end{subfigure}
    \caption{Stellar density of the \texttt{SigMA} selection of stars in Sco-Cen. 
    A 3D interactive version of this plot is available \href{https://shutsch.github.io/sco_cen_data.html}{online}.}
    \label{fig:density}
\end{figure}

\begin{table}[!t]
\begin{small}
\caption{The prior hyper-parameters for the correlation structure model of the $v_x$, $v_y$, $v_z$ components of \texttt{main} and $\delta$ vector fields.}
\label{tab:prior_parameters}
\centering
\renewcommand{\arraystretch}{1.2}
\begin{tabular}{lccc}
\hline 
\hline
\multicolumn{1}{l}{Parameter Name} &
\multicolumn{1}{c}{Unit} &
\multicolumn{1}{c}{\texttt{main}} &
\multicolumn{1}{c}{$\delta$} \\
\hline
offset mean $x$ & km\,s$^{-1}$ & $-6.7$  & $0$ \\ 
offset mean $y$ & km\,s$^{-1}$ & $-19.7$ & $0$ \\ 
offset mean $z$ & km\,s$^{-1}$ & $-5.5$  & $0$ \\ 
\hline 
offset std$^\star$ ($x,y,z$) & km\,s$^{-1}$ & \multicolumn{2}{c}{$15 \pm 15$} \\
fluctuations$^\star$  ($x,y,z$) & km\,s$^{-1}$ & \multicolumn{2}{c}{$25 \pm 25$} \\
loglogavgslope ($x,y,z$) & & \multicolumn{2}{c}{$-7.3 \pm 3.0$} \\ 
asperity$^\star$  ($x,y,z$) & & \multicolumn{2}{c}{$0.5 \pm 0.5$} \\ 
flexibility$^\star$  ($x,y,z$) & & \multicolumn{2}{c}{$0.5 \pm 0.5$} \\ 
\hline
\end{tabular}
\renewcommand{\arraystretch}{1}
\tablefoot{
The parameters marked with a $^\star$ have log-normal priors, with the given mean and standard deviations moment matched to the log-normal distribution. 
Units are given if the prior is applicable; otherwise, the quantities are unitless.
The prior is the same for both fields, apart from the three offset mean parameters.
}
\end{small}
\end{table}

\begin{figure*}
    \centering
    \begin{subfigure}{0.85\textwidth}
        \includegraphics[trim=200 0 80 600, clip, width=\linewidth]{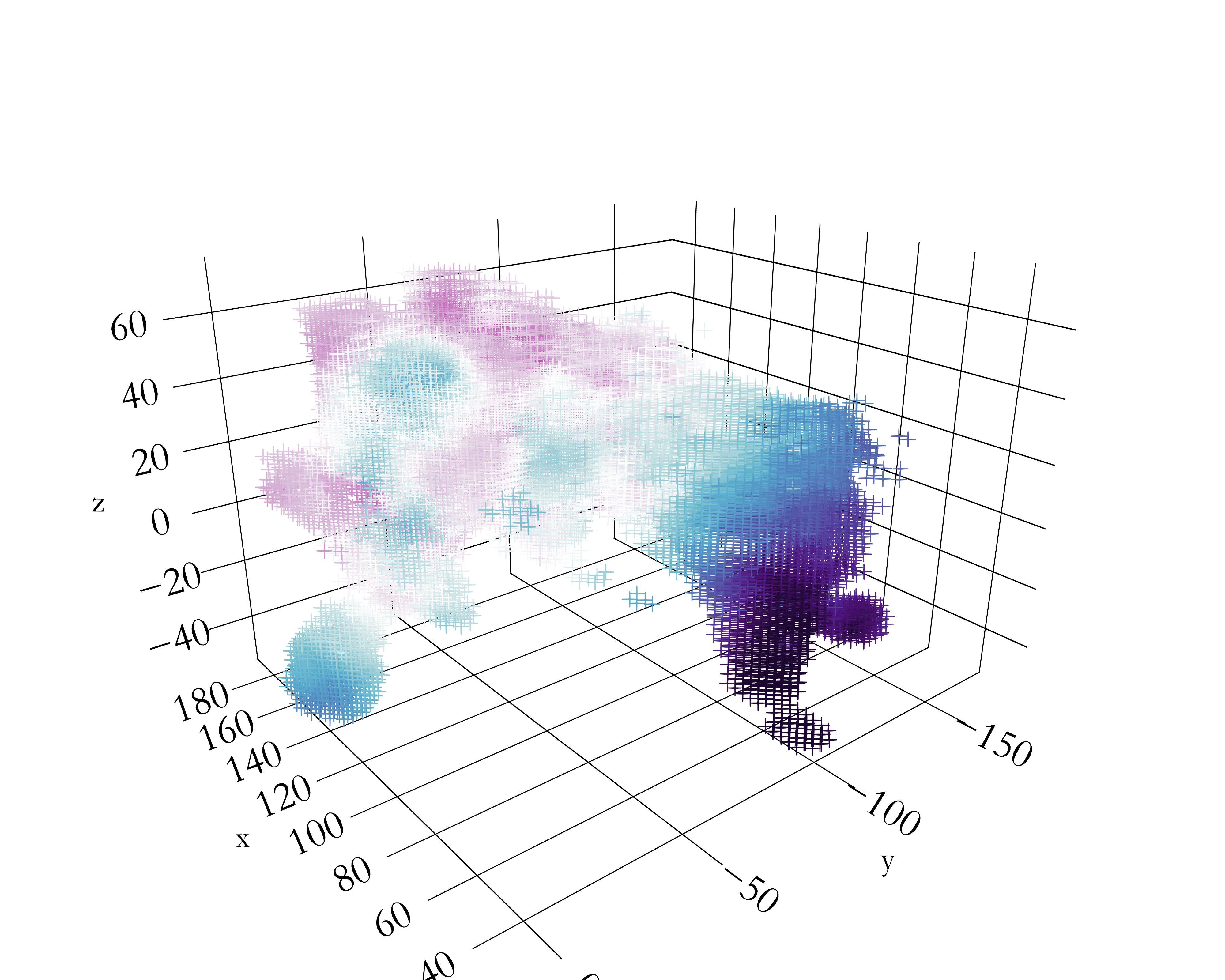}
    \end{subfigure}
    \begin{subfigure}{0.1\textwidth}
        \includegraphics[width=\linewidth]{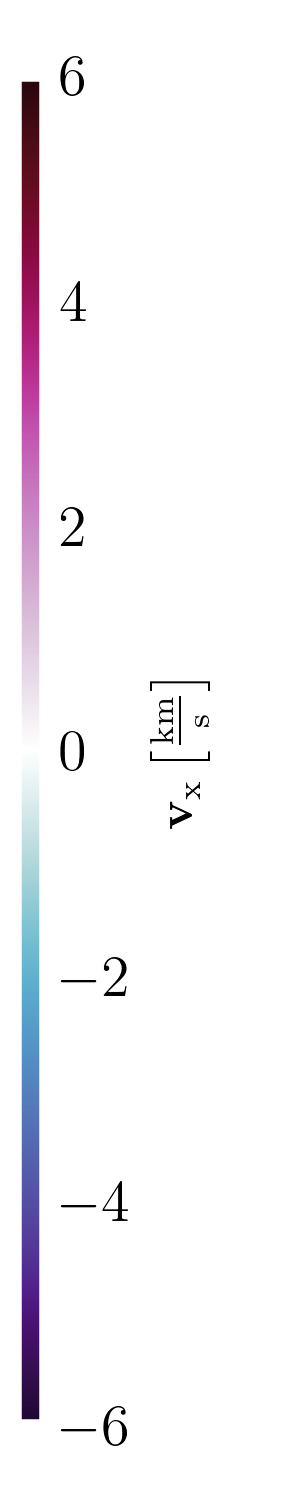}
    \end{subfigure}
    \caption{\label{fig:x_component} Posterior mean x-component \texttt{main} flow field. 
    A 3D interactive version of this plot is available \href{https://shutsch.github.io/sco_cen_data.html}{online}.}
\end{figure*}

\begin{figure*}
    \centering
    \begin{subfigure}{0.85\textwidth}
        \includegraphics[trim=200 0 80 600, clip, width=\linewidth]{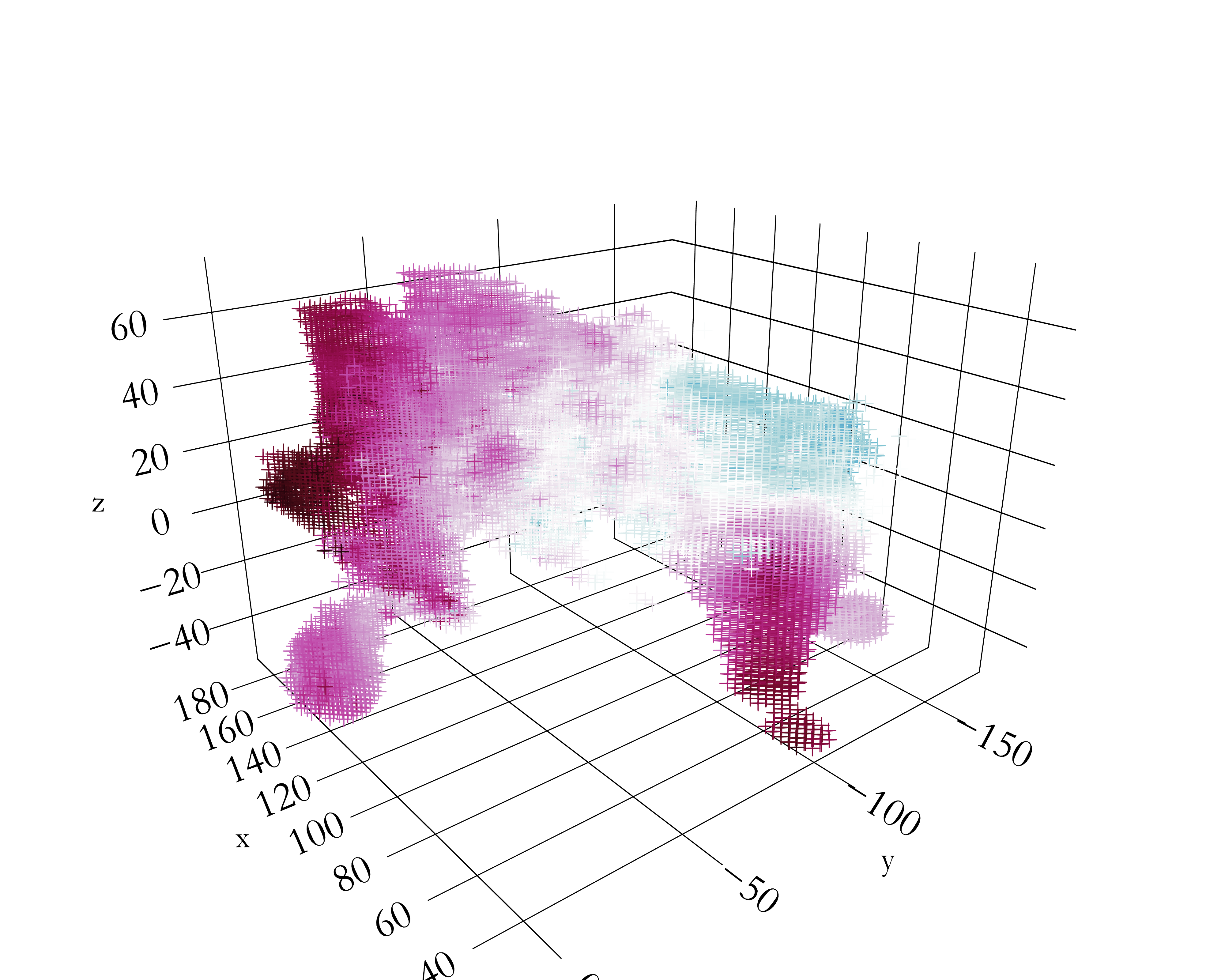}
    \end{subfigure}
    \begin{subfigure}{0.1\textwidth}
        \includegraphics[width=\linewidth]{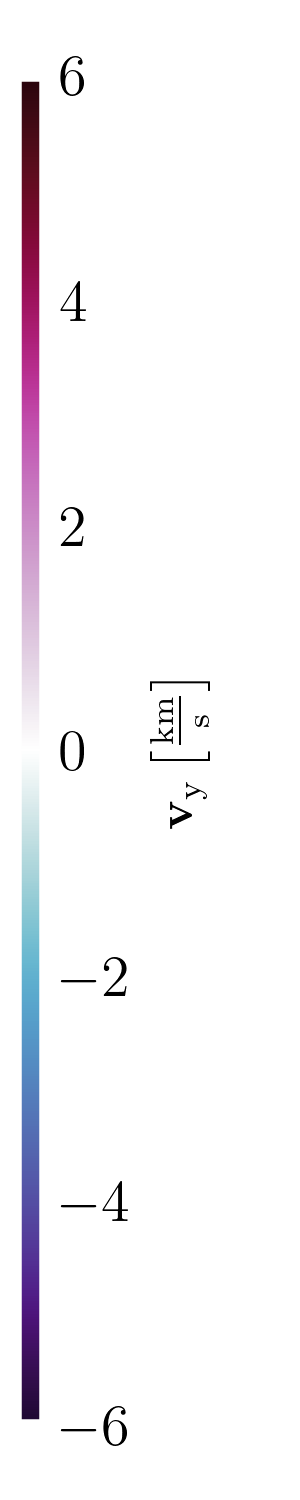}
    \end{subfigure}
    \caption{\label{fig:y_component} Posterior mean y-component \texttt{main} flow field. 
    A 3D interactive version of this plot is available \href{https://shutsch.github.io/sco_cen_data.html}{online}.}
\end{figure*}

\begin{figure*}
    \centering
    \begin{subfigure}{0.85\textwidth}
        \includegraphics[trim=200 0 80 600, clip, width=\linewidth]{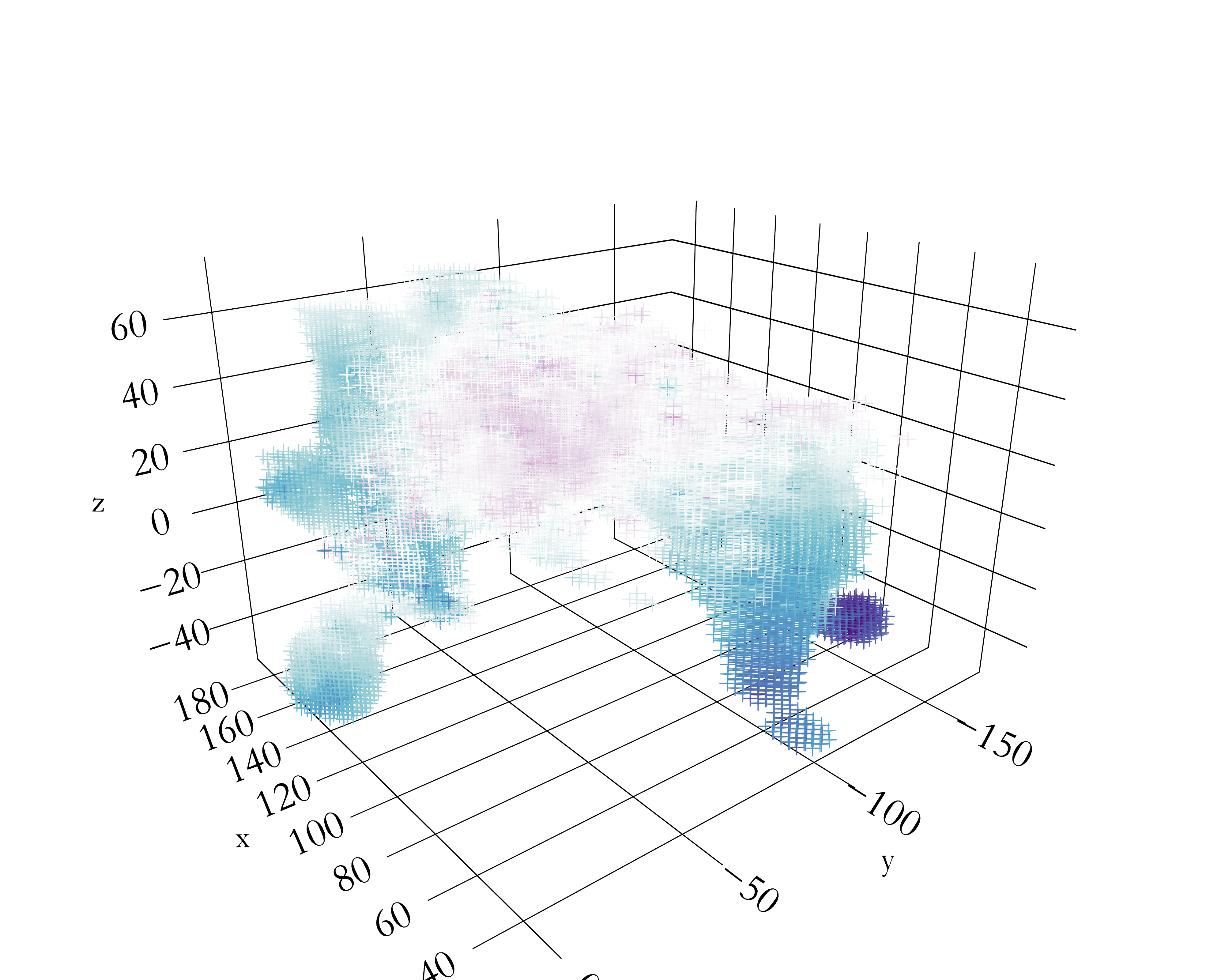}
    \end{subfigure}
    \begin{subfigure}{0.1\textwidth}
        \includegraphics[width=\linewidth]{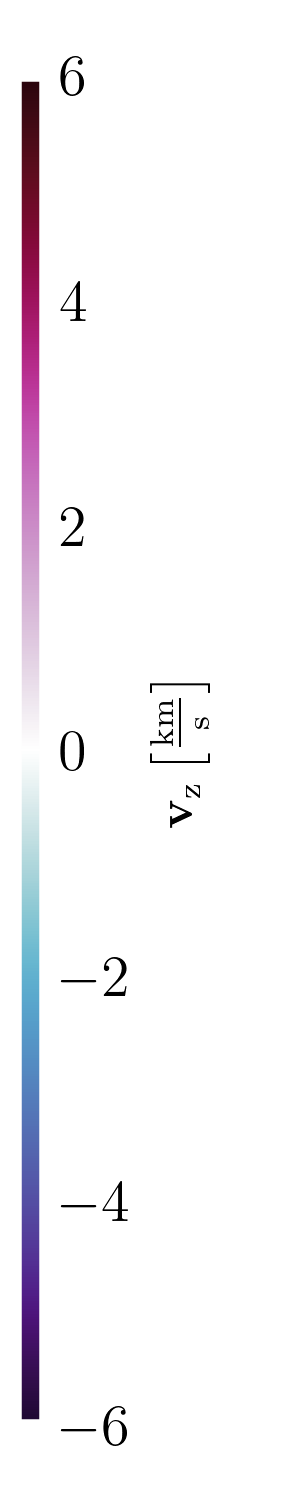}
    \end{subfigure}
    \caption{\label{fig:z_component} Posterior mean z-component \texttt{main} flow field. 
    A 3D interactive version of this plot is available \href{https://shutsch.github.io/sco_cen_data.html}{online}.}
\end{figure*}

\begin{figure*}
    \centering
    \begin{subfigure}{0.85\textwidth}
        \includegraphics[trim=200 0 80 600, clip, width=\linewidth]{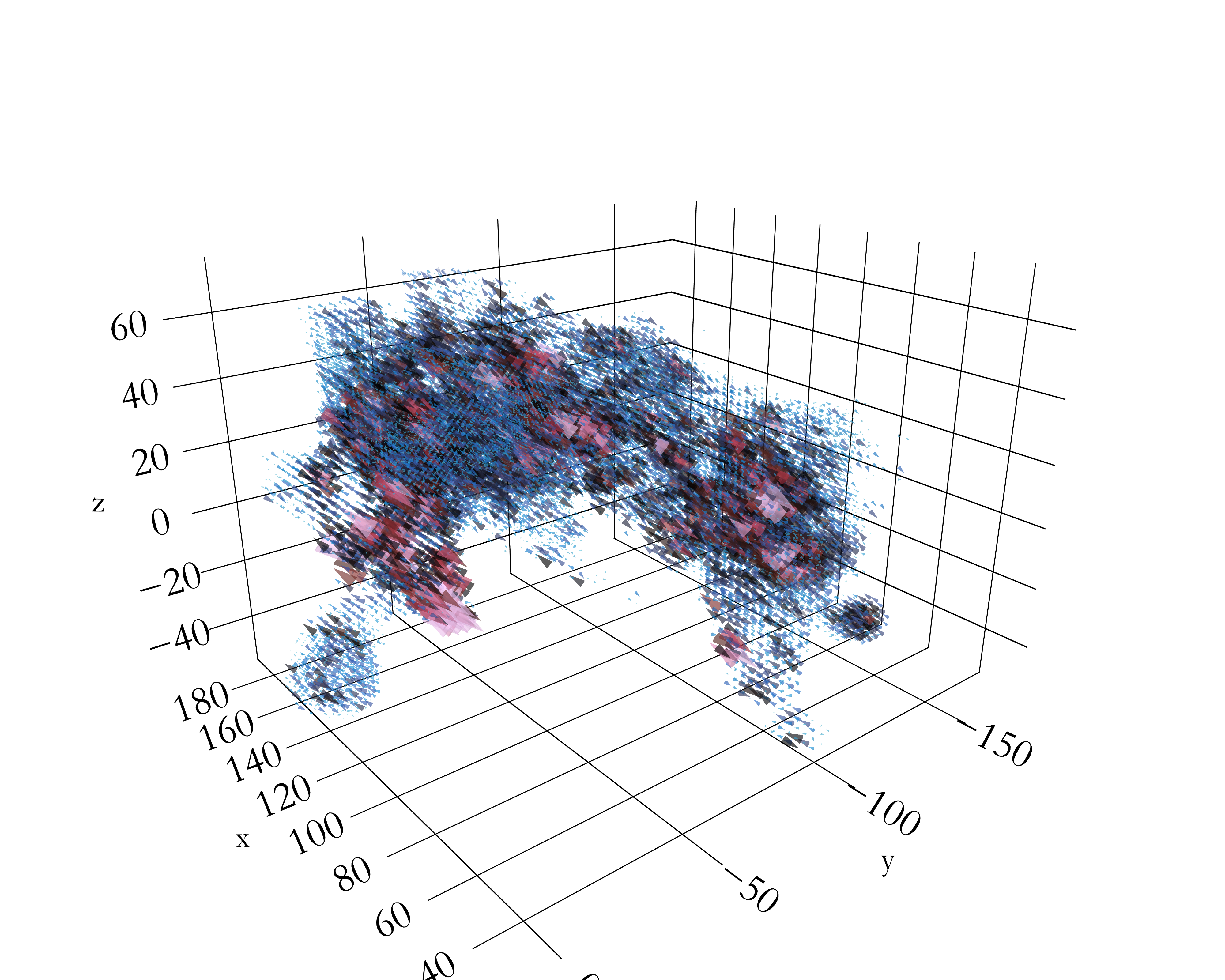}
    \end{subfigure}
    \begin{subfigure}{0.1\textwidth}
        \includegraphics[width=\linewidth]{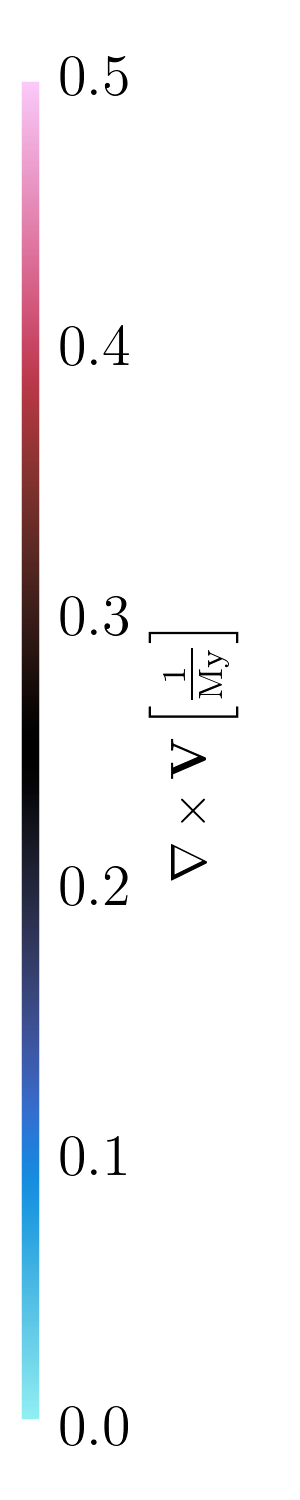}
    \end{subfigure}
    \caption{\label{fig:vorticity} Posterior mean vorticity of the \texttt{main} flow field. 
    A 3D interactive version of this plot is available \href{https://shutsch.github.io/sco_cen_data.html}{online}.}
\end{figure*}

\begin{figure*}
    \centering
    \begin{subfigure}{0.48\textwidth}
        \includegraphics[trim=200 0 80 600, clip, width=\linewidth]{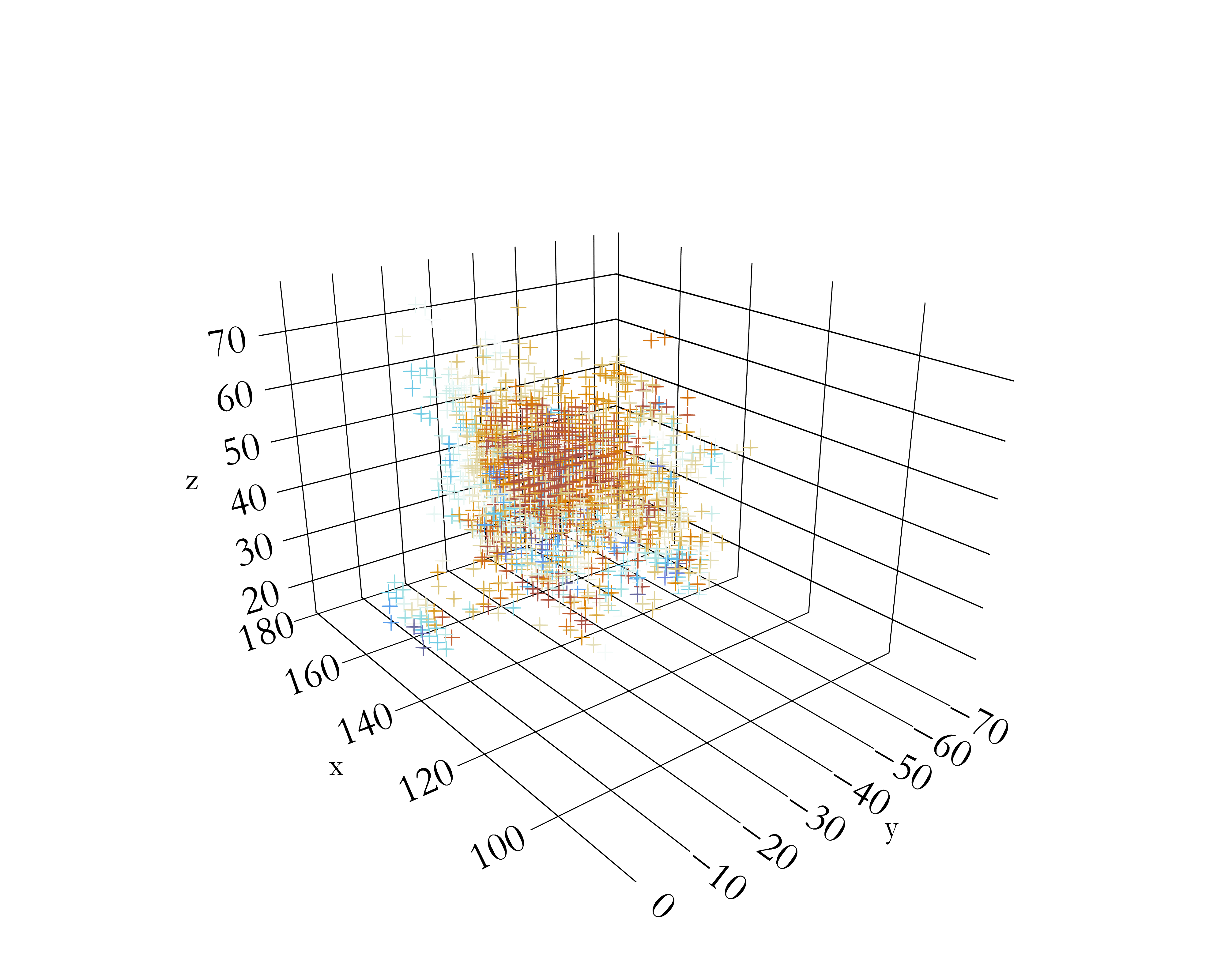}
    \caption{\label{fig:divergence_second} }
    \end{subfigure}
    \begin{subfigure}{0.48\textwidth}
        \includegraphics[trim=200 0 80 600, clip, width=\linewidth]{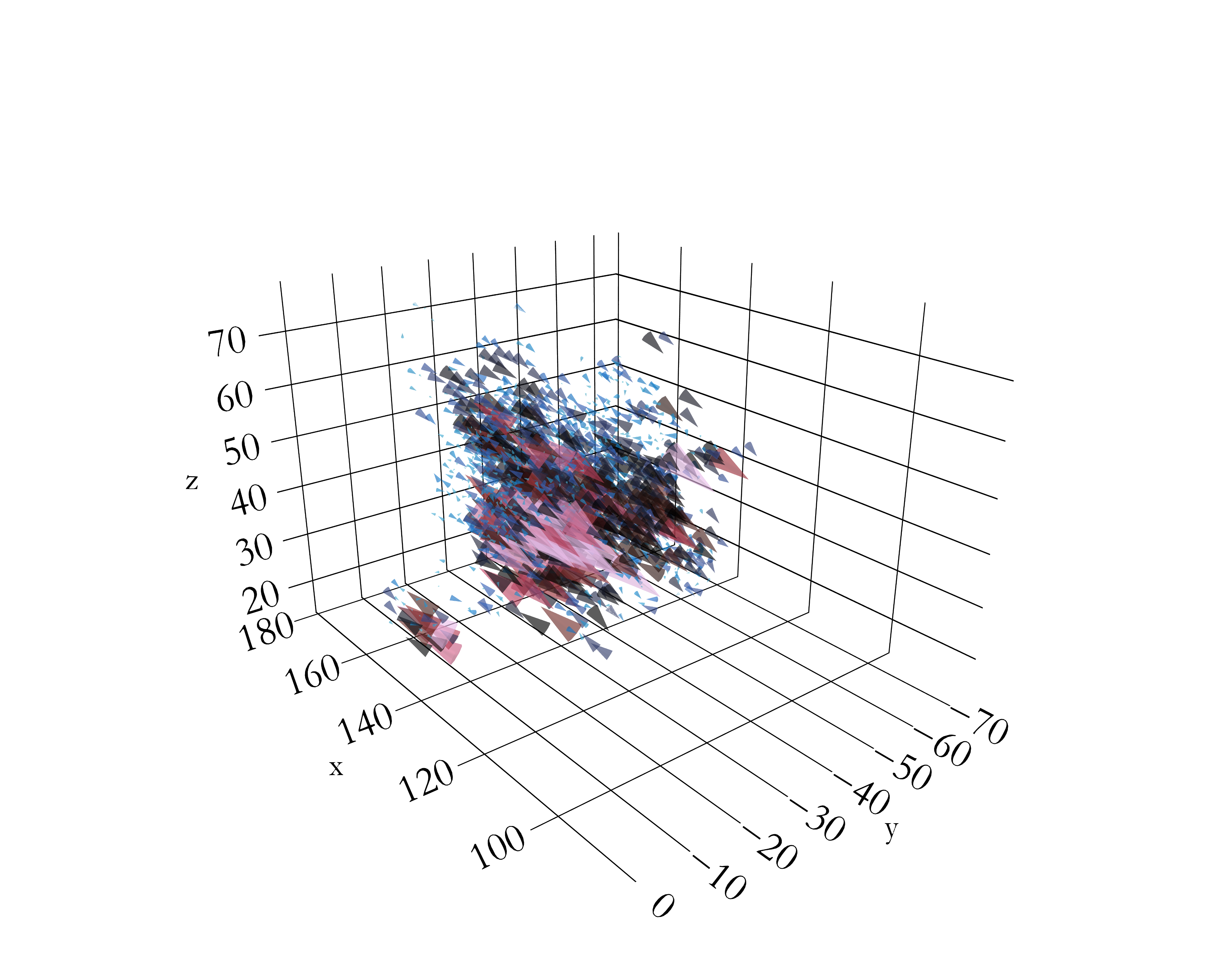}
    \caption{\label{fig:vorticity_second} }
    \end{subfigure}
    \begin{subfigure}{0.48\textwidth}
        \includegraphics[trim=200 0 80 600, clip, width=\linewidth]{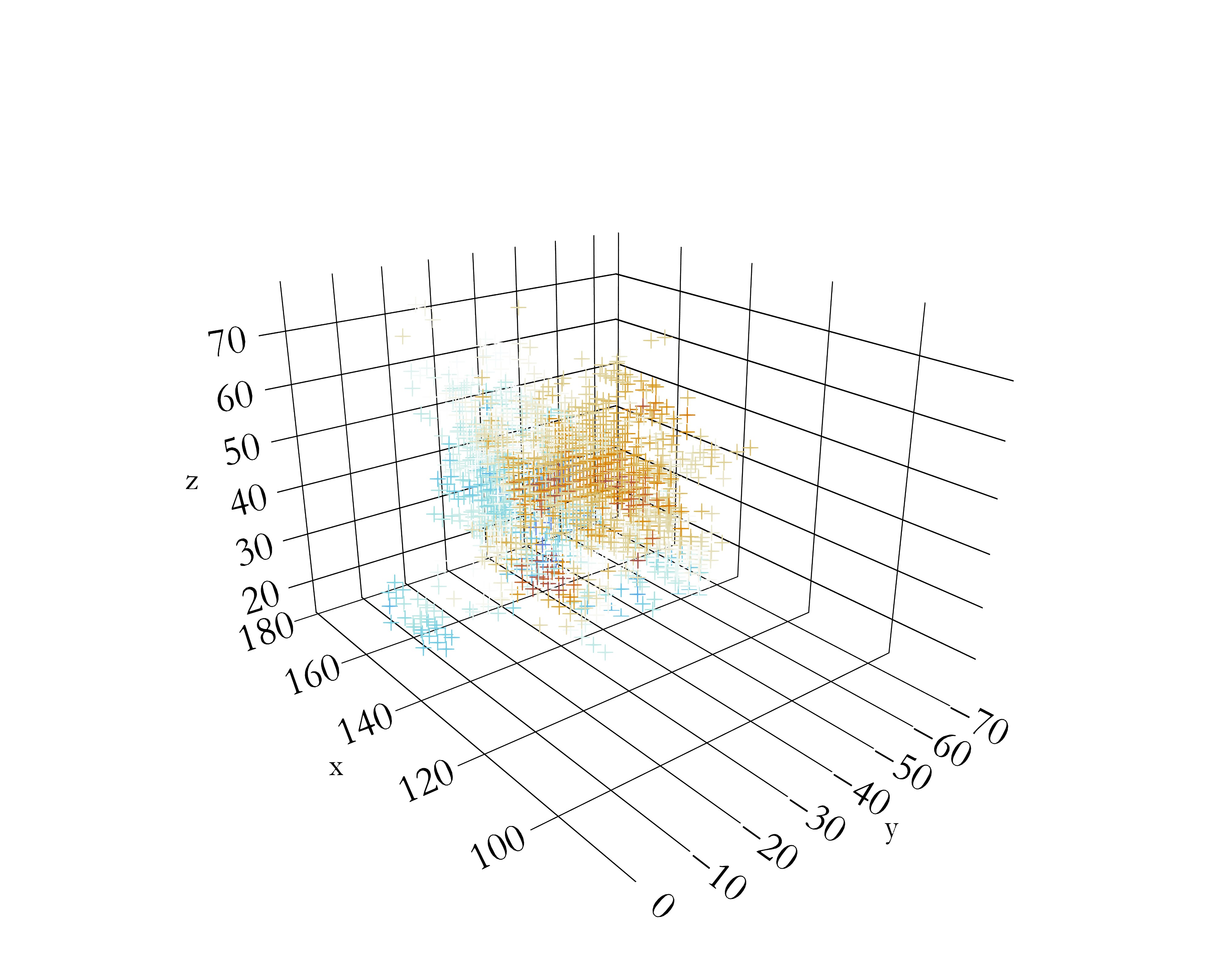}
    \caption{\label{fig:divergence_delta} }
    \end{subfigure}
    \begin{subfigure}{0.48\textwidth}
        \includegraphics[trim=200 0 80 600, clip, width=\linewidth]{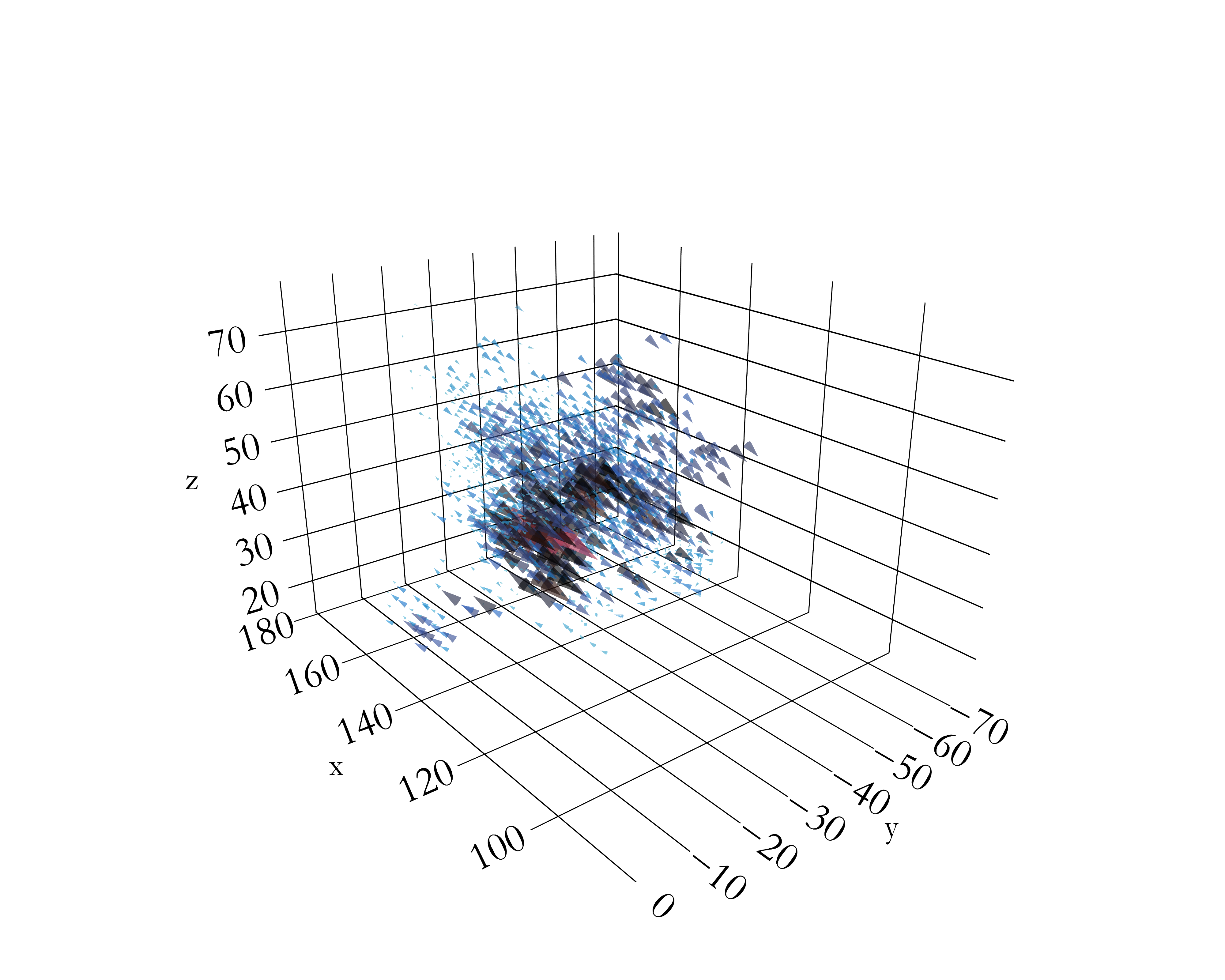}
    \caption{\label{fig:vorticity_delta} }
    \end{subfigure}
    \caption{\label{fig:other_derivatives} Posterior mean divergence vorticity of the \texttt{secondary} and $\delta$ flow fields. 
    The colormaps in these plots are the same as in the respective \texttt{main} field plots.
    3D interactive versions of these plots are available \href{https://shutsch.github.io/sco_cen_data.html}{online}.}
\end{figure*}

We use the RV compilation from \citetalias{2025Grossschedl}, who collected the data from the following 23 spectral surveys or literature catalogs: \citet{1999Wichmann, 2001Joergens, 2006Torres, 2006Gontcharov, 2006Jilinski, 2006James, 2007Guenther, 2011Chen, 2012Biazzo, 2012Dahm, 2012Nguyen, 2012Gilmore, 2013Galli, 2015DeSilva, 2013Murphy, 2017Majewski, 2017Sacco, 2017Kunder, 2017Frasca, 2021Buder, 2021Santana, 2020aSteinmetz, 2022Miret-Roig, 2022Abdurrouf, 2022Jackson, 2023Katz, 2023Fang, 2025MiretRoig}.

In Fig.~\ref{fig:data_S/N} we show the S/N of the used proper motion and RV data. 
This illustrates the high quality of the proper motion data compared to the radial velocities.
 
Figure~\ref{fig:data_distance_error} illustrates the likelihood $\mathcal{P}_{\mathrm{voxel}}$ (as color-scale) for each star, highlighting the probability of the star being the voxel it falls into if using the inverted parallax as a distance estimator. 
The strong heterogeneity of the parallax uncertainty leads to $\mathcal{P}_{\mathrm{voxel}}$ values of almost 0 to almost 1 (i.e., the star is almost certainly in the respective voxel). 

We do not include the clusters labeled `Norma-North`, `Ophiuchus-North-Far`, and `Ophiuchus-South-East` from \citetalias{2023Ratzenboeck1}, as they are likely not connected to the Sco-Cen OB association, but interlopers from adjacent regions based on their kinematic and age properties, following also the analyses of \citetalias{2023Ratzenboeck2} and \citetalias{2025Grossschedl}.
The clusters in the \texttt{main} and \texttt{secondary} selection are summarized in Table~\ref{tab:summary_main_selection} and Table~\ref{tab:summary_secondary_selection}, respectively, including their ages and the number of stars contained in each cluster.

\section{Physical field assumption}
\label{app:field_assumption}

In fitting a flow field to the observed 3D space motion of stars, we take an analogy to fluid dynamics where continuous fields describe the discrete motion of particles with dynamics governed by the continuity equation.
This comes with several implied assumptions, which we try to make explicit in this section.

The first is the continuum hypothesis and the associated notion of smoothness, i.e., the idea that neighboring volume elements will have similar field values, which is an important ingredient to our prior model. 
While the degree of smoothness is uncertain and a global free parameter of the model, it can never be absent, as this is the only way to regularize the many parameters.
Since we work on a discretized grid, we assume that typical structures of the field will have dimensions larger than the voxel size of 3 pc.

Furthermore, the stellar flow is likely inviscid (since star-star interactions are negligible). 
The lack of viscosity implies that stellar flows can overlap easily, in which case a single flow field would at best describe the net flow between volume elements. 
In this case, the field would lose much of its descriptive value as derived quantities such as energy or momentum densities become meaningless, as they only make sense in the absence of co-spatiality of flows.  
This can be remedied by introducing more flow components to the model.

\section{Stellar density} 
\label{app:stellar_density}

We show a kernel density estimation of the stellar number density field in Fig.~\ref{fig:density}.
This field was calculated using the mean Cartesian positions as input to \texttt{scipy.stats.gaussian\_kde} function in python, with the \texttt{bw\_method} parameter set to 0.1. 
The main purpose of this field is the definition of the boundaries of Sco-Cen for visualization; we have refrained from incorporating parallax uncertainties in the estimation. 
In future work, the calculation of, e.g., a stellar momentum field will require dealing with this source of uncertainty.

\section{Prior} 
\label{app:prior}
All three components of both the \texttt{main} and $\delta$ vector fields are modeled as Gaussian random fields with unknown correlation structure, using the model by \citet{2022Arras}.  
The hyperparameters are listed in Table~\ref{tab:prior_parameters}. 
These describe the prior assumptions for the correlation structure of Gaussian random fields representing the three velocity components ($v_x, v_y, v_z$) of the \texttt{main} and \texttt{add} vector fields in the barycentric velocity frame. 
The `offset' parameters specify the expected value of the global mean of the component fields, where the \texttt{main} field shows a significant non-zero prior offset centered on the \citetalias{2025Grossschedl} reference frame, while the \texttt{add} field is centered at zero. 
The `offset std'  reflects the prior uncertainty around the offset mean, and is set uniformly across components at $15\pm15$~km\,s$^{-1}$, implying a broad and weakly informative prior that encompasses the frame difference between the barycenter and Sco-Cen.
The `fluctuations' term models the standard deviation of the variations around the mean, also set with a wide uncertainty band of $25\pm25$~km\,s$^{-1}$, capturing the magnitude of random local deviations, and easily covering the expected velocity range of Sco-Cen \citepalias[e.g.,][]{2025Grossschedl}.

The remaining parameters shape the spectral behavior of the Gaussian random fields. 
`loglogavgslope' determines the slope of the power spectrum in log-log space, with steeper slopes correspond to smoother fields. 
We choose a rather smooth a-prior value of $7.3$, with a standard deviation of 3, chosen souchthat a wide range of slope values is covered within the 1-$\sigma$ bound. 
`asperity' controls the degree of spikiness in deviations from the power-law, while `flexibility' regulates the amplitude of these variations. 
Both are given non-informative priors centered at 0.5 with wide uncertainties of 0.5, reflecting agnosticism about how structured or smooth the power spectra might be. 
Together, these parameters define a flexible yet physically plausible model for the underlying spatial structure in the vector fields, following the hierarchical approach of \citet{2022Arras}.

\section{Details on the likelihood derivation}
\label{app:likelihood}

Here we explicitly outline several calculation steps that were omitted in the description of the likelihood in Sect.~\ref{subsubsec:likelihood}. 
We focus on a subset of stars with unobserved RVs and uncertain distance estimates, since this case allows us to demonstrate the necessary steps while still allowing for a more compact notation. 
The case including RVs is completely analogous. 
A large part of this calculation follows roughly similar approaches as in \citet{2019Leike} and \citet{2023Edenhofer} and is quite general.

Following the notation laid out in Sect.~\ref{sec:method}, we denote the likelihood variables, in this case the proper motion data set, as $\mu$, and the observed parallaxes as $\varpi$.
To connect the velocity field and data in the following calculation, we introduce the true proper motions 
\begin{equation}
    \label{eq:true_mu}
    \widetilde{\mu} = \mu_\star(\mathbf{V}, \widetilde{\varpi})
\end{equation}
i.e. the true velocity field $\mathbf{V}$ evaluated at the true distances $1/\widetilde{\varpi}$. 
The true proper motions and parallaxes are related to the data via 
\begin{align}
    \label{eq:true_data_mu}
    \mu &= \widetilde{\mu} + n_\mu = \mu_\star(\mathbf{V}, \widetilde{\varpi}) + n_\mu \\\label{eq:true_data_varpi}
    \varpi &= \widetilde{\varpi} + n_\varpi,
\end{align}
with correlated noise terms $n_\mu$ and $n_\varpi$, with observational covariance $C_{\mu, \varpi}$.
Expanding the likelihood in all of the (unknown) quantities introduced in Eqs.~\eqref{eq:true_mu}, \eqref{eq:true_data_mu}, and \eqref{eq:true_data_varpi} gives:
\begin{align}
    & \mathcal{P}\left(\mu | \vec{V}, \varpi, C_{\mu, \varpi} \right) = \nonumber \\ 
    &= \int \mathrm{d}\widetilde{\varpi}\, \mathrm{d}n_\mu\, \mathrm{d}n_\varpi\, \mathcal{P}\left(\mu, \widetilde{\varpi}, n_\mu, n_\varpi  | \vec{V}, \varpi, C_{\mu, \varpi}  \right) = \nonumber
\end{align}
We refactor the joint probability density function above piece by piece using the product rule.
We start by exploiting the fact that the proper motion data are fully explained by Eq.~\eqref{eq:true_data_mu}  
\begin{align}
    &= \int \mathrm{d}\widetilde{\varpi}\,\mathrm{d}n_\mu\,\mathrm{d}n_\varpi\, \mathcal{P}\left(\mu |  \vec{V}, \widetilde{\varpi}, n_\mu\right)\,
    \,\mathcal{P}\left(\widetilde{\varpi}, n_\mu, n_\varpi  | \vec{V}, \varpi, C_{\mu, \varpi}  \right) =\nonumber  
\end{align}
and similarly using Eq.~\eqref{eq:true_data_varpi} for the parallaxes
\begin{align}
    &= \int \mathrm{d}\widetilde{\varpi}\,\mathrm{d}n_\mu\,\mathrm{d}n_\varpi\, \mathcal{P}\left(\mu |  \vec{V}, \widetilde{\varpi}, n_\mu\right)\,
     \mathcal{P}\left(\widetilde{\varpi} | n_{\varpi}, \varpi \right) \,\mathcal{P}\left(n_\mu, n_\varpi  | C_{\mu, \varpi}  \right)
\end{align}
Since the only information we have on the noise statistics is the covariance, the appropriate noise prior distribution to choose is a bi-variate zero-centered Gaussian. 
This distribution maximizes the information entropy in case one only has information on the first and second moments, i.e., it makes the least additional assumptions \citep{2019Ensslin}.
The first two distributions are delta distributions, since they are completely determined by Eqs.~\eqref{eq:true_data_mu} and \eqref{eq:true_data_varpi}.
This allows us to perform the marginalization over the noise terms analytically, giving us
\begin{align}
        &\mathcal{P}\left(\mu | \vec{V}, \varpi, C_{\mu, \varpi} \right) = \int \mathrm{d}\widetilde{\varpi}\, \mathcal{G}\left(\begin{array}{l}
             \mu - \mu_\star(\mathbf{V}, \widetilde{\varpi}) \\
             \widetilde{\varpi} - \varpi
        \end{array},  C_{\varpi, \mu}   \right)  
\end{align}
The final step, namely the marginalization over $\widetilde{\varpi}$, is not analytically tractable in general, since the variation of $\mu_\star$ with distance depends on the unknown structure of the velocity field along the line of sight.  
We hence introduce the following numerical approximation

\begin{align}
    \mathcal{P}\left(\mu | \vec{V}, \varpi, C_{\mu, \varpi} \right) &= \int \mathrm{d}\widetilde{\varpi}\,  \mathcal{P}\left(\mu, \widetilde{\varpi}| \vec{V}, \varpi, C_{\mu, \varpi} \right) \nonumber \\  &\approx \sum_i \mathcal{P}\left(\mu | \widetilde{\varpi}_i, \vec{V}, \varpi, C_{\mu, \varpi} \right)   \nonumber \\  &\approx  
    \mathcal{G}\left(
             \mu - \langle \mu_\star\rangle_\varpi  ,  C_{\mu}  +  \langle C_{\mu}  \rangle_{\varpi}   \right)  
\end{align}

with 

\begin{align}
     \langle \mu_\star\rangle_\varpi &=  \frac{1}{N} \sum_i^{N} \mu_\star(\varpi_i) \\
     \langle C_{\mu}  \rangle_{\varpi} &= \begin{pmatrix} \sigma^2_{\mu_{\mathrm{RA}} | \varpi}   & \sigma^2_{\mu_{\mathrm{RA}}, \mu_{\mathrm{DEC}}  | \varpi}   \\ * & \sigma^2_{\mu_{\mathrm{DEC}} | \varpi}  \end{pmatrix} \\
     \sigma^2_{x, y | \varpi}  &= \frac{N}{N - 1} \left(\langle  x_\star y_\star \rangle_\varpi - \langle x_\star\rangle_\varpi\langle y_\star\rangle_\varpi\right)
\end{align}    
These sample estimates were calculated from 10 parallax samples per star. 
We note that this estimation needs to be evaluated at each step of the optimization, making this the main computational bottleneck.
The number of parallax samples was chosen to strike a balance between computational feasibility and statistical precision.

\section{Noise estimation results}
\label{app:noise_estimation}

\begin{figure}
    \centering
    \includegraphics[width=0.99\linewidth]{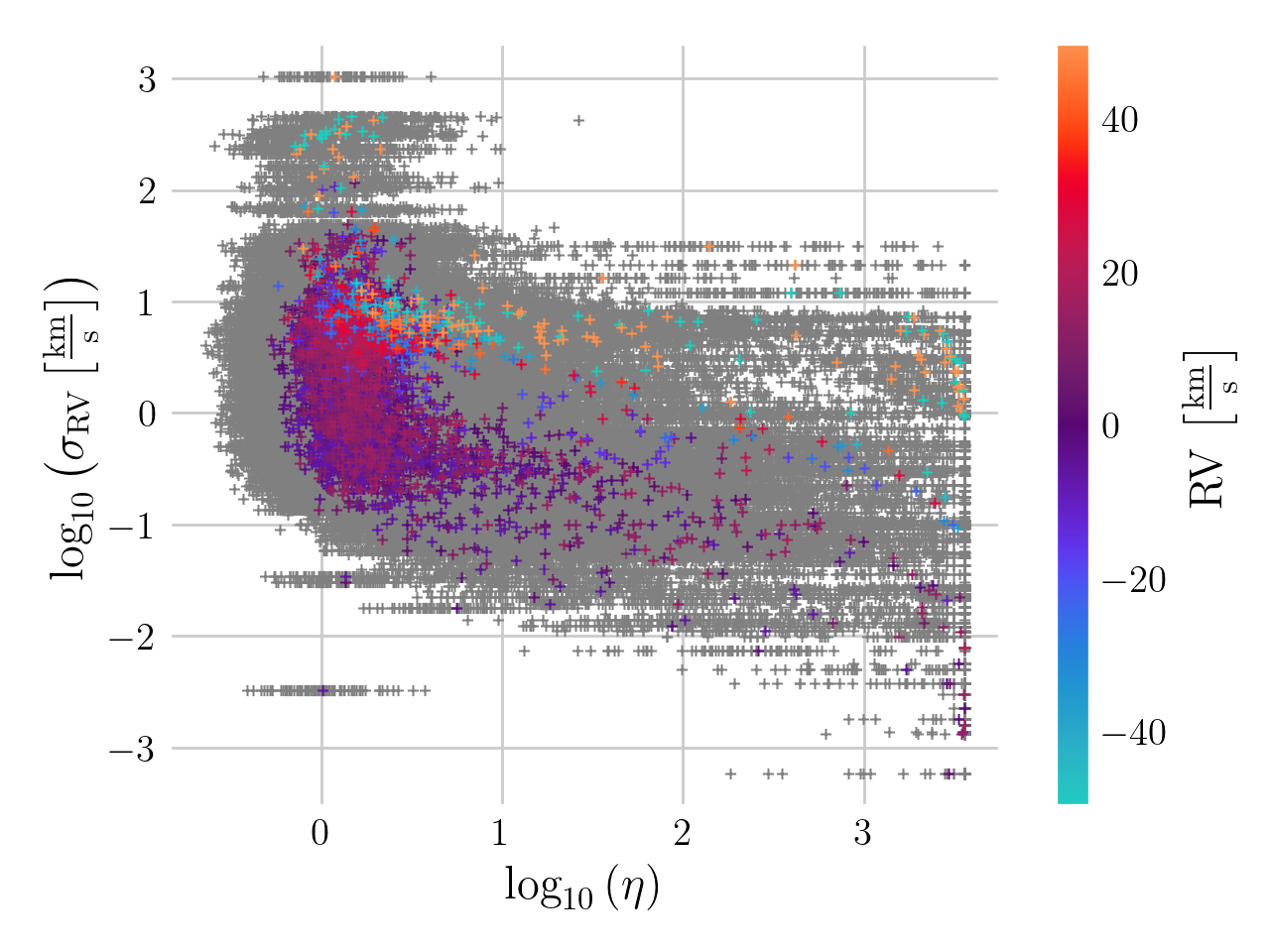}
    \caption{Noise estimation results. The plot shows the inferred noise estimation factors $\eta$ per RV measurement versus the log-scaled observed uncertainties.
    The observed RV data are shown as a color scale. 
    The scale is saturated at $\pm 50\, \mathrm{km\,s^{-1}}$.
    The posterior samples of $\eta$ are in gray, thereby illustrating the statistical uncertainties of the result.}
    \label{fig:noise}
\end{figure}

As discussed in Sect.~\ref{subsubsec:likelihood}, we have introduced a noise estimation factor $\eta$ to the observational noise to correct for the possible systematic error introduced by binaries.
In Fig.~\ref{fig:noise}, we relate the inferred $\eta$ factors with the observed RVs and the respective observational noise. 
The plot reveals several interesting patterns. 
For most stars, the noise statistics do not change, i.e., they have an $\eta$ value of the order of 1. 
Data points with very to extremely small uncertainties (from about $0.5~\mathrm{km\,s^{-1}}$ to $ 0.001~\mathrm{km\,s^{-1}}$) are almost always down-weighted, indicating that the RV values are inconsistent with the surrounding RV field at the precision given by the noise. 
However, these data points are still informative, as the $\eta$ factors only bring these error bars to the level of the bulk of the RV measurements in Sco-Cen.  
Most stars have an observational noise between $0.5~\mathrm{km\,s^{-1}}$ to $10~\mathrm{km\,s^{-1}}$. 
In this regime, we see an increasing trend in that stars with high $\eta$ values also have very high absolute RV values, which likely indicates that they are kinematic outliers. 
This gives good evidence that the automatic down-weighting using the noise estimation works.
In the high noise regime, almost no stars are down-weighted with high $\eta$ values. 
These stars likely have almost no impact on the analysis due to the already small likelihood value, making further outlier rejection unnecessary.

\section{Component fields} \label{app:components}

We show the component fields of the \texttt{main} vector field in Figs.~\ref{fig:x_component}, \ref{fig:y_component} and \ref{fig:z_component}.  
These plots give a supplementary viewpoint to the power spectrum plots in Fig.~\ref{fig:power} and the full vector field plotted in Fig.~\ref{fig:flow_field_main}. 
The inside-out acceleration pattern of Sco-Cen is again nicely visible in all components.

\section{Additional vorticity and divergence fields}
\label{app:additional_derivatives}

Figure~\ref{fig:vorticity} shows the vorticity of the \texttt{main} flow field of Sco-Cen. 
This plot reveals many small-scale structures.
In general, it can be noted that many of the vorticity vectors seem to be parallel or anti-parallel to the direction connecting negative $x$, $y$, $z$, with the more positive values on the respective axes, which, at least for $x$ and $y$, is the direction of Galactic rotation. 
A more detailed analysis of the vorticity will follow in Paper~III.
We furthermore show the divergence and vorticity maps of the \texttt{secondary} and $\delta$ fields in Fig.~\ref{fig:other_derivatives}.

\end{document}